\documentclass[twocolumn]{aastex701}

\usepackage{natbib}
\makeatletter
\newcommand*{\rom}[1]{\expandafter\@slowromancap\romannumeral #1@}
\usepackage[version=4,arrows=pgf-filled]{mhchem}
\makeatother
\usepackage{graphicx}
\usepackage{txfonts}
\usepackage{siunitx}
\usepackage{xcolor}
\usepackage{enumitem}
\usepackage{subfigure}
\usepackage{mathrsfs}
\usepackage{booktabs}
\usepackage{amsmath}
\usepackage{bm}
\usepackage{morefloats}
\usepackage{multirow}
\usepackage{gensymb}
\usepackage{threeparttable} 

\newcommand{\xray}{\hbox{X-ray}}
\newcommand{\aox}{$\alpha_{\rm OX}$}
\newcommand{\ltkf}{$L_{\rm 2500~\textup{\AA}}$} 

\newcommand{\Rmnum}[1]{\expandafter\@slowromancap\romannumeral #1@}

\shorttitle{AASTeX v6.3.1 Sample article}
\shortauthors{Huang et al.}

\graphicspath{{./}{draft_figures/}}

\begin{document} 
\title{Strong \xray\ Variability of I~Zwicky~1: Obscuration from Clumpy Accretion-Disk Winds}

\author[orcid=0000-0002-9335-9455]{Jian Huang}
\affiliation{School of Astronomy and Space Science, Nanjing University, Nanjing, Jiangsu 210093, People's Republic of China; bluo@nju.edu.cn}
\affiliation{Key Laboratory of Modern Astronomy and Astrophysics (Nanjing University), Ministry of Education, Nanjing 210093, People's Republic of China}
\email{jhuang@nju.edu.cn}

\author[orcid=0000-0002-9036-0063]{Bin Luo}
\affiliation{School of Astronomy and Space Science, Nanjing University, Nanjing, Jiangsu 210093, People's Republic of China; bluo@nju.edu.cn}
\affiliation{Key Laboratory of Modern Astronomy and Astrophysics (Nanjing University), Ministry of Education, Nanjing 210093, People's Republic of China}
\email{bluo@nju.edu.cn}

\author[orcid=0000-0002-0167-2453]{W. N. Brandt}
\affiliation{Department of Astronomy \& Astrophysics, 525 Davey Lab,
	The Pennsylvania State University, University Park, PA 16802, USA}
\affiliation{Institute for Gravitation and the Cosmos,
	The Pennsylvania State University, University Park, PA 16802, USA}
\affiliation{Department of Physics, 104 Davey Lab, The Pennsylvania State University, University Park, PA 16802, USA}
\email{ }


\author[orcid=0000-0001-6947-5846]{Luis C. Ho}
\affiliation{Kavli Institute for Astronomy and Astrophysics, Peking University, Beijing 100871, People's Republic of China}
\affiliation{Department of Astronomy, School of Physics, Peking University, Beijing 100871, People's Republic of China}
\email{ }

\author[orcid=0000-0002-8577-2717]{Qingling Ni}
\affiliation{Max-Planck-Institut f\"{u}r extraterrestrische Physik (MPE), Gie{\ss}enbachstra{\ss}e 1, D-85748 Garching bei M\"ucnchen, Germany}
\email{ }

\begin{abstract}
    Obscuration from clumpy \hbox{accretion-disk} winds has been invoked to explain the extreme \xray\ weakness and \xray\ variability observed in a substantial fraction of \hbox{super-Eddington} accreting quasars.
    We present a comprehensive study of the strong \xray\ variability of the \hbox{super-Eddington} accreting active galactic nucleus (AGN) I~Zwicky~1 (I~Zw~1), a prototypical \hbox{narrow-line} Seyfert 1 galaxy (NLS1), with the aims of testing the \hbox{disk-wind} obscuration scenario as the underlying mechanism and characterizing the \hbox{disk-wind} absorber properties.
    We focus on spectral and temporal analyses of simultaneous \hbox{XMM-Newton} and NuSTAR observations in 2020, and a \hbox{100-day} NICER monitoring campaign in 2022.
    Despite strong \xray\ variability by factors of $\approx3$ and $\approx6$ on \hbox{short-term} and \hbox{long-term} timescales, respectively, the \hbox{XMM-Newton} Optical Monitor observations do not show contemporaneous significant UV variability, and archival data reveal only mild \hbox{long-term} optical/infrared variability ($\approx30\%$), indicating a stable accretion process in I~Zw~1.
    The strong \xray\ variability thus likely arises from variable absorption of relatively stable coronal emission.
    We perform \hbox{time-resolved} \xray\ spectroscopy utilizing a \hbox{partial-covering} absorption model with a stable corona and  varying ionized absorbers.
    We identify three distinct absorbers whose variations in the column density and covering factor successfully explain the observed \xray\ ``flares'' in 2020 and the \hbox{longer-term} spectral evolution in 2022.
    Our results support a unified scenario in which obscuration from clumpy disk winds produces the strong \xray\ variability observed in \hbox{super-Eddington} accreting AGNs.
    This scenario may be applicable to other NLS1s exhibiting strong \xray\ variability to better characterize the disk winds driven by \hbox{super-Eddington} accretion.
\end{abstract}

\keywords{\uat{High Energy astrophysics}{739} --- \uat{Quasars}{1319} --- \uat{X-ray Active Galactic Nuclei}{2035}}

\section{Introduction} \label{sec:intro}
Active galactic nuclei (AGNs) are powered by accretion onto supermassive black holes (SMBHs). 
Their optical/UV emission is dominated by thermal radiation from the accretion disk (e.g., \citealt{Shakura1973}), 
while the \xray\ emission originates primarily in the accretion-disk corona through inverse Compton scattering of the optical/UV photons (e.g., \citealt{Sunyaev1980,Haardt1993}).
Observations have revealed significant correlations between the \xray\ and optical/UV radiation in \hbox{radio-quiet} type 1 AGNs, such as the negative correlation between the \hbox{\xray-to-optical} \hbox{power-law} slope parameter ($\alpha_{\rm OX}$)\footnote{$\alpha_{\rm OX}$ is defined as $\alpha_{\rm OX}=-0.3838~{\rm log}(f_{2500~\textup{\AA}}/f_{\rm 2~keV})$, where $f_{2500~\textup{\AA}}$ and $f_{\rm 2~keV}$ are the flux densities at \hbox{2500~\AA} and \hbox{2 keV}, respectively.} and 2500~\AA\ monochromatic luminosity ($L_{2500~\textup{\AA}}$; e.g., \citealt{Steffen2006, Just2007, Vagnetti2013, Chiaraluce2018, Pu2020, Huang2025}) and the non-linear correlation between the \xray\ and optical/UV monochromatic luminosities ($L_{\rm 2~keV}\textrm{--}L_{2500~\textup{\AA}}$; e.g., \citealt{Risaliti&Lusso2019}).
These correlations suggest a physical connection between the accretion disk and the corona.
Deviations from these correlations occasionally occur in \hbox{radio-quiet} type 1 AGNs, typically shown as weaker-than-expected \xray\ emission (e.g., \citealt{Luo2015,Nardini2019,Pu2020,Timlin2020,Laurenti2022,Ni2022}).
\xray\ deficits often arise from obscuration by \hbox{dust-free} absorbers that minimally affect optical/UV radiation.
An extreme example is the weak \hbox{emission-line} quasars (WLQs), which are characterized by exceptionally weak or absent broad UV emission lines \citep[e.g.,][]{Fan1999, Diamond-Stanic2009, Plotkin2010, Shemmer2010}.
The fraction of \xray\ weak objects among WLQs reaches $\approx50\%$ \citep[e.g.,][]{Luo2015, Ni2018, Ni2022}.
They are considered to have substantial \xray\ absorption from a thick accretion disk and/or its associated disk wind.
Alternatively, intrinsic \xray\ weakness has been proposed for some \hbox{super-Eddington} accreting AGNs showing extreme \xray\ weakness but no obscuration signatures (e.g., \citealt{Leighly2007b, Leighly2007a, Dong2012, Laurenti2022, Trefoloni2023}).
The ``little red dots'' (LRDs) recently discovered by the James Webb Space Telescope (JWST) at $z\gtrsim4$ are almost universally \xray\ weak \citep[e.g.,][]{Maiolino2024, Yue2024}, which is considered due to either \hbox{Compton-thick} obscuration or intrinsic \xray\ weakness \citep[e.g.,][]{Inayoshi2025, Maiolino2025}.

X-ray variability is a characteristic feature of AGNs.
In \hbox{radio-quiet} type 1 AGNs, typical \xray\ flux variability amplitudes range from $\approx20\%$ to $50\%$, generally attributed to instabilities or fluctuations in the accretion disk and corona \citep[e.g.,][]{Ulrich1997, McHardy2006, Kelly2011, Gibson2012, Yang2016, Paolillo2017, Zheng2017, Kara2025, Paolillo2025}.
However, there is a growing sample of type 1 quasars that exhibit extreme (maximum flux variability amplitudes $f_{\rm var}\gtrsim10$) and sometimes rapid (down to hours) \xray\ variability, including sources like PHL~1092 (\citealt{Miniutti2012}), SDSS J075101.42+291419.1 (\citealt{Liu2019}), SDSS J153913.47+395423.4 (\citealt{Ni2020}), SDSS J135058.12+261855.2 (hereafter SDSS J1350+2618; \citealt{Liu2022}), SDSS J081456.10+532533.5 (hereafter SDSS J0814+5325; \citealt{Huang2023}), and SDSS J152156.48$+$520238.5 (\citealt{Wang2024}).
These quasars transition between \xray\ nominal-strength states and multiple \xray\ weak states in the \hbox{\aox--\ltkf} plane.
Crucially, they lack contemporaneous optical/UV continuum or \hbox{emission-line} variability, and they also lack strong \hbox{long-term} infrared (IR) variability, suggesting relatively stable accretion rates.
They typically exhibit high or \hbox{super-Eddington} accretion rates.
Some of these quasars display apparently absorbed \xray\ spectra (i.e., small effective \hbox{power-law} photon indices) in their \xray\ weak states.
Together, these characteristics support a scenario of obscuration from \hbox{small-scale} \hbox{accretion-disk} winds.
\hbox{High-velocity}, \hbox{high-density}, and clumpy winds may be radiatively driven from the thick disk in \hbox{super-Eddington} accreting systems \citep[e.g.,][]{Proga2000, Jiang2014, Sadowski2014, Jiang2019, Nomura2020, Hu2025}.
In principle, spectral analysis of the \hbox{multi-epoch} \xray\ spectra may reveal properties of the \xray\ absorbers.
However, the limited photon statistics of the \hbox{weak-state} spectra of these quasars prevent detailed spectral modeling, and simplified models (e.g., one neutral absorber) are often able to describe the observed spectra reasonably well.

In the local universe, AGNs with high accretion rates may also exhibit strong \xray\ variability ($f_{\rm var}\gtrsim5$) without contemporaneous optical/UV variability. 
These variability amplitudes may be not as extreme as those of higher mass/luminosity quasars that exhibit extreme \xray\ variability ($f_{\rm var}\gtrsim10$), but mechanisms beyond typical disk/corona instability are still required for interpretation.
Many of these AGNs are \hbox{narrow-line} Seyfert 1 galaxies (NLS1s), which are considered to have \hbox{super-Eddington} accretion rates in general \citep[e.g.,][]{Boller1996, Boller2021, Parker2021, Jin2022}.
Historically, studies of their \xray\ variability focused on the \hbox{X-rays} themselves, without considering the relative \xray\ emission strengths (e.g., in the \hbox{\aox--\ltkf} plane) and the lack of simultaneous strong optical/UV/IR variability.
A frequently adopted model is the relativistic reflection model in a \hbox{lamp-post} setup.
The \xray\ variability is driven by changes in the coronal height (controlling the reflection strength) as well as the normalization of the intrinsic coronal \hbox{power-law} continuum.
The freely varying latter parameter actually contradicts the stable accretion process inferred from the stable optical/UV/IR light curves.
On the other hand, several studies have proposed that \hbox{partial-covering} absorption could be an alternative interpretation of the \xray\ spectra \citep[e.g.][]{Tanaka2004, Turner2009, Midooka2023}, although the normalization of the intrinsic \hbox{power-law} continuum is still allowed to vary between the \hbox{multi-epoch} spectra in these studies.
An additional advantage of the \hbox{partial-covering} absorption scenario is that it unifies the \hbox{strong--extreme} \xray\ variability of local NLS1s and their higher mass/luminosity quasar counterparts that we have been studying extensively.
Compared to the more distant \hbox{super-Eddington} accreting quasars, these local strongly \xray\ variable AGNs are much brighter.
Therefore, additional insights into the absorber properties may be obtained via spectral analyses; the varying obscuration may also reveal dynamics of the absorbers.

Among the NLS1s exhibiting strong \xray\ variability, I~Zwicky~1 (I~Zw~1; $z=0.0611$) presents a compelling case.
I~Zw~1, also known as \hbox{PG~0050+124} or \hbox{Mrk~1502}, is a prototypical NLS1.
With optical \ion{Fe}{2} emission equivalent width (EW) of $\approx75~\AA$ and [\ion{O}{3}] $\lambda5007$ EW of $\approx22~\AA$ \citep{Boroson1992}, I~Zw~1 displays typical Eigenvector 1 features for NLS1s, i.e., showing stronger optical \ion{Fe}{2} emission (above the 75th percentile) and weaker [O III] emission (below the 32nd percentile) compared to SDSS quasars with similar luminosities \citep{Wu2022}.
Its \ion{C}{4} $\lambda1549$ line (EW$\approx29~\AA$; \citealt{Laor1997}) is also weak (below the 6th percentile), similar to WLQs.
Previous reverberation mapping (RM) observations revealed that I~Zw~1 is a \hbox{super-Eddington} accreting AGN, with a SMBH mass of $M_{\rm BH}={9.3}^{+1.26}_{-1.38}\times10^{6}~M_{\odot}$ \citep{Huang2019} and an Eddington ratio of $\lambda_{\rm Edd}\sim2.7$ ($\lambda_{\rm Edd}=L_{\rm bol}/L_{\rm Edd}$, assuming $L_{\rm bol}=10L_{\rm 5100~\textup{\AA}}$).
There are extensive archival \xray\ observations of I~Zw~1, including four deep \hbox{XMM-Newton} \citep{Jansen2001} exposures in 2002, 2005, 2015, and 2020 with effective exposures of 20--140~ks per epoch and one NuSTAR \citep{Harrison2013} observation in 2020.
I~Zw~1 exhibited strong \hbox{long-term} \xray\ variability with $f_{\rm var}\approx6$ across the four \hbox{XMM-Newton} observation epochs \citep[e.g.,][]{Gallo2007, Silva2018, Wilkins2021, Wilkins2022, Rogantini2022, Ding2022}.
During its 2020 \hbox{XMM-Newton} observation, I~Zw~1 exhibited strong \hbox{short-term} variability: the \hbox{0.3--10~keV} light curves displayed two \hbox{flare-like} events with $f_{\rm var}\approx3$, accompanied by complex spectral variations evidenced by significant changes in the \xray\ hardness ratio \citep[e.g.,][]{Wilkins2022}.

Based on the characteristics of I~Zw~1 described above, we consider the possibility that the observed \xray\ flares and variability are explained by obscuration from clumpy disk winds.
In this study, we reanalyze the archival 2020 \hbox{XMM-Newton} and NuSTAR observations of I~Zw~1 to interpret its strong \xray\ variability within a pure obscuration scenario, explicitly incorporating the \hbox{$\alpha_{\rm OX}$--$L_{\rm 2500~\textup{\AA}}$} relation.
We also present a \hbox{100-day} monitoring campaign from 14 September to 23 December 2022 using the Neutron Star Interior Composition Explorer (NICER; \citealt{Gendreau2016}), and perform a temporal analysis of the resulting data.
We aim to interpret the observed \xray\ variability from \hbox{XMM-Newton}, NuSTAR, and NICER solely via evolving obscuration.
A further goal of this study is to investigate in more detail (compared to the higher mass/luminosity quasar counterparts) the \hbox{disk-wind} absorber properties under the absorption scenario.
We organize our study as follows.
In Section~\ref{sec:observations}, we describe the \xray\ and multiwavelength observations of I~Zw~1.
In Section~\ref{sec:multiwave_property}, we describe the multiwavelength properties.
In Section~\ref{sec:xray_spec_analyses}, we present \xray\ temporal spectral analyses and explain the \xray\ variability utilizing a \hbox{partial-covering} absorption model with a stable corona and three varying ionized absorbers.
In Section~\ref{sec:Discussion}, we examine the \hbox{disk-wind} absorber properties of I~Zw~1, and we discuss unifying the \hbox{strong--extreme} \xray\ weakness and \xray\ variability of \hbox{super-Eddington} accreting AGNs and quasars under the wind obscuration scenario.
We summarize our results in Section~\ref{sec:conclusion_future_work}.
Throughout this paper, we use a cosmology of \hbox{$H_{0}=67.4~{\rm km}~{\rm s}^{-1}~{\rm {Mpc}}^{-1}$}, $\Omega_{\rm M} = 0.315$, and $\Omega_{\Lambda} = 0.685$ \citep{Planck2020}.
The spectral analyses were carried out using XSPEC (v.12.14.1; \citealt{Arnaud1996}).
To evaluate the goodness of fit, we used the $\chi^2$ statistic.
The reported uncertainties were calculated at the $1\sigma$ significance level.

\section{\xray\ and Multiwavelength Observations}\label{sec:observations}

\begin{deluxetable}{llll}
\tablewidth{0pt}
\tablecaption{\xray\ Observation Log}
\tablehead{
\colhead{Observatory}  &
\colhead{Observation}  &
\colhead{Observation} &
\colhead{Exposure} \\
\colhead{ }  &
\colhead{ID}  &
\colhead{Start Time} &
\colhead{Time (ks)} \\
\colhead{(1)} &
\colhead{(2)} &
\colhead{(3)} &
\colhead{(4)}
}
\startdata
NuSTAR & 60501030002 & 2020-01-11 & $463.2$ \\
XMM-Newton & 0851990101& 2020-01-12 & $45.6$ \\
XMM-Newton & 0851990201& 2020-01-14 & $45.9$ \\
NICER & 4565010101--& 2022-09-13--& $0.1$--$2.3$ \\
 & 4565020101& 2022-12-22 &   \\
\enddata
\tablecomments{
Column (1): \xray\ observatory.
Column (2): observation ID.
Column (3): observation start time.
Column (4): cleaned exposure time.
}
\label{tbl:observation}
\end{deluxetable}

\subsection{XMM-Newton Observations}\label{sec:xmm_obs}
The basic information for the two 2020 \hbox{XMM-Newton} observations is presented in Table~\ref{tbl:observation}.
The Science Analysis System (SAS; v21.0.0) was used to reduce the \hbox{XMM-Newton} data.
Following the approach adopted in prior studies of the 2020 \hbox{XMM-Newton} observations \citep[e.g.,][]{Wilkins2021, Wilkins2022, Rogantini2022}, we focus our analysis on the pn data, which provide the highest sensitivity of the three EPIC detectors.
We followed the standard procedure in the SAS Data Analysis Threads\footnote{\url{https://www.cosmos.esa.int/web/xmm-newton/sas-threads}.} to process the EPIC-pn \citep{Struder2001} data.
We obtained calibrated and concatenated event lists using the
task \textsc{epproc} for the pn detector. 
A \hbox{count-rate} threshold of \hbox{0.4 ${\rm cts}~{\rm s}^{-1}$} was used to filter background flares, and the task \textsc{tabgtigen} was used to create \hbox{good-time-interval} files.
The cleaned exposures for the pn data in the two observations are 45.6 ks and 45.9 ks, respectively.

\begin{figure}
\centering
\includegraphics[scale=0.18]{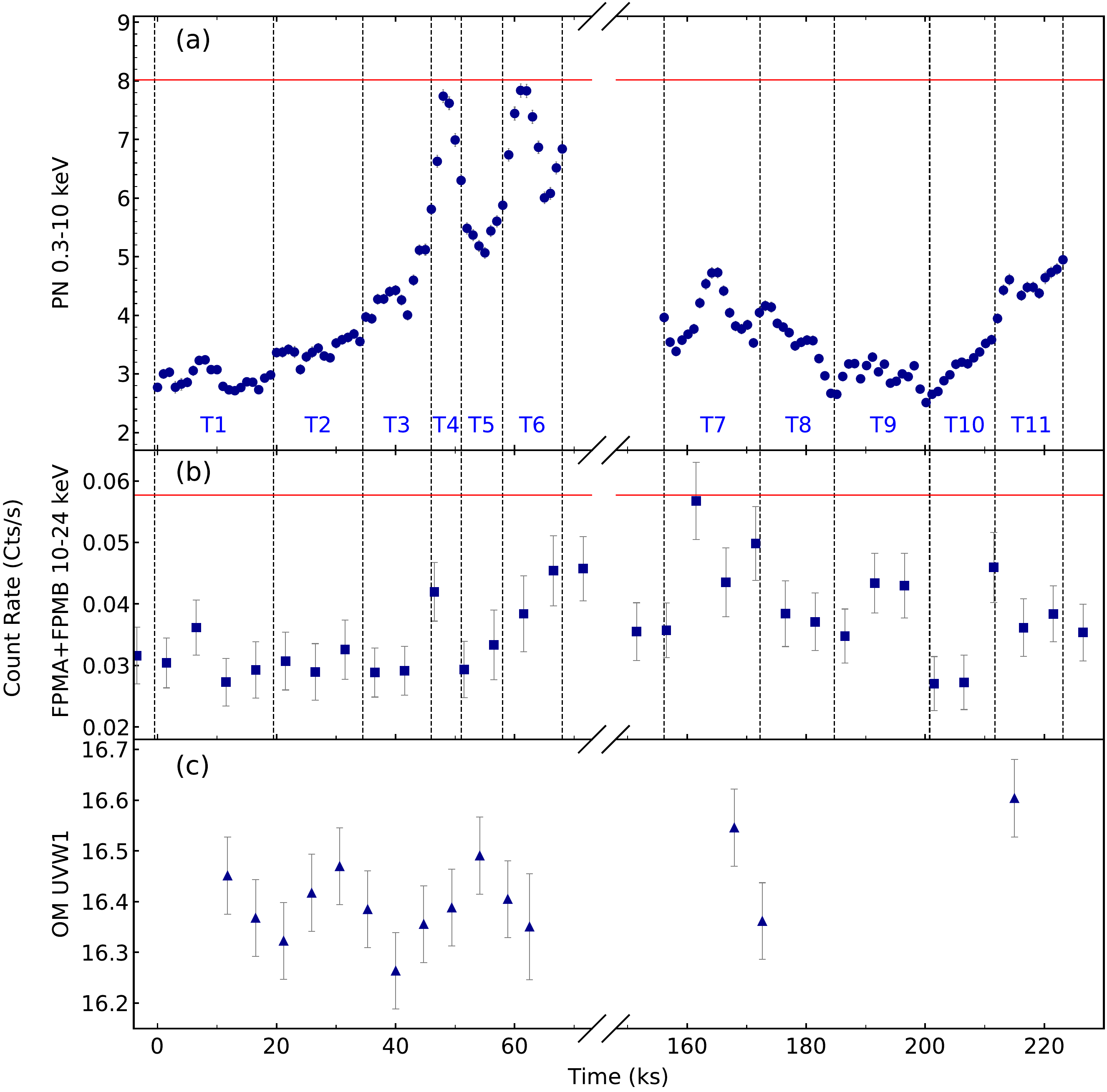}
\caption{Light curves of I~Zw~1 from the 2020 \hbox{XMM-Newton} and NuSTAR observations: (a) EPIC-pn in the \hbox{0.3--10~keV} band, (b) NuSTAR FPMA+FPMB in the \hbox{10--24~keV} band, and (c) OM in the UVW1 band.
The \hbox{XMM-Newton} and NuSTAR light curves have bin sizes of 1~ks and 5~ks, respectively.
The vertical dashed lines mark the 11 segments (\hbox{T1--T11}) where \xray\ spectra were extracted for the \hbox{time-resolved} spectral analysis, with the corresponding segment names labeled in blue.
The OM UVW1 light curve exhibits no significant variability, suggesting stable intrinsic coronal emission and a steady accretion rate.
The red line represent the expected \hbox{0.3--10~keV} and \hbox{10--24~keV} count rates of I~Zw~1, which were estimated from the intrinsic $f_{2500~\textup{\AA}}$ value derived in Section~\ref{sec:uv_ext_sed}, $\Gamma=2.2$, and the \cite{Just2007} \hbox{\aox--\ltkf} relation.
}
\label{fig:lc_xmm_nustar_om}
\end{figure}

We extracted source and background light curves from the two \hbox{XMM-Newton} observations using the task \textsc{evselect} with a \hbox{35\arcsec-radius} circular source region and a \hbox{40\arcsec-radius} circular \hbox{source-free} background region on the same CCD chip.
We used the \textsc{epiclccorr} task to account for corrections of the dead time and exposure variation.
The \hbox{0.3--10~keV} \hbox{background-subtracted} and \hbox{aperture-corrected} light curve is shown in Figure~\ref{fig:lc_xmm_nustar_om}a.
The count rates range from 7.84 to 2.51~${\rm counts}~{\rm s}^{-1}$, with a variability amplitude of $f_{\rm var}\approx3.1$.
We also checked the \hbox{soft-band} (\hbox{0.3--2~keV}) and \hbox{hard-band} (\hbox{2--10~keV}) light curves, and found variability amplitudes of $f_{\rm var}\approx3.3$ and $f_{\rm var}\approx2.9$, respectively.
The slightly stronger variability observed in the soft \xray\ band is consistent with an \hbox{absorption-driven} scenario.
Assuming that I~Zw~1 emits an intrinsically nominal level of \xray\ emission, we estimated an expected \hbox{0.3--10~keV} XMM-Newton pn count rate based on the intrinsic \hbox{rest-frame} 2500~\AA\ flux density derived in Section~\ref{sec:uv_ext_sed} below, a \hbox{power-law} photon index of $\Gamma=2.2$ from the \hbox{$\Gamma$--$\lambda_{\rm Edd}$} relation of \cite{Huang2020}, and the \hbox{\aox--\ltkf} relation from \cite{Just2007}. The expected \hbox{0.3--10~keV} count rate is shown in Figure~\ref{fig:lc_xmm_nustar_om}a as the red line.
Although the observed pn light curve exhibits two \hbox{flare-like} features, the peak count rates are consistent with the expectation from an \xray\ nominal AGN.
This suggests that the two apparent flares may correspond to periods of reduced obscuration.

The substantial photon counts from the two observations allow us to extract \hbox{time-resolved} spectra and model the temporal evolution of the absorbers through spectral fitting.
As shown in Figure~\ref{fig:lc_xmm_nustar_om}, we partitioned the two observations into 11 segments, ensuring that the spectrum in each segment contains at least 20,000 net counts.
Segments T4 and T6 cover the two flares.
Source and background spectra in each segment were extracted using the task \textsc{evselect} with the same source and background regions as employed for the light curve extraction.
We grouped the source spectra to ensure a minimum of 25 counts per bin and to avoid oversampling the intrinsic energy resolution by more than a factor of 3 (i.e., oversample=3 in the \textsc{specgroup} tool).

Five Optical Monitor (OM; \citealt{Mason2001}) optical/UV filters were used in the \hbox{XMM-Newton} observations,\footnote{We present the \hbox{imaging-mode} data analysis here. We have verified that the \hbox{fast-mode} light curves (e.g., Figure 10 of \citealt{Wilkins2022}) are consistent with those from the imaging mode and exhibit similar variability.} including UVW2, UVW1, U, B, and V with effective wavelengths of 2120~\AA, 2910~\AA, 3440~\AA, 4500~\AA, and 5430~\AA, respectively.
In both observations, the numbers of exposures differ across the OM filters, and the UVW1 filter has the longest exposure times.
For the first observation, the OM exposure times range from 2.9~ks to 50.7~ks, while for the second observation, the exposure times range from 2.9~ks to 13.2~ks.
We used the pipeline task \textsc{omichain} to process the OM data of each \hbox{XMM-Newton} observation. 
Source count rates, flux densities, and magnitudes for each filter were extracted from the generated SWSRLI files.
The OM UVW1 \hbox{count-rate} light curve is shown in Figure~\ref{fig:lc_xmm_nustar_om}c, with a small variability amplitude of $f_{\rm var}\approx1.02$.
Therefore, the UVW1 light curve shows no coordinated variability with the \hbox{0.3--10~keV} \xray\ light curve.
For each filter in each observation, we adopted the mean magnitude and flux density of all the exposures.
We then \hbox{de-reddened} the flux densities by adopting the \cite{Fitzpatrick2019} Milky Way \hbox{$R_V$-dependent} extinction model with $R_V=3.1$ and Galactic extinction \hbox{$E(B-V)=0.057$} \citep{Schlegel1998}, and they are listed in Table~\ref{tbl:om_prop}.
The two sets of flux densities differ at the \hbox{$\approx0.4$--$2.6\sigma$} levels, where $\sigma$ is the combined standard deviation of two measurements.
Given the generally larger OM exposure times in the first \hbox{XMM-Newton} observation, we interpolated its UVW2 and UVW1 flux densities to derive an observed $f_{2500~\textup{\AA}}$ value, which was then corrected for the intrinsic extinction derived in Section~\ref{sec:uv_ext_sed} below.

\begin{deluxetable*}{lllllll}
\tablewidth{0pt}
\tablecaption{XMM-Newton OM Optical/UV Measurements}
\tablehead{
\colhead{Observation}  &
\colhead{Observation}  &
\colhead{$f_{\rm UVW2}$} & 
\colhead{$f_{\rm UVW1}$}  &
\colhead{$f_{\rm U}$}  &
\colhead{$f_{\rm B}$}  &
\colhead{$f_{\rm V}$}  \\
\colhead{ID}  &
\colhead{Start Time}  &
\colhead{ } & 
\colhead{ }  &
\colhead{ }  &
\colhead{ }  &
\colhead{ }  \\
\colhead{(1)}  &
\colhead{(2)}  &
\colhead{(3)}  & 
\colhead{(4)}  &
\colhead{(5)}  &
\colhead{(6)}  &
\colhead{(7)}
}
\startdata
0851990101 & 2020-01-12 22:46 & $17.08\pm0.24$ & $31.63\pm0.04$ & $36.10\pm0.16$ & $47.22\pm0.21$ & $79.37\pm0.38$ \\
0851990201 & 2020-01-14 18:08 & $16.62\pm0.35$ & $31.85\pm0.08$ & $36.18\pm0.15$ & $47.98\pm0.21$ & $78.36\pm0.38$ \\
\enddata
\tablecomments{Column (1): observation ID.
Column (2): observation start time.
Columns (3)--(7): Galactic-extinction-corrected \hbox{XMM-Newton} OM flux-density measurements in units of $10^{-27}~{\rm erg}~{\rm cm}^{-2}~{\rm s}^{-1}~{\rm Hz}^{-1}$.}
\label{tbl:om_prop}
\end{deluxetable*}

\subsection{NuSTAR Observation}\label{sec:nustar_obs}
The basic information for the 2020 NuSTAR observation is presented in Table~\ref{tbl:observation}.
For NuSTAR data reduction, we used HEASoft (v6.34; \citealt{Blackburn1995}) and the NuSTAR Data Analysis Software (NuSTARDAS; v2.1.4) with NuSTAR CALDB (Calibration Database) 20240325.
Calibrated event files for the FPMA and FPMB detectors are generated using \textsc{nupipeline}.
We created NuSTAR images in the 3--24~keV band for each detector using the Chandra Interactive Analysis of Observation (CIAO; v4.17) tool \textsc{dmcopy}, and we used the CIAO tool \textsc{wavdetect} \citep{Freeman2002} with a \hbox{false-positive} probability of $10^{-5}$ to search for \xray\ sources and determine the \xray\ position of I~Zw~1 in each image.

We used the \textsc{nuproducts} task to extract source and background light curves in the \hbox{10--24~keV} band, adopting a source region with a radius of 70\arcsec\ centered on the \xray\ position and an annular background region centered on the \xray\ position with an inner radius of 100\arcsec\ and an outer radius of 140\arcsec.
The \hbox{background-subtracted} and \hbox{aperture-corrected} light curve is shown in Figure~\ref{fig:lc_xmm_nustar_om}b.
Assuming that I~Zw~1 is intrinsically \xray\ nominal, we estimated an expected \hbox{10--24~keV} NuSTAR count rate following the same approach as in Section~\ref{sec:xmm_obs}, shown as the red line in Figure~\ref{fig:lc_xmm_nustar_om}b.
The generally lower observed count rates are also suggestive of \xray\ absorption.
We estimate that a neutral hydrogen column density of $N_{\rm H}\approx1.5\times10^{24}~{\rm cm}^{-2}$ (i.e., \hbox{Compton-thick} level) is required to reduce the NuSTAR count rate by a factor of 2, assuming a simple $\Gamma=2.2$ \hbox{power-law} continuum attenuated by intrinsic neutral absorption.

Compared to the \hbox{lower-energy} \hbox{XMM-Newton} light curve, the NuSTAR light curve exhibits a lower variability amplitude ($f_{\rm var}\approx2.1$) and its \xray\ weakness is less pronounced relative to the expected count rate.
Both properties are naturally explained by strong intrinsic \xray\ absorption.
We then examined whether the \hbox{XMM-Newton} and NuSTAR light curves are correlated, as would be expected if the variability in both bands is driven either by global changes in coronal properties or by variations in the column density of a simple uniform neutral absorber.
We ran an interpolated \hbox{cross-correlation} function (ICCF; \citealt{Gaskell1986}, \citealt{Peterson1998}) analysis of the \hbox{XMM-Newton} and NuSTAR light curves shown in Figure~\ref{fig:lc_xmm_nustar_om},\footnote{We also tested multiple binning schemes for the NuSTAR light curve, and the resulting maximum correlation coefficients ($r_{\rm max}$) are all $\lesssim0.5$.} and found no significant correlation between them, with a maximum correlation coefficient $r_{\rm max}=0.51$.
This result favors a scenario of complex, variable obscuration in I~Zw~1, which decouples the soft and hard \xray\ variability.

We extracted NuSTAR source and background spectra with \textsc{nuproducts} using the same regions for extracting the light curve and the same time segments defined for the \hbox{XMM-Newton} pn \hbox{time-resolved} spectra.
For each segment, we merged the source spectra, background spectra, and response files for FPMA and FPMB using the HEASoft tool \textsc{addspec}.
We grouped the merged source spectra to ensure a minimum of 25 counts per bin using the \textsc{ftgrouppha} tool.

\begin{figure}
\centering
\includegraphics[scale=0.32]{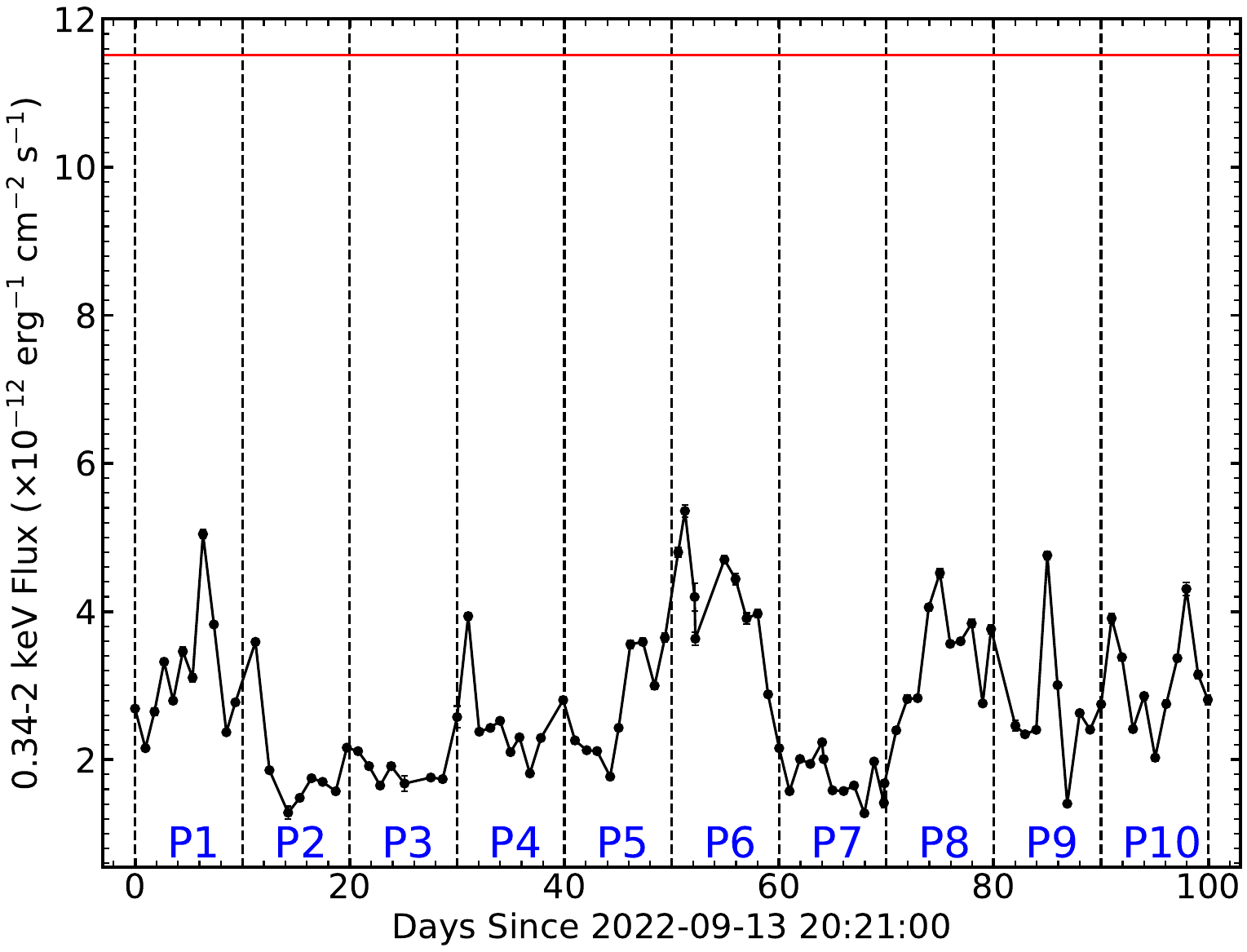}
\caption{
The \hbox{0.34--2~keV} flux light curve for the 2022 NICER observations.
The flux for each observation is calculated from the \hbox{best-fit} simple \hbox{power-law} model.
The vertical dashed lines mark the 10 segments (\hbox{P1--P10}) where spectra were extracted for the \hbox{time-resolved} spectral analysis, with the corresponding segment names labeled in blue.
The red line represents the expected \hbox{0.34--2~keV} flux of I~Zw~1, which were estimated from the intrinsic $f_{2500~\textup{\AA}}$ value derived in Section~\ref{sec:uv_ext_sed}, $\Gamma=2.2$, and the \cite{Just2007} \hbox{\aox--\ltkf} relation.
}
\label{fig:nicer_segment}
\end{figure}

\subsection{NICER Observations}\label{sec:nicer_obs}
We monitored I~Zw~1 with NICER over a 100-day period from 14 September to 23 December 2022, obtaining 102 observations.
For each NICER observation, level 2 data products were generated using the HEASoft \textsc{nicerl2} task with the CALDB version xti20240206, ensuring full calibration and screening for \hbox{non-\xray} events and problematic time intervals.
Two observations with zero effective exposure after filtering were excluded.
Four additional observations exhibiting strong particle background flares (\hbox{post-filtering} \hbox{10--12~keV} count rate exceeding $1.0~{\rm counts}~{\rm s}^{-1}$) were also removed.
The final dataset consists of 96 observations, and the basic information is presented in Table~\ref{tbl:observation}.

NICER spectra were extracted using the \textsc{nicerl3-spect} task.
The SCORPEON background model was applied, which consists of mainly \xray\ sky background components (e.g., the cosmic \xray\ background) and \hbox{non-\xray} background components (e.g., from high energy particles), and the latter components dominate for our spectra.
Since background counts generally dominate the spectra above 2~keV, we modeled the \hbox{0.34--2}~keV spectrum of each observation with a simple \hbox{power-law}\footnote{$N(E)=A[E(1+z)]^{-\Gamma}$, where $\Gamma$ is the \hbox{power-law} photon index and $A$ is the normalization at 1~keV in units of ${\rm photons}~{\rm keV}^{-1}~{\rm cm}^{-2}~{\rm s}^{-1}$.} continuum to determine the observed flux and then construct a light curve.
The resulting \hbox{0.34--2~keV} flux light curve is shown in Figure~\ref{fig:nicer_segment}.
I~Zw~1 shows strong and rapid \xray\ variability during this period, reaching a maximum amplitude of $f_{\rm var}\approx4.2$.
We estimated the expected \hbox{0.34--2~keV} flux (red line in Figure~\ref{fig:nicer_segment}) using the same methodology applied to the 2020 \hbox{XMM-Newton} observations.
The observed soft \xray\ fluxes are \hbox{$\approx 2.1$--9.0} times lower than the expected value, again suggestive of significant \xray\ absorption.

For time-resolved spectral analysis and to improve the \hbox{signal-to-noise} ratios of the spectra, we divided the observation period into 10 contiguous \hbox{10-day} segments.
All observations within each segment were combined to produce a merged spectrum, which was then grouped using the \textsc{ftgrouppha} task with optimal binning \citep{Kaastra2016} and a minimum of 25 counts per bin.
The \hbox{0.34--9}~keV energy range was adopted for fitting the merged NICER spectra, as it is optimal for mitigating systematic errors.\footnote{\url{https://heasarc.gsfc.nasa.gov/docs/nicer/analysis_threads/spectrum-systematic-error/}.}

\subsection{Multiwavelength Photometric Data}\label{sec:multiwave_photo}

To investigate optical and IR variability of I~Zw~1, we collected its \hbox{multi-epoch} multiwavelength measurements from the Near-Earth Object Wide-field Infrared Survey Explorer Reactivation Mission (\hbox{NEOWISE}, \citealt{Mainzer2011}), All-Sky Automated Survey for Supernovae (ASAS-SN, \citealt{Shappee2014}), and Zwicky Transient Facility (ZTF, \citealt{Masci2019}) catalogs.
We also collected IR measurements of I~Zw~1 from the {Wide-field~Infrared~Survey~Explorer} (WISE; \citealt{Wright2010}) and the Two Micron All Sky Survey (2MASS; \citealt{Skrutskie2006}) catalogs, along with optical measurements from the SDSS catalog, for spectral energy distribution (SED) construction.
The optical and IR measurements were corrected for the Galactic extinction.



\begin{figure}
\centering
\includegraphics[scale=0.3]{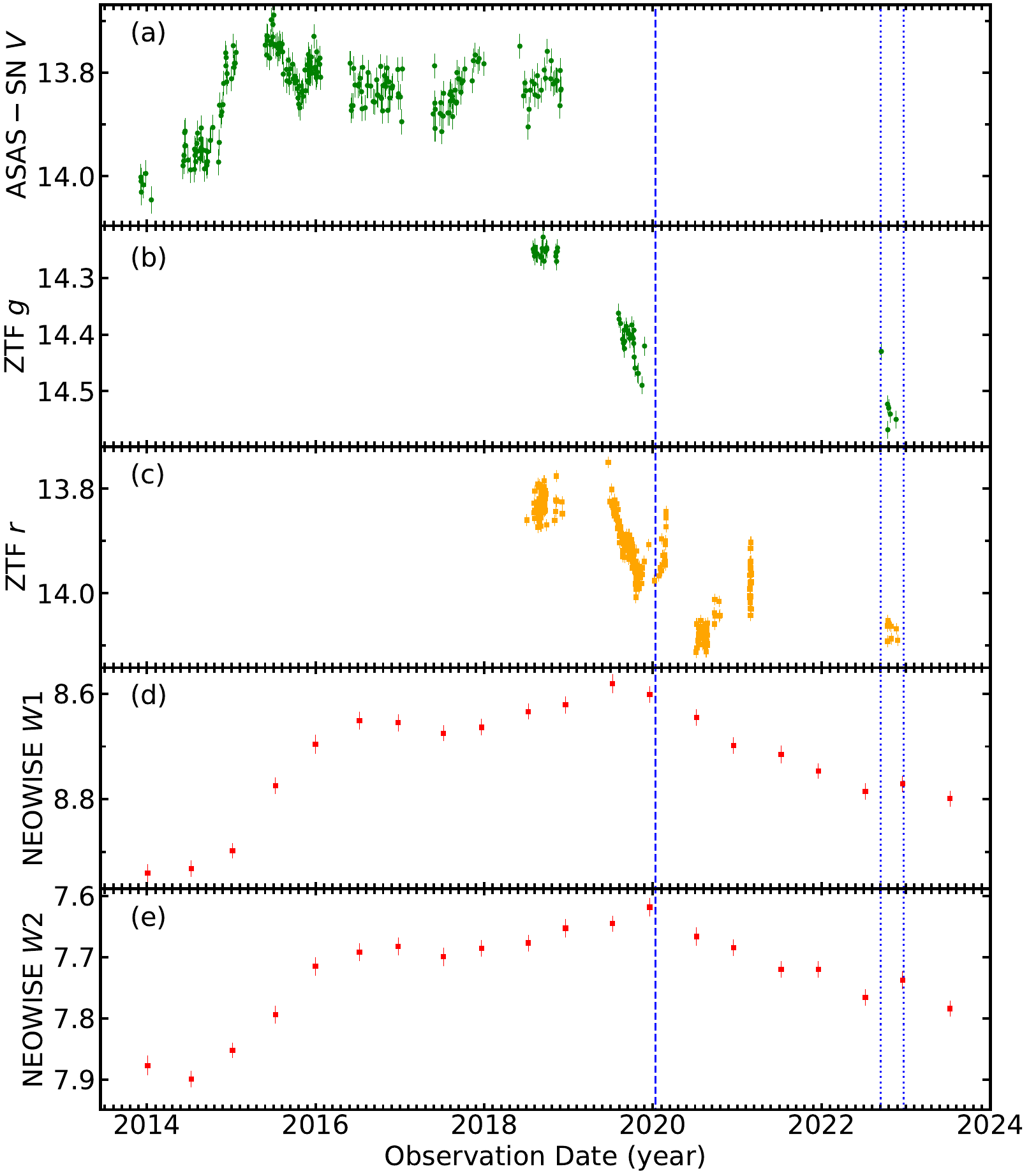}
\caption{
Light curves (magnitudes) in the (a) ASAS-SN V band, ZTF (b) g and (c) r bands, and NEOWISE (d) W1 and (e) W2 bands; we grouped intraday measurements.
The blue dashed line represents the start time of the \hbox{XMM-Newton} observation in 2020.
The blue dotted lines represent the start and end dates of the 2022 NICER monitoring campaign.
These light curves indicate that I~Zw~1 does not show any substantial long-term variability in the IR and optical bands, suggesting a relatively stable accretion process.
}
\label{fig:lc_asassn_ztf_neowise}
\end{figure}

\section{Multiwavelength Properties}\label{sec:multiwave_property}
\subsection{Multiwavelength Variability}\label{sec:multiwave_var}
We constructed \hbox{optical--IR} light curves using the archival ASAS-SN ($V$ band), ZTF ($g$ and $r$ bands), and NEOWISE ($W1$ and $W2$ bands) data, as shown in Figure~\ref{fig:lc_asassn_ztf_neowise}.
The ZTF $r$-band magnitudes range from 13.75 to 14.11, with a median magnitude of 13.92, while the ASAS-SN $V$-band magnitudes range from 13.69 to 14.05, with a median magnitude of 13.82.
The NEOWISE $W1$-band magnitudes range from 8.58 to 8.94, with a median magnitude of 8.70.
These optical/IR light curves exhibit a maximum variability amplitude of approximately 0.36 mag ($\approx30\%$), suggesting a relatively stable accretion process over the \hbox{$\sim$10-year} timescale.
The variability amplitudes between the 2020 \hbox{XMM-Newton} and 2022 NICER observation windows were notably smaller, at only $\approx$0.1 mag in the ZTF $g$ band ($\approx10\%$).
This mild optical/IR variability suggests the accretion rate in I Zw 1 did not change significantly between these two epochs, and thus changes of accretion rate cannot account for the strong \xray\ variability observed between the \hbox{XMM-Newton} and NICER observations (a maximum variability amplitude of $f_{\rm var}\approx 6$ in the \hbox{0.3--2~keV} band; see Section~\ref{sec:nicer_spec_analyses} below).

During the 2022 NICER monitoring campaign, I~Zw~1 was also monitored in the optical bands by the Las Cumbres Observatory (LCO) at a $\sim0.5~{\rm days}$ cadence \citep{Drewes2026}.
Among the LCO $uBgVriz$ bands, the $u$ band exhibited the strongest variability ($\sim20\%$), with the flux densities ranging from $2.2\times10^{-26}$ to $2.8\times10^{-26}~{\rm erg}~{\rm cm}^{-2}~{\rm s}^{-1}~{\rm Hz}^{-1}$.
This mild variability is again insufficient to account for the much stronger \xray\ variability ($f_{\rm var}\sim4$) observed simultaneously by NICER.

\begin{figure}
\centering
\includegraphics[scale=0.32]{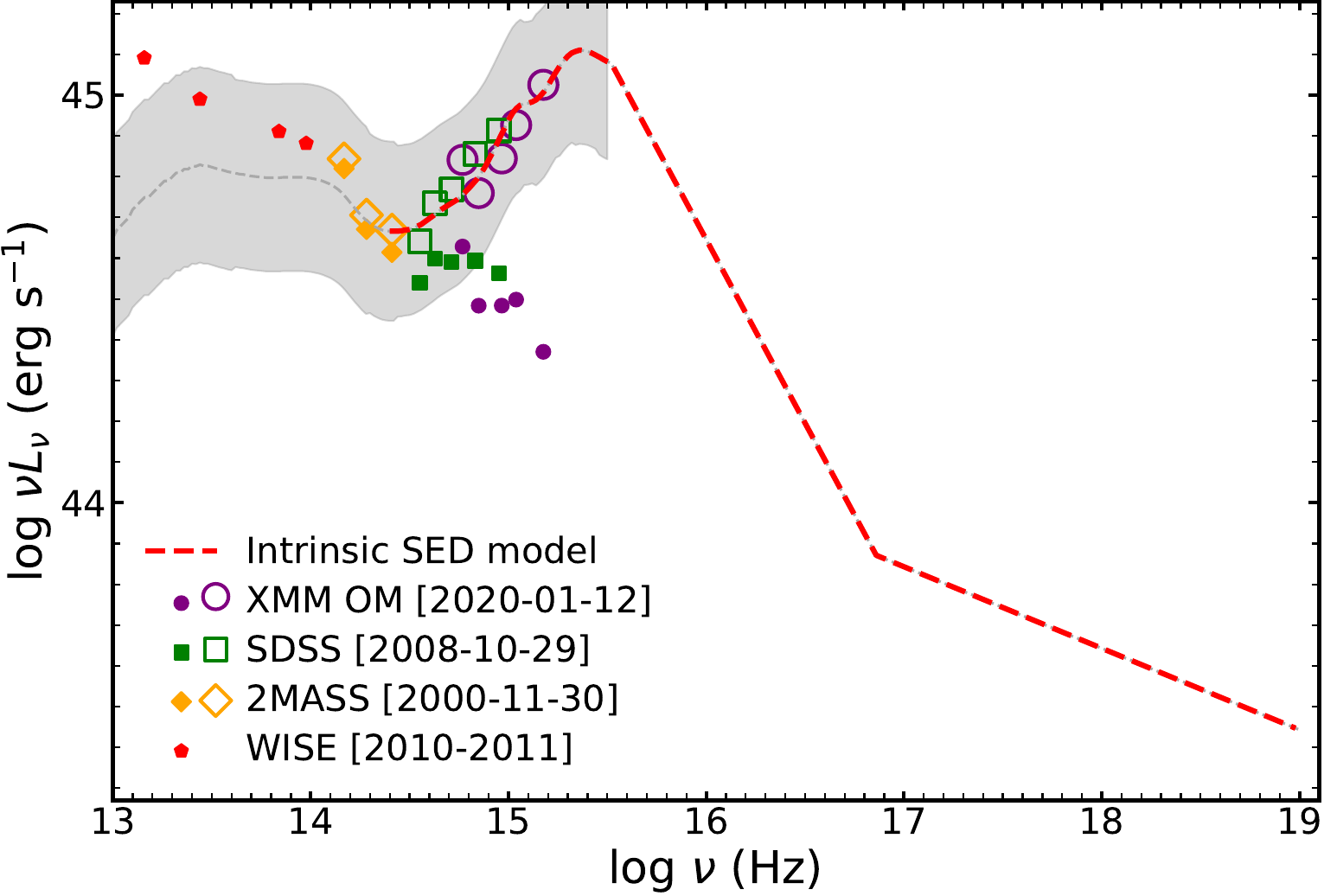}
\caption{WISE, 2MASS, SDSS, and
XMM-Newton OM photometric measurements (filled symbols) of I~Zw~1.
The $1\sigma$ uncertainties of these measurements are small ($\lesssim0.04$ magnitude), and they are not displayed.
The two rightmost OM data points (UVW2 and UVW1) were used to derive an intrinsic $E(B-V)$ value of 0.185 by comparing these measurements to the \cite{Krawczyk2013} mean quasar SED.
The \hbox{extinction-corrected} 2MASS, SDSS, and OM measurements are shown as the open diamonds, squares, and circles, respectively.
We note that the \hbox{near-IR-to-optical} data might have minor \hbox{host-galaxy} contamination.
The gray shaded region illustrates the mean quasar SED from \cite{Krawczyk2013} scaled to the dereddened 5100~\AA\ luminosity and its scatter.
The extinction corrected SED agrees with the mean quasar SED overall.
The red dashed curve represents the 1~eV to 40~keV intrinsic SED model inputted into XSTAR.
}
\label{fig:sed_dereddened}
\end{figure}

\subsection{UV Extinction and Spectral Energy Distribution}\label{sec:uv_ext_sed}

We constructed the \hbox{IR-to-UV} SED of I~Zw~1 using \hbox{Galactic-extinction-corrected} WISE, 2MASS, SDSS, and OM measurements, shown in Figure~\ref{fig:sed_dereddened}.
The red \hbox{optical--UV} SED shape suggests intrinsic dust extinction.
To estimate the intrinsic extinction, we adopted the method used in \cite{Huang2023}.
The total 5100~\AA\ flux density was determined by linear interpolation between the \hbox{XMM-Newton} OM $B$ (4500~\AA) and $V$ (5430~\AA) band measurements, yielding a value of $7.72\times10^{-26}~{\rm erg}~{\rm cm}^{-2}~{\rm s}^{-1}~{\rm Hz}^{-1}$.
The AGN continuum at 5100~\AA\ was isolated by subtracting the \hbox{host-galaxy} contribution from the total observed flux density.
We adopted the \hbox{host-galaxy} contribution of $1.30\times10^{-26}~{\rm erg}~{\rm cm}^{-2}~{\rm s}^{-1}~{\rm Hz}^{-1}$ from \cite{Huang2019}.
This value ($\approx17\%$ of our total flux density) was derived by decomposing Hubble Space Telescope (HST) images of I~Zw~1 within a spectral extraction aperture of approximately 8\arcsec.5$~\times~$2\arcsec.5.
The resulting AGN continuum flux density at 5100~\AA\ is $6.42\times10^{-26}~{\rm erg}~{\rm cm}^{-2}~{\rm s}^{-1}~{\rm Hz}^{-1}$.
We then assumed that the measurements in the two bluest bands of OM, UVW2 and UVW1 (2120~\AA\ and 2910~\AA), have negligible \hbox{host-galaxy} contamination, and we estimated the extinction by comparing the UVW2, UVW1, and 5100~\AA\ measurements to the mean quasar SED from \cite{Krawczyk2013}.
We adopted the Small Magellanic Cloud (SMC) extinction model (\citealt{Gordon2003}; $R_V=2.74$), which is commonly used to model the intrinsic extinction of AGNs \citep[e.g.,][]{Hopkins2004,Glikman2012}.
We derived $E(B-V) = 0.185$ that minimizes the differences between the dereddened UVW1 and UVW2 luminosities and the mean quasar SED scaled to the dereddened 5100~\AA\ luminosity (Figure~\ref{fig:sed_dereddened}).
This $E(B-V)$ value is moderately higher than the $E(B-V)=0.13$ reported by \cite{Rudy2000} using the \hbox{\ion{O}{1} $\lambda1304$/\ion{O}{1} $\lambda8446$} \hbox{emission-line} ratio method, and it is slightly lower than the $E(B-V)=0.206$ reported by \cite{Juranova2024} from their SED fitting of the 2015 HST data.
We show the 2MASS, SDSS, and OM measurements corrected for the intrinsic extinction in Figure~\ref{fig:sed_dereddened}.
We did not subtract \hbox{host-galaxy} contributions from these measurements as the contributions in the \hbox{near-IR-to-optical} bands are minor\footnote{We estimated the \hbox{host-galaxy} contributions using the S\'ersic profiles for the HST F438W (4326~\AA) and F105W (10550~\AA) bands  presented in Figure~3 of \cite{Huang2019}.
For the $5.7$\arcsec\ OM aperture, the host contribution is estimated to be approximately $13\%$ of the total flux density at 4326~\AA.
For the $1.5$\arcsec\ SDSS aperture, the host contribution is estimated to be approximately $6\%$ of the total flux density at 4326~\AA.
For the $4$\arcsec\ 2MASS aperture, the host contribution is estimated to be approximately $16\%$ of the total flux density at 10550~\AA.}
and they do not affect our following analyses.
The extinction corrected SED agrees with the scaled mean quasar SED overall (Figure~\ref{fig:sed_dereddened}).

We then constructed an intrinsic optical-to-\xray\ (1~eV to 40~keV) SED model of I~Zw~1 based on the dereddened 5100~\AA\ luminosity and the assumption that I~Zw~1 is intrinsically \xray\ normal.
We first adopted the scaled mean quasar SED (Figure~\ref{fig:sed_dereddened}) between 1~eV and 13.6~eV.
We then estimated the expected $f_{\rm 2~keV}$ value ($1.2\times10^{-29}~{\rm erg}~{\rm cm}^{-2}~{\rm s}^{-1}~{\rm Hz}^{-1}$) from the extinction corrected $f_{2500~\textup{\AA}}$ value ($8.7\times10^{-26}~{\rm erg}~{\rm cm}^{-2}~{\rm s}^{-1}~{\rm Hz}^{-1}$),\footnote{Contribution from the UV Fe \hbox{pseudo-continuum} is typically not subtracted from the photometrically derived $f_{2500~\textup{\AA}}$ values when computing \aox\ and studying the \hbox{\aox--\ltkf} relation.
For I Zw 1, we estimated that this contribution is $\approx19\%$ given its HST spectrum from \citet{Laor1997} and the Fe emission template from \citet{Vestergaard2001}.
This small contamination (changing \aox\ by 0.03 and its deviation from the \hbox{\aox--\ltkf} relation by 0.02) should not affect our following analysis.} using the \hbox{\aox--\ltkf} relation of \cite{Just2007}.
From the expected $f_{\rm 2~keV}$, we constructed the \hbox{0.3--40~keV} intrinsic SED by assuming a \hbox{power-law} continuum with $\Gamma=2.2$; this $\Gamma$ value was estimated using the \hbox{$\Gamma$--$\lambda_{\rm Edd}$} relation in \cite{Huang2020}.
The intrinsic SED between 13.6~eV and 0.3~keV was adopted as a \hbox{power-law} connecting the two end points.
The intrinsic SED model (1~eV to 40~keV) is shown in Figure~\ref{fig:sed_dereddened} as the red dashed curve, which was used as the input to XSTAR in Section~\ref{sec:xray_spec_analyses} below.


\begin{figure*}
\centering
\includegraphics[scale=0.35]{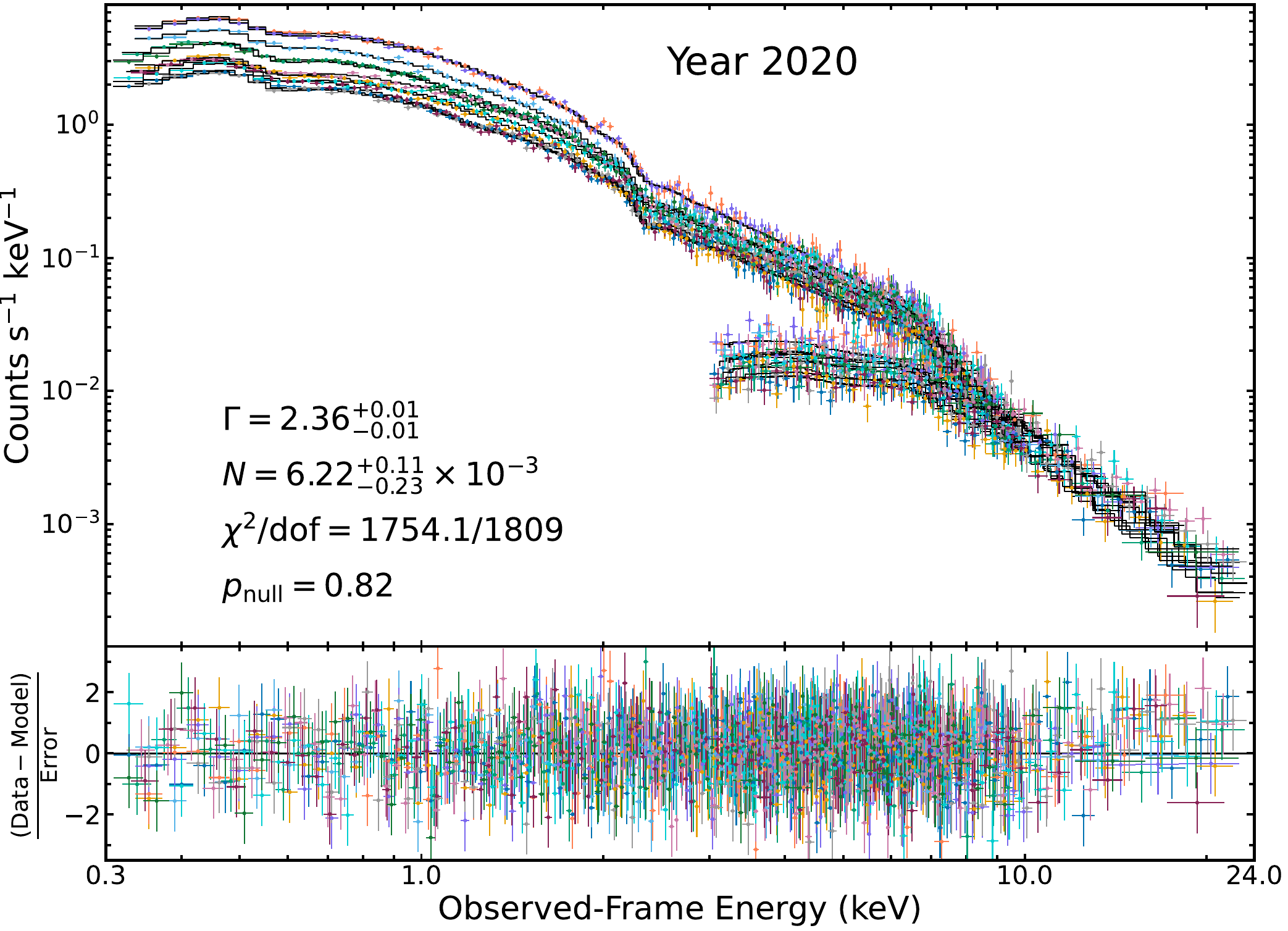}
\caption{The time-resolved \hbox{XMM-Newton} and NuSTAR spectra overlaid with the \hbox{best-fit} \hbox{partial-covering} absorption model.
All spectra were grouped with at least 25 counts per bin, and the \hbox{XMM-Newton} pn spectra were additionally grouped to avoid oversampling the intrinsic energy resolution
by more than a factor of 3.
The \hbox{best-fit} model curves are shown in black.
The bottom panel shows the fitting residuals.
}
\label{fig:xmm_nustar_joint_fitting}
\end{figure*}

\begin{figure}
\includegraphics[scale=0.35]{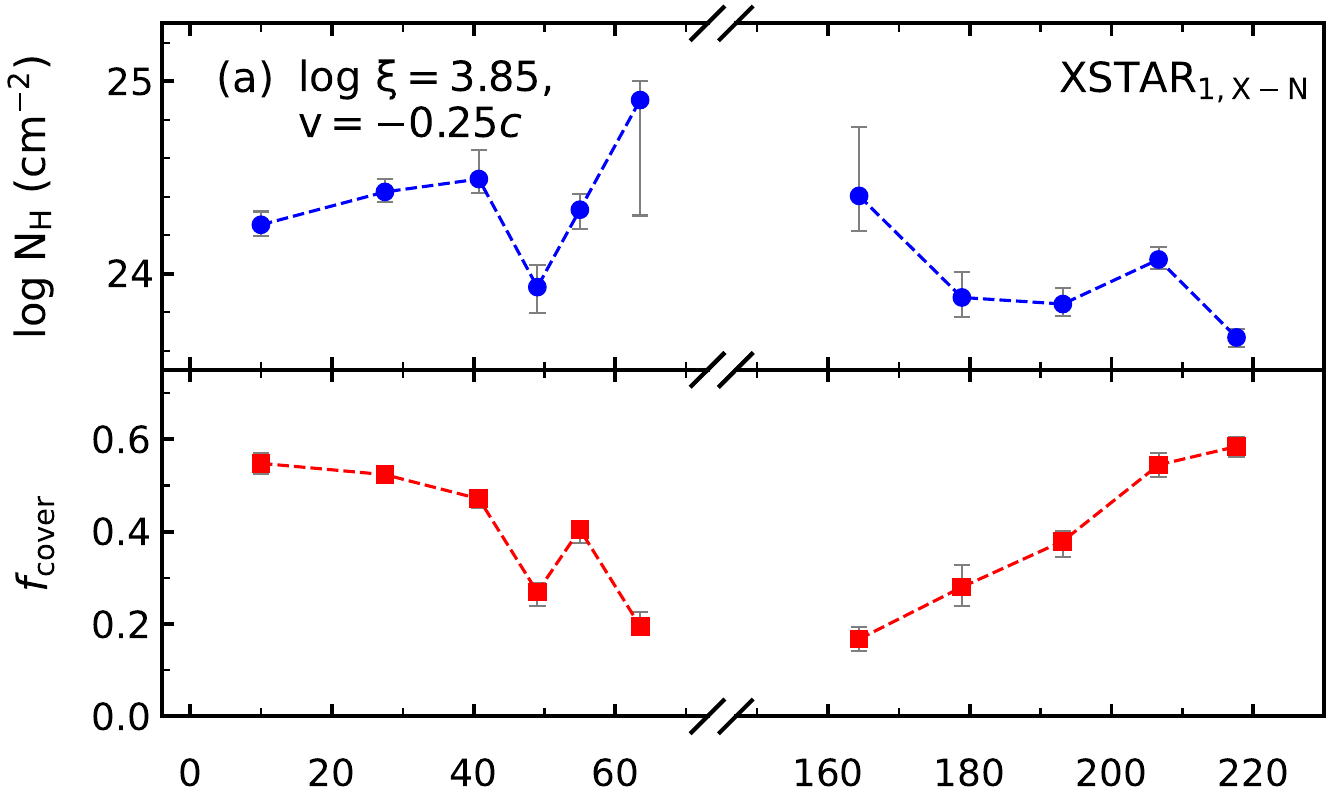}
\includegraphics[scale=0.35]{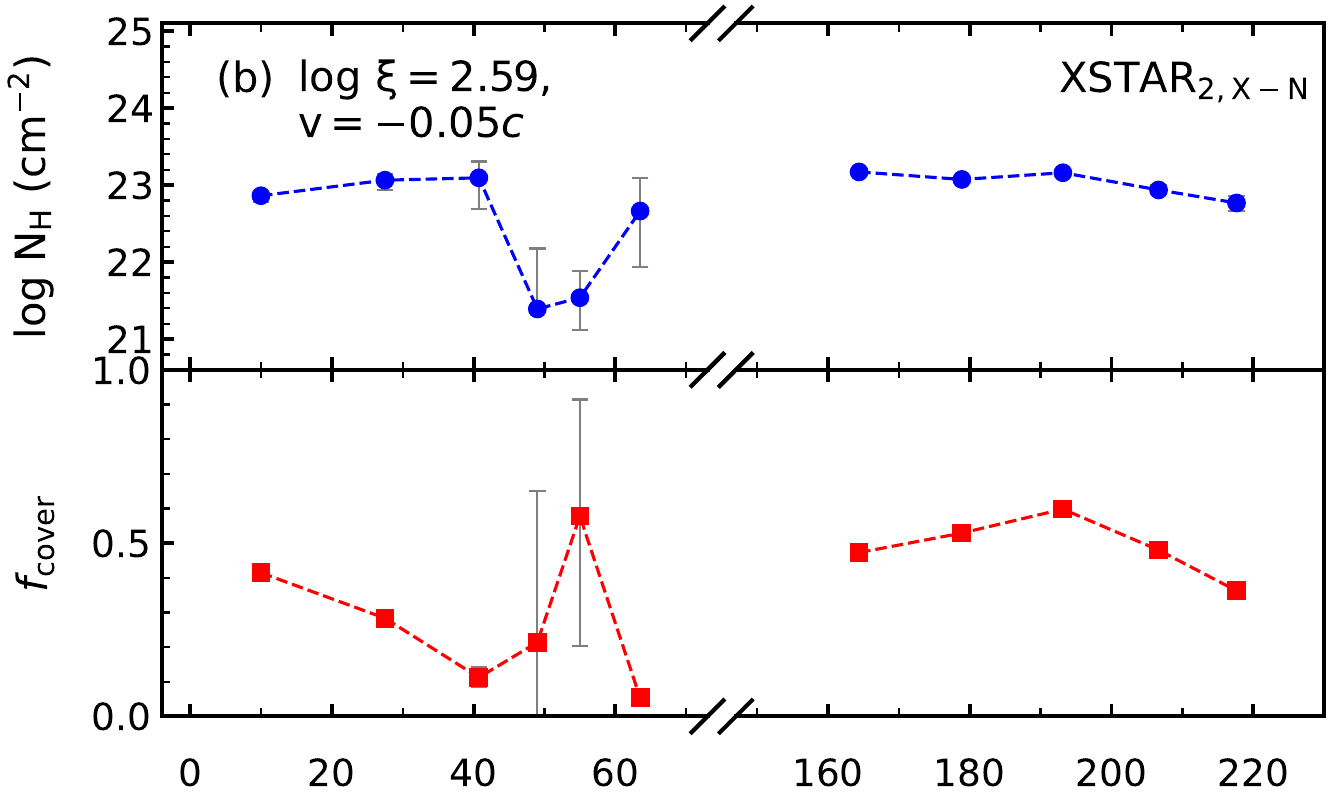}
\includegraphics[scale=0.35]{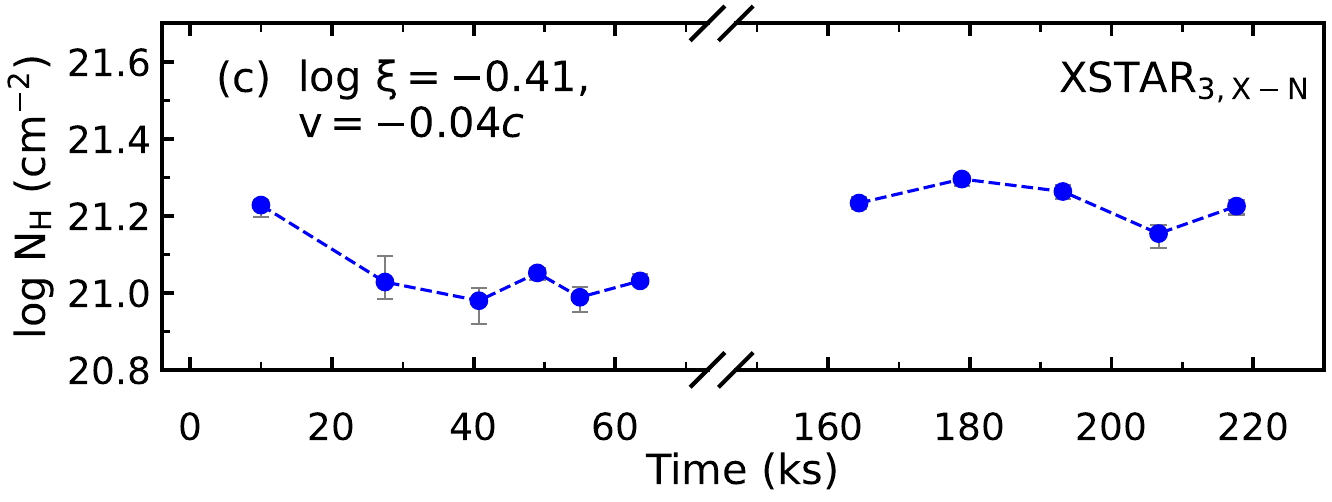}
\caption{
Temporal evolution of the column densities and covering factors for (a) ${\rm XSTAR}_{1,\rm X-N}$, (b) ${\rm XSTAR}_{2,\rm X-N}$, and (c) ${\rm XSTAR}_{3,\rm X-N}$ identified from the \hbox{time-resolved} spectral analysis of the 2020 \hbox{XMM-Newton} and NuSTAR observations.
}
\label{fig:xmm_wind_variation}
\end{figure}

\section{X-Ray Spectral Analyses}\label{sec:xray_spec_analyses}

\begin{deluxetable*}{c l l l l l l l c c} 
\tablewidth{0pt}
\tablecaption{Best-Fit Parameters for the \hbox{Time-Resolved 2020 \hbox{XMM-Newton} and NuSTAR Spectra}}
\tablehead{
\multicolumn{10}{c}{Intrinsic Power-law Continuum$^{\rm a}$} \\
\cline{1-10}
\colhead{ } & \colhead{ } & \multicolumn{3}{l}{$\Gamma=2.36\pm0.01$} & \multicolumn{3}{l}{$A=6.22^{+0.11}_{-0.23}\times10^{-3}$} & \colhead{ } & \colhead{ } \\
\cline{1-10}
\multicolumn{10}{c}{Broad Fe K Emission$^{\rm b}$} \\
\cline{1-10}
\colhead{ } & \multicolumn{2}{c}{${E}_{\rm rest}=6.72\pm0.09~{\rm keV}$} & \multicolumn{2}{c}{${\sigma}=522^{+78}_{-66}~{\rm eV}$} & \multicolumn{3}{c}{${\rm Norm}=1.04^{+0.11}_{-0.10}\times10^{-5}~{\rm photons}~{\rm cm}^{-2}~{\rm s}^{-1}$} & \colhead{ } & \colhead{ } \\
\cline{1-10}
\colhead{ } & \multicolumn{2}{c}{${\rm XSTAR}_{1,~\rm X-N}$} & \multicolumn{2}{c}{${\rm XSTAR}_{2,~\rm X-N}$} & \multicolumn{2}{c}{${\rm XSTAR}_{3,~\rm X-N}$} & \colhead{ } & \colhead{ } & \colhead{ } \\
\cline{1-10}
${\rm log}~{\xi}~({\rm erg}~{\rm cm}^{}~{\rm s}^{-1})^{\rm c}$& \multicolumn{2}{c}{$3.85^{+0.04}_{-0.03}$} & \multicolumn{2}{c}{$2.59^{+0.02}_{-0.01}$} & \multicolumn{2}{c}{$-0.41^{+0.03}_{-0.03}$} & \colhead{ } & \colhead{ } & \colhead{ } \\
$v_{\rm abs}$ ${(c)}^{\rm c}$ & \multicolumn{2}{c}{$-0.245^{+0.004}_{-0.005}$} & \multicolumn{2}{c}{$-0.049^{+0.007}_{-0.005}$} & \multicolumn{2}{c}{$-0.035^{+0.004}_{-0.004}$} & \colhead{ } & \colhead{ } & \colhead{ } \\
\cline{1-10}
\colhead{Segment} & \colhead{${\rm log}~{N}_{\rm H}$} & \colhead{$f_{\rm cov}$} & \colhead{${\rm log}~{N}_{\rm H}$} & \colhead{$f_{\rm cov}$} & \colhead{${\rm log}~{N}_{\rm H}$} & \colhead{$f_{\rm cov}$} &  \colhead{$\chi^{2}/{\rm dof}$} & \colhead{$F_{\rm 0.3-2~keV}$} & \colhead{$F_{\rm 2-10~keV}$} \\
\colhead{(1)} & \colhead{(2)} & \colhead{(3)} & \colhead{(4)} & \colhead{(5)} & \colhead{(6)} & \colhead{(7)} &  \colhead{(8)} & \colhead{(9)} & \colhead{(10)}
}
\startdata
T1 & $24.25^{+0.07}_{-0.06}$ & $0.55^{+0.02}_{-0.02}$ & $22.86^{+0.06}_{-0.08}$ & $0.41^{+0.02}_{-0.02}$ & $21.23^{+0.01}_{-0.03}$ & $ 1 $ &188.8/186 & $3.4$ & $3.6$ \\ 
T2 & $24.42^{+0.07}_{-0.05}$ & $0.52^{+0.01}_{-0.02}$ & $23.06^{+0.08}_{-0.12}$ & $0.28^{+0.02}_{-0.02}$ & $21.03^{+0.07}_{-0.04}$ & $ 1 $ &150.4/173 & $4.1$ & $3.6$ \\ 
T3 & $24.49^{+0.15}_{-0.07}$ & $0.47^{+0.02}_{-0.02}$ & $23.09^{+0.21}_{-0.40}$ & $0.11^{+0.03}_{-0.03}$ & $20.98^{+0.03}_{-0.06}$ & $ 1 $ &139.8/159 & $5.4$ & $4.2$ \\ 
T4 & $23.93^{+0.11}_{-0.13}$ & $0.27^{+0.02}_{-0.03}$ & $21.39^{+0.78}_{-0.01}$ & $0.21^{+0.44}_{-0.21}$ & $21.05^{+0.01}_{-0.02}$ & $ 1 $ &142.0/133 & $8.5$ & $6.6$ \\ 
T5 & $24.33^{+0.08}_{-0.10}$ & $0.40^{+0.01}_{-0.03}$ & $21.54^{+0.34}_{-0.42}$ & $0.58^{+0.34}_{-0.37}$ & $20.99^{+0.03}_{-0.04}$ & $ 1 $ &109.3/134 & $6.5$ & $5.0$ \\ 
T6 & $24.90^{+0.10}_{-0.60}$ & $0.20^{+0.03}_{-0.02}$ & $22.66^{+0.43}_{-0.73}$ & $0.05^{+0.02}_{-0.02}$ & $21.03^{+0.02}_{-0.01}$ & $ 1 $ &167.2/166 & $8.3$ & $6.5$ \\ 
T7 & $24.40^{+0.36}_{-0.18}$ & $0.17^{+0.03}_{-0.03}$ & $23.17^{+0.03}_{-0.04}$ & $0.47^{+0.01}_{-0.01}$ & $21.23^{+0.02}_{-0.02}$ & $ 1 $ &195.8/195 & $4.5$ & $5.4$ \\ 
T8 & $23.88^{+0.13}_{-0.10}$ & $0.28^{+0.05}_{-0.04}$ & $23.07^{+0.04}_{-0.04}$ & $0.53^{+0.01}_{-0.02}$ & $21.30^{+0.01}_{-0.02}$ & $ 1 $ &171.0/185 & $4.0$ & $5.2$ \\ 
T9 & $23.84^{+0.08}_{-0.06}$ & $0.38^{+0.02}_{-0.03}$ & $23.16^{+0.03}_{-0.03}$ & $0.60^{+0.01}_{-0.01}$ & $21.26^{+0.02}_{-0.02}$ & $ 1 $ &230.7/190 & $3.3$ & $4.5$ \\ 
T10 & $24.07^{+0.07}_{-0.05}$ & $0.54^{+0.03}_{-0.03}$ & $22.94^{+0.06}_{-0.05}$ & $0.48^{+0.02}_{-0.02}$ & $21.15^{+0.02}_{-0.04}$ & $ 1 $ &136.1/153 & $3.6$ & $3.9$ \\ 
T11 & $23.67^{+0.05}_{-0.05}$ & $0.58^{+0.02}_{-0.02}$ & $22.77^{+0.09}_{-0.10}$ & $0.36^{+0.02}_{-0.02}$ & $21.23^{+0.02}_{-0.02}$ & $ 1 $ &123.0/135 & $5.2$ & $5.3$ \\
\enddata
\tablecomments{Column (1): time segment as defined in Figure~\ref{fig:lc_xmm_nustar_om}.
Columns (2)--(3): hydrogen column density and covering factor of the absorber ${\rm XSTAR}_{1,~\rm X-N}$.
Columns (4)--(5): hydrogen column density and covering factor of the absorber ${\rm XSTAR}_{2,~\rm X-N}$.
Columns (6)--(7): hydrogen column density and covering factor of the absorber ${\rm XSTAR}_{3,~\rm X-N}$; the covering factor was fixed at 1.
Column (8): ${\chi}^{2}$ statistic value over the degrees of freedom (dof).
Columns (9)--(10): observed flux in the soft and hard bands derived from the \hbox{best-fit} model, in units of $10^{-12}~{\rm erg}~{\rm cm}^{-2}~{\rm s}^{-1}$.\\
a. The photon index and normalization parameters were tied across all the \hbox{time-resolved} spectra. \\
b. The \hbox{rest-frame} energy, line width, and normalization parameters were tied across all the \hbox{time-resolved} spectra. \\
c. The ionization parameters and blueshifted velocities were tied for all the spectra.}
\label{tbl:best_fit_xmm_nustar}
\end{deluxetable*}

\begin{deluxetable*}{c l l l l l l l c c} 
\tablecaption{Best-Fit Parameters for the Time-Resolved 2022 NICER Spectra}
\tablehead{
\multicolumn{10}{c}{Intrinsic Power-law Continuum$^{\rm a}$} \\
\cline{1-10}
\colhead{ } & \colhead{ } & \multicolumn{3}{l}{$\Gamma=2.09^{+0.03}_{-0.02}$} & \multicolumn{3}{l}{$A=6.20^{+0.17}_{-0.20}\times10^{-3}$} & \colhead{ } & \colhead{ } \\
\cline{1-10}
\colhead{ } & \multicolumn{2}{c}{${\rm XSTAR}_{1,~\rm NI}$} & \multicolumn{2}{c}{${\rm XSTAR}_{2,~\rm NI}$} & \multicolumn{2}{c}{${\rm XSTAR}_{3,~\rm NI}$} & \colhead{ } & \colhead{ } & \colhead{ } \\
\cline{1-10}
${\rm log}~{\xi}~({\rm erg}~{\rm cm}^{}~{\rm s}^{-1})^{\rm b}$ & \multicolumn{2}{c}{$2.66^{+0.07}_{-0.04}$} & \multicolumn{2}{c}{$0.07^{+0.06}_{-0.07}$} & \multicolumn{2}{c}{$-0.70^{+0.04}_{-0.05}$} & \colhead{ } & \colhead{ } & \colhead{ } \\
$v_{\rm abs}$ $(c)^{\rm b}$ & \multicolumn{2}{c}{$-0.317^{+0.021}_{-0.011}$} & \multicolumn{2}{c}{$-0.224^{+0.009}_{-0.014}$} & \multicolumn{2}{c}{$-0.057^{+0.003}_{-0.004}$} & \colhead{ } & \colhead{ } & \colhead{ } \\
\cline{1-10}
\colhead{Segment} & \colhead{${\rm log}~{N}_{\rm H}$} & \colhead{$f_{\rm cov}$} & \colhead{${\rm log}~{N}_{\rm H}$} & \colhead{$f_{\rm cov}$} & \colhead{${\rm log}~{N}_{\rm H}$} & \colhead{$f_{\rm cov}$} &  \colhead{$\chi^{2}/{\rm dof}$} & \colhead{$F_{\rm 0.34-2~keV}$} & \colhead{$F_{\rm 2-9~keV}$} \\
\colhead{(1)} & \colhead{(2)} & \colhead{(3)} & \colhead{(4)} & \colhead{(5)} & \colhead{(6)} & \colhead{(7)} &  \colhead{(8)} & \colhead{(9)} & \colhead{(10)}
}
\startdata
P1 & $23.50^{+0.03}_{-0.03}$ & $0.61^{+0.02}_{-0.04}$ & $21.75^{+0.10}_{-0.14}$ & $0.43^{+0.07}_{-0.05}$ & $21.25^{+0.03}_{-0.03}$ & 1 & 96.4/98 & $2.7$ & $5.9$ \\ 
P2 & $23.67^{+0.07}_{-0.01}$ & $0.75^{+0.01}_{-0.01}$ & $21.29^{+0.20}_{-0.07}$ & $1.00^{}_{-0.08}$ & $21.22^{+0.04}_{-0.06}$ & 1 & 112.6/104 & $1.6$ & $4.0$ \\ 
P3 & $23.85^{+0.19}_{-0.10}$ & $0.78^{+0.01}_{-0.01}$ & $21.33^{+0.11}_{-0.18}$ & $0.85^{+0.15}_{-0.23}$ & $21.16^{+0.09}_{-0.14}$ & 1 & 102.1/98 & $1.5$ & $3.2$ \\ 
P4 & $23.88^{+0.27}_{-0.11}$ & $0.70^{+0.01}_{-0.01}$ & $21.33^{+0.18}_{-0.21}$ & $0.62^{+0.28}_{-0.18}$ & $21.31^{+0.05}_{-0.07}$ & 1 & 100.0/103 & $2.0$ & $3.9$ \\ 
P5 & $23.61^{+0.04}_{-0.04}$ & $0.65^{+0.01}_{-0.02}$ & $20.79^{+0.34}_{-0.14}$ & $1.00^{}_{-0.49}$ & $21.43^{+0.02}_{-0.02}$ & 1 & 87.7/99 & $2.4$ & $5.2$ \\ 
P6 & $23.87^{+0.33}_{-0.19}$ & $0.56^{+0.04}_{-0.01}$ & $21.33^{+0.09}_{-0.16}$ & $0.55^{+0.15}_{-0.11}$ & $21.25^{+0.04}_{-0.06}$ & 1 & 106.0/103 & $3.2$ & $5.4$ \\ 
P7 & ${25.00_{-0.68}}^{c}$ & $0.72^{+0.01}_{-0.01}$ & $22.05^{+0.04}_{-0.06}$ & $0.50^{+0.03}_{-0.02}$ & $21.16^{+0.03}_{-0.03}$ & 1 & 98.7/98 & $1.6$ & $3.0$ \\ 
P8 & $23.58^{+0.05}_{-0.06}$ & $0.60^{+0.01}_{-0.01}$ & $21.33^{+0.11}_{-0.18}$ & $0.68^{+0.16}_{-0.41}$ & $21.29^{+0.03}_{-0.03}$ & 1 & 113.6/106 & $2.8$ & $5.8$ \\ 
P9 & $23.66^{+0.09}_{-0.05}$ & $0.68^{+0.01}_{-0.01}$ & $21.65^{+0.12}_{-0.60}$ & $0.44^{+0.33}_{-0.08}$ & $21.33^{+0.07}_{-0.04}$ & 1 & 103.4/104 & $2.1$ & $4.7$ \\ 
P10 & $23.79^{+0.11}_{-0.09}$ & $0.61^{+0.01}_{-0.02}$ & $21.87^{+0.11}_{-0.10}$ & $0.44^{+0.05}_{-0.05}$ & $21.19^{+0.05}_{-0.04}$ & 1 & 103.7/97 & $2.7$ & $5.0$ \\ 
\enddata
\tablecomments{Column (1): time segment as defined in Figure~\ref{fig:nicer_segment}.
Columns (2)--(3): hydrogen column density and covering factor of the absorber ${\rm XSTAR}_{1,~\rm NI}$.
Columns (4)--(5): hydrogen column density and covering factor of the absorber ${\rm XSTAR}_{2,~\rm NI}$.
Columns (6)--(7): hydrogen column density and covering factor of the absorber ${\rm XSTAR}_{3,~\rm NI}$; the covering factor was fixed at 1.
Column (8): ${\chi}^{2}$ statistic value over the degrees of freedom (dof).
Columns (9)--(10): observed flux in the \hbox{0.34--2~keV} and \hbox{2--9~keV} bands derived from the \hbox{best-fit} model, in units of $10^{-12}~{\rm erg}~{\rm cm}^{-2}~{\rm s}^{-1}$.\\
a. The photon index and normalization parameters were tied across all the \hbox{time-resolved} spectra. \\
b. The ionization parameters and blueshifted velocities were tied for all the spectra. \\
c. No upper bound is available if the spectrum is not sensitive to very large $N_{\rm H}$ values.}
\label{tbl:best_fit_nicer}
\end{deluxetable*}

\begin{figure*}
\centering
\includegraphics[scale=0.35]{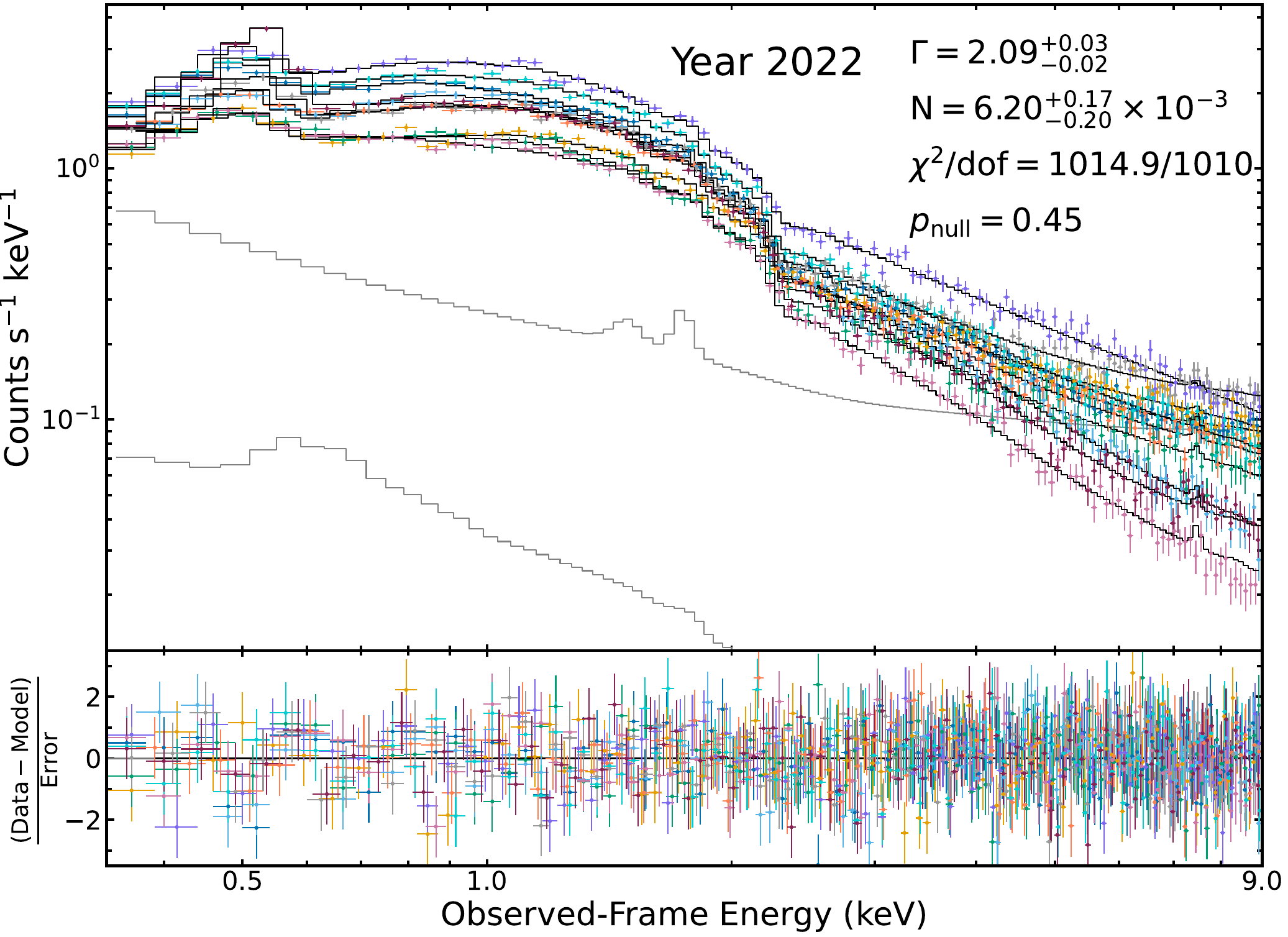}
\caption{The NICER spectra overlaid with the \hbox{best-fit} \hbox{partial-covering} absorption model.
The spectra were grouped using the optimal binning method described in \cite{Kaastra2016} and have at least 25 counts per bin.
The \hbox{best-fit} model curves are shown in black.
The two gray curves, from top to bottom, represent the average \hbox{non-\xray} and \xray\ background models of the NICER spectra.
The bottom panel shows the fitting residuals.
}
\label{fig:nicer_joint_fitting}
\end{figure*}

\begin{figure}
\includegraphics[scale=0.35]{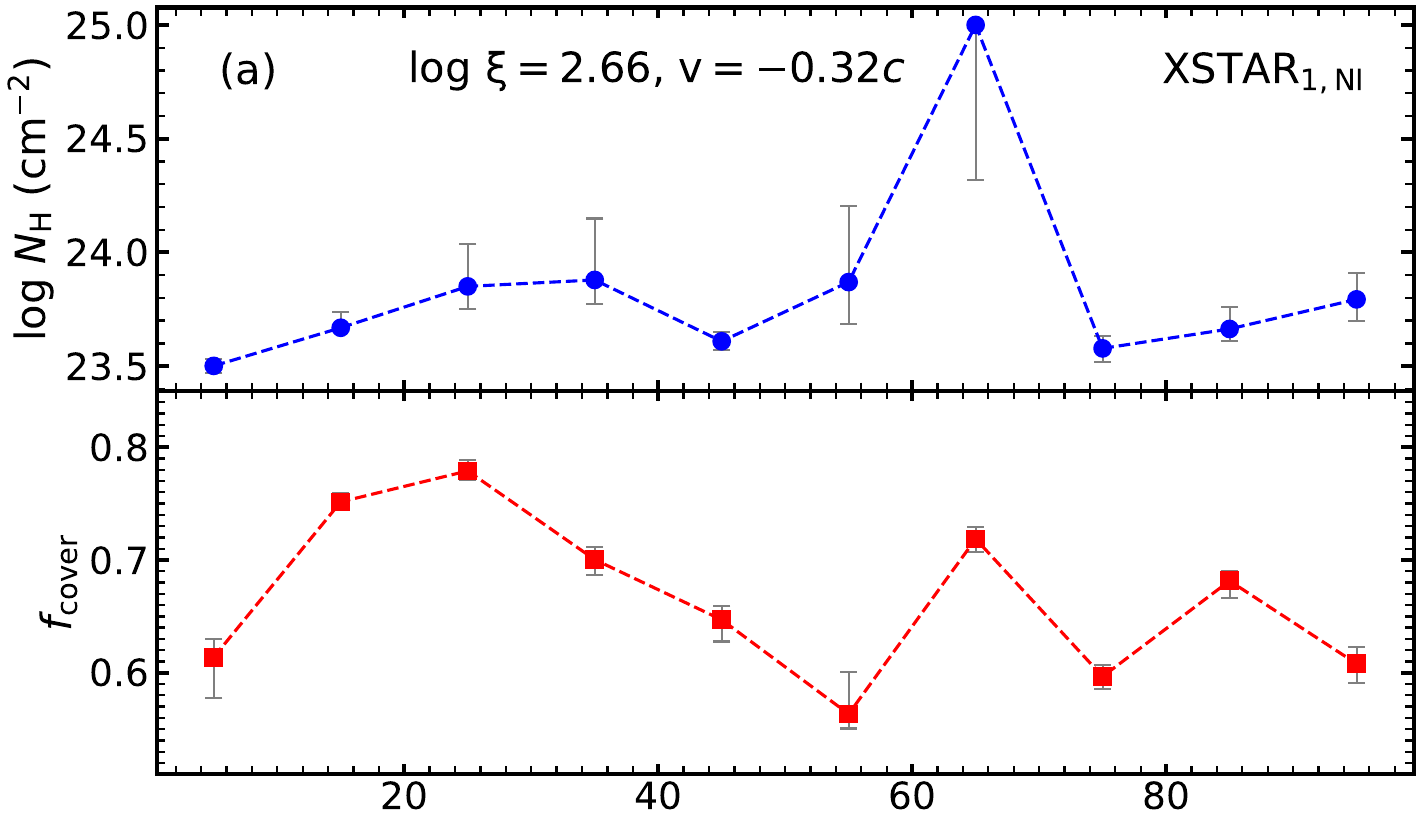}
\includegraphics[scale=0.35]{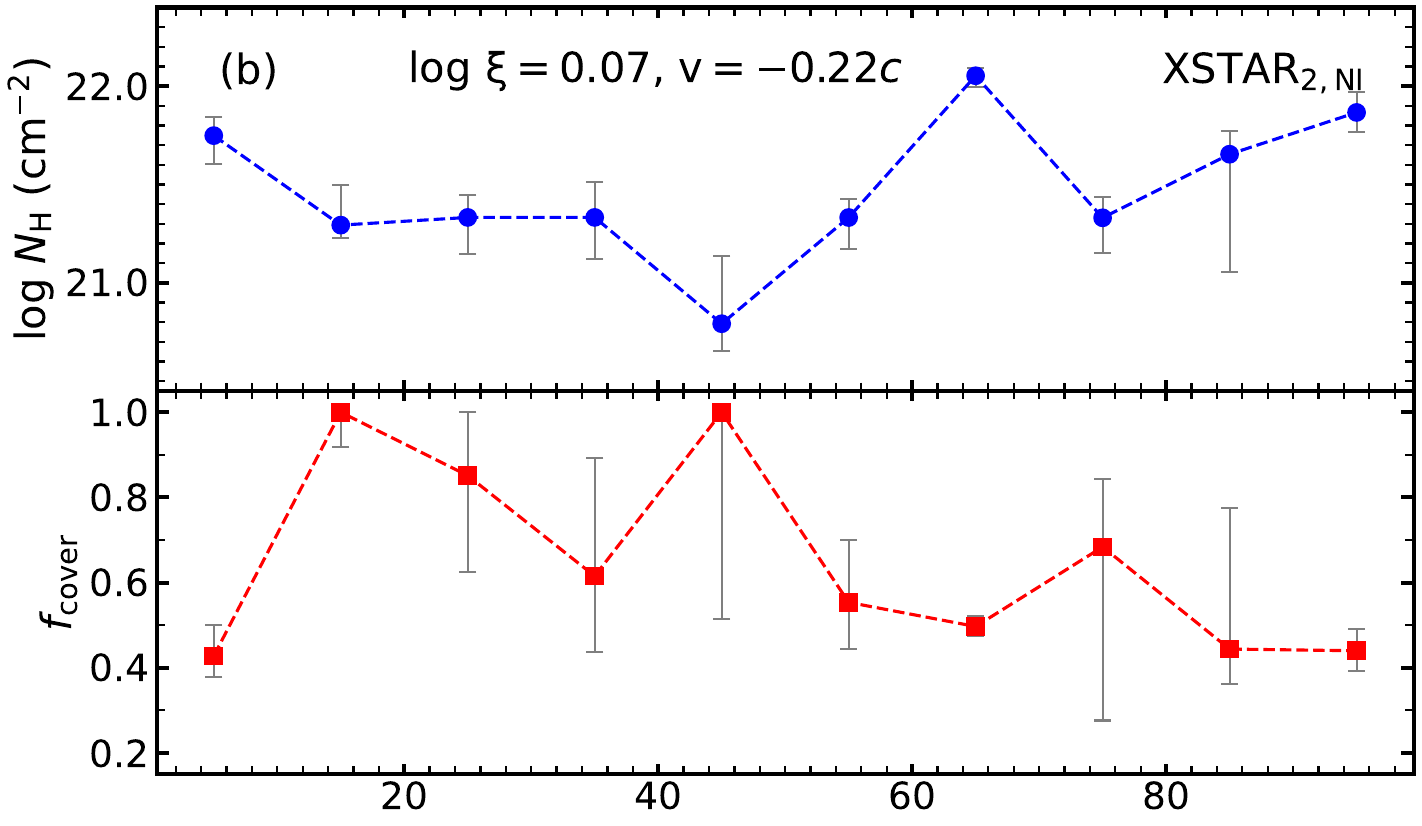}
\includegraphics[scale=0.35]{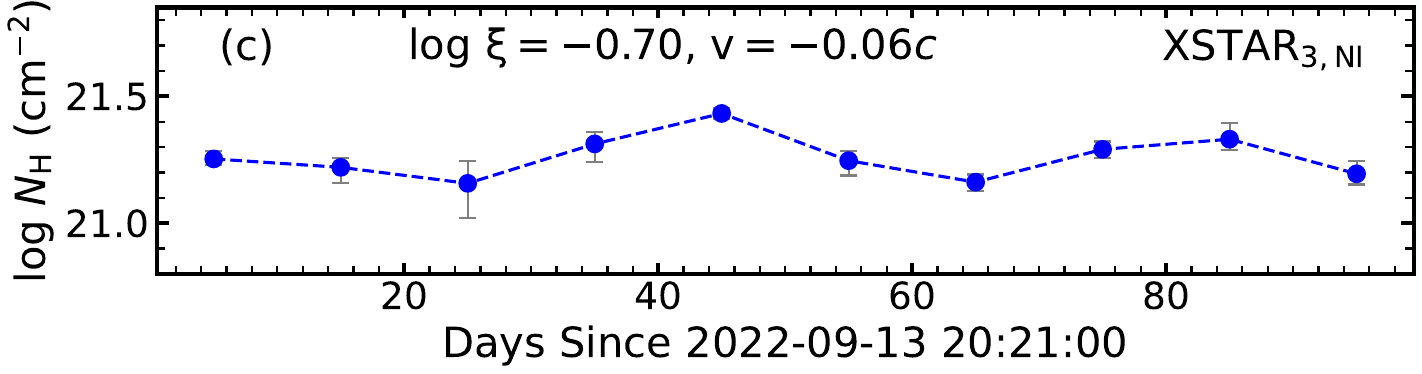}
\caption{
Temporal evolution of the column densities and covering factors for (a) ${\rm XSTAR}_{1,~\rm NI}$, (b) ${\rm XSTAR}_{2,~\rm NI}$, and (c) ${\rm XSTAR}_{3,~\rm NI}$ identified from the time-resolved spectral analysis of the 2022 NICER observations.
}
\label{fig:nicer_wind_variation}
\end{figure}

\subsection{XMM-Newton and NuSTAR Spectral Analyses}\label{sec:xmm_spec_analyses}
We sought to explain the complex \xray\ variability observed in the 2020 \hbox{XMM-Newton} and NuSTAR data entirely through obscuration, employing an absorbed \hbox{power-law} model with multiple absorbers.
To model \xray\ absorption by ionized absorbers in I~Zw~1, we used a grid of XSTAR \citep{Kallman2004} photoionization models based on the publicly available tabulated ``grid 25'', which was also used to generate the model \textsc{zxipcf} in XSPEC by \cite{Reeves2008}.
This grid of models adopt the Fe K treatment of \cite{Kallman2004}, a turbulent velocity of $\sigma=200~{\rm km}~{\rm s}^{-1}$, and solar abundance \citep{Grevesse&Sauval1998}.
We explored a wide parameter space in the hydrogen column density (${\rm log}~N_{\rm H}=19$--$25~{\rm cm}^{-2}$; 19 data points) and ionization parameter (${\rm log}~{\xi}=-3$--$6$, 19 data points).
The input photoionizing continuum was the intrinsic \hbox{1~eV--40~keV} SED model derived in Section~\ref{sec:uv_ext_sed}.

Previous studies have found that the soft \xray\ spectra, including pn and Reflection Grating Spectrometer (RGS) spectra, for the 2020 \hbox{XMM-Newton} epoch are significantly affected by intrinsic absorption, often modeled with \hbox{2--5} ionized absorbers \citep[e.g.,][]{Wilkins2022, Rogantini2022,Ding2022}.
Therefore, we initially adopted a simple \hbox{power-law} model (\textsc{zpowerlw}) modified by Galactic absorption (\textsc{phabs}) and two \hbox{partial-covering} XSTAR absorbers.
The normalization ($A$) and the photon index ($\Gamma$) of the \hbox{power-law} continuum were free parameters but were tied across all the \hbox{time-resolved} spectra to reflect intrinsically stable coronal emission.
For each XSTAR absorber, we added a \textsc{cabs} component to account for the Compton scattering effect.
The Galactic neutral hydrogen column density $N_{\rm H,~Gal}$ was fixed at $4.63\times10^{20}~{\rm cm}^{-2}$ \citep{HI4PI2016}.
We assumed that the basic properties of the absorbers remain constant throughout the observations that span $\approx2$~days.
Therefore, we tied the ionization parameter and velocity of individual absorbers and allowed only the hydrogen column densities and covering factors of the absorbers to vary across the spectra (a consequence of their clumpy nature).
We added a normalization parameter to the NuSTAR spectra to account for any \hbox{inter-instrument} calibration offset between \hbox{XMM-Newton} and NuSTAR.
This \hbox{cross-calibration} constant was tied across all the \hbox{time-resolved} spectra, and its \hbox{best-fit} value is close to unity ($1.09^{+0.01}_{-0.02}$).
We note that there is also an empirical correction to the \hbox{XMM-Newton} \hbox{3--12~keV} effective area of the order of \hbox{$6$--$8\%$} to align the spectral shapes from \hbox{XMM-Newton} and NuSTAR observations in this energy range \cite[e.g.,][]{Kang2023}.\footnote{The correction can be applied by setting applyabsfluxcorr=yes during spectral extraction; see \url{https://xmmweb.esac.esa.int/docs/documents/CAL-SRN-0388-1-4.pdf}.}
We opted not to apply this correction here, as this correction, limited to the \hbox{3--12~keV} band, would artificially bend the broader \hbox{0.3--10~keV} \hbox{XMM-Newton} spectra, introducing biases to these spectra that have higher \hbox{signal-to-noise} ratios than the NuSTAR spectra.

We then performed joint fitting of the \hbox{XMM-Newton} (\hbox{0.3--10~keV}) and NuSTAR (\hbox{3--24~keV}) spectra from all the 11 segments.
The \hbox{best-fit} results show that the \hbox{two-absorber} model cannot describe the \hbox{XMM-Newton} and NuSTAR spectra well in general, with ${\chi}^{2}/{\rm dof}=2179.3/1825$ (dof representing degrees of freedom) and a low \hbox{null-hypothesis} probability of $1.6\times10^{-8}$.
We thus added a third \hbox{partial-covering} XSTAR absorber.
Adding the third absorber significantly improves the quality of fit with $\Delta\chi^{2}/{\rm dof}=360.4/13$ and an \hbox{$F$-test} probability of $2.5\times10^{-62}$.
This model yields a good fit to the \xray\ spectra, with ${\chi}^{2}/{\rm dof}=1818.9/1812$ and a \hbox{null-hypothesis} probability of 0.45.
There appear to be systematic residuals around the \hbox{6--7~keV} band, and previous studies of the 2020 I~Zw~1 \hbox{XMM-Newton} spectra have often invoked a broad Fe K emission line (either a Gaussian profile or inherited from the relativistic reflection model; e.g., \citealt{Ding2022}; \citealt{Rogantini2022}; \citealt{Wilkins2022}).
Therefore, we added a {\sc zgauss} component to the model; this component is assumed to be constant with the three parameters tied across the spectra (see discussion in Section~\ref{sec:clumpy_winds_in_AGNs} below).
This Fe K line further improves the fit, with $\Delta\chi^{2}/\mathrm{dof}=64.8/3$ and an \hbox{$F$-test} probability of $3.7\times10^{-14}$.
The \hbox{best-fit} line parameters are $E_{\rm rest}=6.72\pm0.09~{\rm keV}$ and $\sigma=522^{+78}_{-66}~{\rm eV}$, with equivalent widths ranging from $174~{\rm eV}$ to $307~{\rm eV}$ (an average value of $229~{\rm eV}$).
The final model could be written as 
\begin{equation}\label{eq:model}
\left\{
\begin{aligned}
&{\rm OBS}={\rm PHABS}*({\rm TRAN_3+ZGAUSS}), \\
&{\rm TRAN_3}={C_3}\times\ {\rm CABS_3}\times\ {\rm XSTAR_3}\times\ {\rm TRAN_2}+\\
&(1-C_3)\times\ {\rm TRAN_2}~(C_3=1), \\
&{\rm TRAN_2}={C_2}\times\ {\rm CABS_2}\times\ {\rm XSTAR_2}\times\ {\rm TRAN_1}+\\
&(1-C_2)\times\ {\rm TRAN_1},~{\rm and}\\
&{\rm TRAN_1}={C_1}\times\ {\rm CABS_1}\times\ {\rm XSTAR_1}\times\ {\rm ZPOW_0}+\\
&(1-C_1)\times\ {\rm ZPOW_0}, 
\end{aligned}
\right.
\end{equation}
where ${\rm ZPOW_0}$ is the intrinsic \xray\ continuum, ${\rm TRAN}_i$ denotes the continuum transmitted through each absorber (${\rm XSTAR}_{i,~\rm X-N}$ for XMM-Newton $+$ NuSTAR analyses, where $i=1,~2,~3$), and $C_i$ is the covering factor of each absorber.
The covering factor of ${\rm XSTAR}_{3,~\rm X-N}$ remains consistent with unity within 1$\sigma$ uncertainties across all time segments.
We therefore consider that ${\rm XSTAR}_{3,~\rm X-N}$ fully covered the corona throughout the observations and fixed its covering factor at unity for all the 11 segments ($C_3=1$).
The \hbox{best-fit} results have ${\chi}^{2}/{\rm dof}=1754.1/1809$ and a high \hbox{null-hypothesis} probability of 0.82, indicating that the model describes well the observed \xray\ spectra.
We show the \hbox{XMM-Newton} and NuSTAR spectra overlaid with the \hbox{best-fit} model in Figure~\ref{fig:xmm_nustar_joint_fitting}, and list the \hbox{best-fit} parameters in Table~\ref{tbl:best_fit_xmm_nustar}.

In Table~\ref{tbl:best_fit_xmm_nustar}, we list the observed \xray\ fluxes for the 11 segments in the soft band (\hbox{$0.3$--$2~{\rm keV}$}) and hard band (\hbox{$2$--$10~{\rm keV}$}) derived from the \hbox{best-fit} models .
The \hbox{soft-band} \xray\ fluxes range from $3.3\times10^{-12}~{\rm erg}~{\rm cm}^{-2}~{\rm s}^{-1}$ to $8.5\times10^{-12}~{\rm erg}~{\rm cm}^{-2}~{\rm s}^{-1}$, with a maximum variability amplitude of $f_{\rm var}\approx2.6$.
The \hbox{hard-band} \xray\ fluxes range from $3.6\times10^{-12}~{\rm erg}~{\rm cm}^{-2}~{\rm s}^{-1}$ to $6.6\times10^{-12}~{\rm erg}~{\rm cm}^{-2}~{\rm s}^{-1}$, with a maximum variability amplitude of $f_{\rm var}\approx1.8$.
These variability amplitudes are smaller than those (3.3 and 2.9) derived from the \hbox{count-rate} light curve in Section~\ref{sec:xmm_obs}, as \hbox{short-term} features are
smoothed out.
We then calculated the observed flux densities at \hbox{rest-frame} 2~keV and the corresponding \aox\ and $\Delta\alpha_{\rm OX}$\footnote{$\Delta\alpha_{\rm OX}=\alpha_{\rm OX}-\alpha_{\rm OX,~exp}$, where $\alpha_{\rm OX,~exp}$ is the expected \aox\ value derived from the \cite{Just2007} \hbox{\aox--\ltkf} relation.
The \xray\ weakness factor ($f_{\rm weak}$) is related to $\Delta\alpha_{\rm OX}$ with $f_{\rm weak}=403^{-\Delta\alpha_{\rm OX}}$.} values for all the 11 segments.
The $\Delta\alpha_{\rm OX}$ values range from $-0.13$ ($f_{\rm weak}=2.2$) to $0.00$ ($f_{\rm weak}=1$, with a mean of $-0.08$ ($f_{\rm weak}=1.6$).
We also computed \hbox{absorption-corrected} (intrinsic) $\alpha_{\rm OX,~corr}$ and $\Delta\alpha_{\rm OX,~corr}$ values from the \hbox{best-fit} intrinsic continuum.
The resulting $\Delta\alpha_{\rm OX,~corr}$ is 0.04, within the typical scatter ($\sim0.14$) of the \hbox{\aox--\ltkf} relation \citep[e.g.,][]{Steffen2006, Just2007, Pu2020, Huang2025}.
Therefore, our joint \hbox{XMM-Newton} and NuSTAR spectral analyses reveal that I~Zw~1 emitted a nominal level of \xray\ emission as expected from the \hbox{\aox--\ltkf} relation, and the observed \xray\ variability was solely driven by variable \hbox{partial-covering} absorption from the three absorbers.

The \hbox{best-fit} absorber properties are displayed in Table~\ref{tbl:best_fit_xmm_nustar}.
The ${\rm XSTAR}_{1,~\rm X-N}$ absorber exhibits the highest ionization parameter (${\rm log}~\xi = 3.85$) and blueshifted velocity ($v_{\rm abs} = -0.245c$) among the three absorbers.
It can be classified as an \hbox{ultra-fast} outflow (UFO) due to its high velocity, following the conventional definition of $|v|\gtrsim0.1c$ (e.g., \citealt{King2015}).
Its column density varied between $4.7\times10^{23}~{\rm cm}^{-2}$ and $7.9\times10^{24}~{\rm cm}^{-2}$, and its covering factor varied between $0.17$ and $0.58$.
The ${\rm XSTAR}_{2,~\rm X-N}$ absorber (${\rm log}~\xi = 2.59$, $v_{\rm abs} = -0.049c$) also shows variability, with a column density that varied between $2.5\times10^{21}~{\rm cm}^{-2}$ and $1.5\times10^{23}~{\rm cm}^{-2}$ and a covering factor that varied between $0.05$ and $0.60$.
The ${\rm XSTAR}_{3,~\rm X-N}$ absorber has the lowest ionization parameter (${\rm log}~\xi = -0.41$) and velocity ($v_{\rm abs}=-0.035c$), and it shows the smallest variation in the column density among the three absorbers, ranging from $9.5\times10^{20}~{\rm cm}^{-2}$ to $2.0\times10^{21}~{\rm cm}^{-2}$.
The temporal evolution of the column densities and covering factors of the three absorbers is shown in Figure~\ref{fig:xmm_wind_variation}.

\subsection{NICER Spectral Analyses}\label{sec:nicer_spec_analyses}
We fitted the \hbox{time-resolved} NICER spectra using the same model (Equation~\ref{eq:model}) minus the {\sc zgauss} component.
The NICER spectra cannot constrain an Fe emission line, likely because the background becomes dominant above $\approx6~{\rm keV}$.
We have verified that adding a {\sc zgauss} component with parameters fixed at the \hbox{best-fit} \hbox{XMM-Newton} + NuSTAR values does not significantly alter the results and our subsequent discussion remains unchanged.
The observed spectra are thus modified by three intrinsic absorbers, denoted as ${\rm XSTAR}_{i,~\rm NI}$ (where $i=1,~2,~3$).
The covering factor of ${\rm XSTAR}_{3,~\rm NI}$ also remains consistent with unity within 1$\sigma$ uncertainties across all time segments.
We therefore consider that ${\rm XSTAR}_{3,~\rm NI}$ fully covered the corona throughout the observations and fixed its covering factor at unity for all the 10 segments ($C_3=1$).
The \hbox{best-fit} results have ${\chi}^{2}/{\rm dof}=1014.8/1010$ and $p_{\rm null}=0.45$, indicating a good fit.
We show the NICER spectra overlaid with the \hbox{best-fit} model in Figure~\ref{fig:nicer_joint_fitting}, and we list the \hbox{best-fit} parameters in Table~\ref{tbl:best_fit_nicer}.

In Table~\ref{tbl:best_fit_nicer}, we list the observed \xray\ fluxes for the 10 segments in the \hbox{0.34--2~keV} and \hbox{2--9~keV} bands derived from the \hbox{best-fit} models.
The \hbox{0.34--2~keV} \xray\ fluxes range from $1.5\times10^{-12}~{\rm erg}~{\rm cm}^{-2}~{\rm s}^{-1}$ to $3.2\times10^{-12}~{\rm erg}~{\rm cm}^{-2}~{\rm s}^{-1}$, with a maximum variability amplitude of $f_{\rm var}\approx 2.1$.
The \hbox{2--9~keV} \xray\ fluxes range from $3.0\times10^{-12}~{\rm erg}~{\rm cm}^{-2}~{\rm s}^{-1}$ to $5.9\times10^{-12}~{\rm erg}~{\rm cm}^{-2}~{\rm s}^{-1}$, with a maximum variability amplitude of $f_{\rm var}\approx 2.0$.
Between the 2020 \hbox{XMM-Newton} and the 2022 NICER observational epochs, I~Zw~1 displays strong \xray\ variability over the $\approx$\hbox{2-year} timescale, with a maximum \hbox{0.3--2~keV} (extrapolating the NICER \hbox{best-fit model to this band}) flux variability amplitude of $f_{\rm var}\approx6$.
This factor substantially exceeds the observed optical/IR variability amplitudes ($\approx$10--30$\%$; see Section~\ref{sec:multiwave_var}), suggesting that changes in the \hbox{accretion-disk} emission cannot account for the strong \xray\ variability between the two epochs.

We calculated the observed $f_{\rm 2~keV}$ values and the corresponding $\Delta\alpha_{\rm OX}$ for all the 10 segments.
The measured $\Delta\alpha_{\rm OX}$ values range from $-0.21$ ($f_{\rm weak}=3.5$) to $-0.09$ ($f_{\rm weak}=1.7$), with a mean of $-0.14$ ($f_{\rm weak}=2.3$).
We also computed \hbox{absorption-corrected} (intrinsic) $\alpha_{\rm OX,~corr}$ and $\Delta\alpha_{\rm OX,~corr}$ values from the \hbox{best-fit} intrinsic continuum.
The resulting $\Delta\alpha_{\rm OX,~corr}$ is 0.07, within the typical scatter ($\approx0.14$) of the \hbox{\aox--\ltkf} relation.
Therefore, our NICER spectral analyses reveal that I~Zw~1 emitted a nominal level of \xray\ emission as expected from the \hbox{\aox--\ltkf} relation during the 2022 observational epoch, and the observed \xray\ variability was solely driven by variable \hbox{partial-covering} absorption from the three absorbers.

The photon index of the intrinsic \xray\ \hbox{power-law} continuum derived from the 2022 NICER spectra differs from that of the
2020 \hbox{XMM-Newton} $+$ NuSTAR spectra ($2.09^{+0.03}_{-0.02}$ vs. $2.37\pm0.01$), while the normalizations are consistent within $1\sigma$ uncertainties ($6.20^{+0.17}_{-0.20}\times10^{-3}$ vs. $6.22^{+0.11}_{-0.23}\times10^{-3}~{\rm photons}~{\rm keV}^{-1}~{\rm cm}^{-2}~{\rm s}^{-1}$).
The difference in the $\Gamma$ values is likely caused by uncertainties in modeling the substantial NICER \hbox{non-\xray} background components (see Figure~\ref{fig:nicer_joint_fitting}).
We have verified that fixing the NICER $\Gamma$ and $A$ values to those obtained from the \hbox{XMM-Newton} $+$ NuSTAR spectra still yields a good fit (${\chi}^{2}/{\rm dof}=1038.7/1012$, $p_{\rm null}=0.27$), and the absorber properties remain unchanged.

The \hbox{best-fit} absorber parameters are presented in Table~\ref{tbl:best_fit_nicer}.
The three absorbers identified from the NICER spectra exhibit \hbox{long-term} variability amplitudes comparable to those of the three absorbers identified from the \hbox{XMM-Newton} and NuSTAR spectra.
The ${\rm XSTAR}_{1,~\rm NI}$ absorber exhibits the highest ionization parameter (${\rm log}~\xi=2.66$) and velocity ($v_{\rm abs} = -0.317c$) among the three absorbers.
It can be classified as a UFO due to its high velocity.
Its column density varied between $3.2\times10^{23}~{\rm cm}^{-2}$ and $1.0\times10^{25}~{\rm cm}^{-2}$, and its covering factor varied between $0.56$ and $0.78$.
The ${\rm XSTAR}_{2,~\rm NI}$ absorber (${\rm log}~\xi = 0.07$ and $v_{\rm abs} = -0.224c$) also shows variability, with a column density varied between $6.2\times10^{20}~{\rm cm}^{-2}$ and $1.1\times10^{22}~{\rm cm}^{-2}$ and a covering factor varied between $0.43$ and $1.00$.
It can also be classified as a UFO.
The ${\rm XSTAR}_{3,~\rm NI}$ absorber has the lowest ionization parameter (${\rm log}~\xi = -0.70$) and velocity $v_{\rm abs} = -0.057c$), and it shows the smallest variation in the column density among the three absorbers, ranging from $1.4\times10^{21}~{\rm cm}^{-2}$ to $2.7\times10^{21}~{\rm cm}^{-2}$.
The temporal evolution of the column densities and covering factors of the three absorbers is shown in Figure~\ref{fig:nicer_wind_variation}.

We compared the physical parameters of the three absorbers with those derived from the joint \hbox{XMM-Newton} $+$ NuSTAR analysis.
The three NICER absorbers all exhibit lower ionization parameters than the corresponding \hbox{XMM-Newton} absorbers.
Compared to ${\rm XSTAR}_{\rm 1,~X-N}$, ${\rm XSTAR}_{\rm 1,~NI}$ has comparable velocity and average column density, while the average covering factor is larger (0.66 vs. 0.40).
${\rm XSTAR}_{\rm 2,~NI}$ exhibits a significantly lower average column density ($4.0\times10^{21}~{\rm cm}^{-2}$) than ${\rm XSTAR}_{\rm 2,~X-N}$ ($8.4\times10^{22}~{\rm cm}^{-2}$), while its average covering factor (0.65 vs. 0.37) and blueshifted velocity ($-0.224c$ vs. $-0.049c$) are higher.
${\rm XSTAR}_{\rm 3,~NI}$ exhibits comparable velocity and average column density compared to ${\rm XSTAR}_{\rm 3,~X-N}$.
As discussed in Section~\ref{sec:properties_of_winds} below, the absorber properties are simplified, averaged estimates of the properties of complex absorption structures over different physical scales, and they are likely dependent on the spectral quality and the time binning scheme.
The NICER spectra have lower \hbox{signal-to-noise} ratios than the \hbox{XMM-Newton} or NuSTAR spectra, and the NICER monitoring observations spanned a longer timescale than the \hbox{XMM-Newton} $+$ NuSTAR observations ($\approx100$~days vs. 2~days); these factors should account for at least part of the differences between the absorber properties derived from the two datasets.
The average absorber properties might also evolve over the $\approx$\hbox{2-year} timescale.

\begin{deluxetable*}{lccccc}
\tablewidth{0pt}
\tablecaption{Physical Properties of the Absorbers
	\label{tbl:wind_physics}}
\tablehead{
\colhead{Absorber}  &
\colhead{${\rm log}~\xi$}  &
\colhead{$v_{\rm abs}$} & 
\colhead{$C_V$}  &
\colhead{$R_{\rm min}$}  &
\colhead{$R_{\rm max}$} \\
\colhead{ }  &
\colhead{(${\rm erg}~{\rm cm}~{\rm s}^{-1}$)}  &
\colhead{($c$)} & 
\colhead{ }  &
\colhead{($R_{\rm g}$)}  &
\colhead{($R_{\rm g}$)} \\
\colhead{(1)}  &
\colhead{(2)}  &
\colhead{(3)}  &
\colhead{(4)}  &
\colhead{(5)}  &
\colhead{(6)}
}
\startdata
\multicolumn{6}{c}{2020} \\
\midrule
${\rm XSTAR}_{1,~\rm X-N}$ & $3.85$ & $-0.245$ & $4.3\times10^{-4}$ & 33 & $44$  \\
${\rm XSTAR}_{2,~\rm X-N}$ & $2.59$ & $-0.049$ & $6.0\times10^{-4}$ & 844 & $1.9\times10^{4}$ \\
${\rm XSTAR}_{3,~\rm X-N}$ & $-0.41$ & $-0.035$ & $1.1\times10^{-6}$ & 1647 & $2.3\times10^{6}$  \\
\midrule
\multicolumn{6}{c}{2022} \\
\midrule
${\rm XSTAR}_{1,~\rm NI}$ & $2.66$ & $-0.317$ & $1.6\times10^{-5}$ & 20 & $72$ \\
${\rm XSTAR}_{2,~\rm NI}$ & $0.07$ & $-0.224$ & $8.4\times10^{-8}$ & 40 & $3.7\times10^{4}$ \\
${\rm XSTAR}_{3,~\rm NI}$ & $-0.70$ & $-0.057$ &  $1.7\times10^{-7}$ & 607 & $6.8\times10^{5}$
\enddata
\tablecomments{
Column (1): absorber name.
Columns (2)--(3): ionization parameter in units of ${\rm erg}~{\rm cm}~{\rm s}^{-1}$ and blueshifted velocity in units of $c$, both adopted from Table~\ref{tbl:best_fit_xmm_nustar} and Table~\ref{tbl:best_fit_nicer}.
Column (4): gas volume filling factor of the absorber.
Column (5): minimum radius of the absorber in units of $R_{\rm g}$.
Column (6): maximum radius of the absorber in units of $R_{\rm g}$.
}
\end{deluxetable*}

\section{Discussion}\label{sec:Discussion}

\subsection{Properties of the Winds}\label{sec:properties_of_winds}
We achieved our first science goal through the analyses in Section~\ref{sec:xray_spec_analyses}, demonstrating that the observed \xray\ variability from \hbox{XMM-Newton}, NuSTAR, and NICER can be well explained solely via evolving obscuration; the \hbox{accretion-disk} optical/UV emission and the coronal \xray\ emission remained stable during the observation periods.
\hbox{Longer-term} optical/IR light curves (Section~\ref{sec:multiwave_var}) also suggest that the accretion process is overall stable.

In this subsection, we aim to constrain the \hbox{disk-wind} properties from the four parameters (ionization parameter, velocity, column density, and covering factor) of the absorbers.
We caution that the absorber properties are simplified, averaged estimates of the properties of complex absorption structures over different physical scales.
The number of absorbers, the \hbox{best-fit} parameters, and the parameter variations are likely dependent on the spectral quality.
To demonstrate this, we considered a simple scenario where I~Zw~1 is placed at a distance 3 times larger, and then the \hbox{XMM-Newton} T1 spectral counts would decrease from $\approx 34,000$ to $\approx 3,700$.
We simulated a \hbox{3,700-count} spectrum based on the \hbox{best-fit} model for the T1 segment.
Instead of the Equation~\ref{eq:model} model with three ionized \hbox{partial-covering} absorbers, this simulated spectrum could be adequately described by a \hbox{partial-covering} absorption model with just a single neutral absorber ($N_{\rm H}=9.1\times10^{21}~{\rm cm}^{-2}$, $f_{\rm cov}=0.50$, $\chi^{2}/{\rm dof}=42.8/41$, and $p_{\rm null}=0.39$).
These absorber constraints are similar to the results reported for the extremely \hbox{X-ray} weak and \hbox{X-ray} variable \hbox{super-Eddington} accreting quasars where the spectral quality is \hbox{low-to-moderate} \citep[e.g.,][]{Liu2019, Huang2023}.
On the other hand, if the spectral statistics improve substantially, more detailed features of the absorbers might be revealed.
Moreover, the parameter variations are clearly dependent on the time resolution of our binning scheme, as I~Zw~1 exhibits obvious variability within individual \hbox{XMM-Newton} or NICER time segments.
Therefore, our current discussion of the I~Zw~1 wind properties is limited by the available datasets and our adoption of the particular \hbox{partial-covering} absorption model.
Nevertheless, this simplified approach might still offer insights into the \hbox{disk-wind} absorbers in \hbox{super-Eddington} accreting AGNs.

We first estimate the locations of the three absorbers derived from the 2020 observations.
We adopted the basic assumptions and formulas presented in \cite{Blustin2005} and \cite{Gofford2015}.
Assuming that the velocity of each wind is its escape velocity, we can calculate the minimum launching radius as
\begin{equation}
    R_{\rm min}\ge2GM_{\rm BH}/{v_{\rm abs}^{2}},
\end{equation}where $G$ is the gravitational constant, $M_{\rm BH}=9.3\times10^{6}M_{\odot}$ is the SMBH mass of I~Zw~1, and $v_{\rm abs}$ is the velocity of the absorber.
The maximum distance of the absorber to the SMBH is estimated as
\begin{equation}
\label{eq:r_max}
    R_{\rm max}\le \frac{L_{\rm ion}C_{V}}{{\xi}N_{\rm H}}, 
\end{equation}where $L_{\rm ion}$ is the 13.6~eV to 13.6~keV ionizing luminosity computed from the intrinsic SED model we construct in Section~\ref{sec:uv_ext_sed}, $C_{V}$ is the gas volume filling factor, and $N_{\rm H}$ is the column density of the absorber.
Assuming a mass outflow rate of $\dot{M}\sim1.23m_{\rm p}L_{\rm ion}C_{V}v_{\rm abs}\Omega/\xi$, where $m_{\rm p}$ is the proton mass and $\Omega$ is the solid angle of the absorber, the volume filling factor $C_{V}$ can be estimated as
\begin{equation}
\label{eq:c_v}
    C_{\rm V}\sim
    \frac{\dot{P}\xi}
    {1.23m_{\rm p}cL_{\rm ion}v_{\rm abs}^{2}\Omega},
\end{equation}where $\dot{P}=\dot{M}v_{\rm abs}$ is the momentum rate that is of the same order of magnitude as the momentum of the radiation it absorbs \citep[e.g.,][]{Gofford2015, King2016}.
To calculate $R_{\rm max}$ using Equations~\ref{eq:r_max} and~\ref{eq:c_v}, we assumed that $\dot{P}\sim L_{\rm ion}/c$ and $\Omega=1.6\pi$ \citep[e.g.,][]{Blustin2005}, and adopted the median value of $N_{\rm H}$ for each absorber.
We list the estimated distances in Table~\ref{tbl:wind_physics}.
Clearly the three absorbers are at different locations: 
\begin{enumerate}
    \item 
    ${\rm XSTAR}_{1,~\rm X-N}$ exhibits the highest ionization parameter (${\rm log}~\xi = 3.85$) and the highest blueshifted velocity ($v_{\rm abs} = -0.245c$) among the three absorbers.
    Its location is tightly restricted between $33R_{\rm g}$ and $44R_{\rm g}$ ($R_{\rm g}=GM_{\rm BH}/c^{2}\sim4.5\times10^{-7}~{\rm pc}$), suggesting a \hbox{dust-free} disk wind launched from the innermost region of the accretion disk in the immediate vicinity of the central SMBH.
    \item 
    ${\rm XSTAR}_{2,~\rm X-N}$ (${\rm log}~\xi = 2.59$) is loosely constrained to locate between $844R_{\rm g}$ and $1.9\times 10^{4}R_{\rm g}$.
    To better interpret these radii, we compared them with the estimated radius of the H$\beta$ \hbox{broad-line} region (BLR).
    Based on RM observations, \cite{Huang2019} measured a time delay of $\tau_{{\rm H}\beta} \approx 37$~days, corresponding to $R_{{\rm H}\beta} \approx 7 \times 10^{4}R_{\rm g}$.
    Therefore, $R_{\rm max}$ of ${\rm XSTAR}_{2,~\rm X-N}$ corresponds to a distance of $\approx 0.2R_{\rm H\beta}$.
    This places the absorber within the BLR, suggesting that it likely originates from a \hbox{dust-free} disk wind launched in the innermost accretion flow.
    \item 
    ${\rm XSTAR}_{3,~\rm X-N}$ (${\rm log}~\xi = -0.41$) is loosely constrained between $1647R_{\rm g}$ and $2.3 \times 10^{6}R_{\rm g}$ ($\sim0.7~{\rm pc}$), corresponding to a range of $\approx 0.02R_{\rm H\beta}$ to $\approx 33R_{\rm H\beta}$.
    The location of ${\rm XSTAR}_{3,~\rm X-N}$ is quite uncertain; the $R_{\rm max}$ value is even larger than the \hbox{mid-IR} torus radius ($\approx 6.2 \times 10^{5}R_{\rm g}$; \citealt{Chen2023}) inferred from the time delay between \hbox{mid-IR} and optical light curves.
    ${\rm XSTAR}_{3,~\rm X-N}$ is found to fully cover the corona, and its column density also varied the least among the three absorbers, suggesting a less clumpy structure.
    It is thus likely the most distant absorber among the three.
    However, considering its substantial blueshifted velocity ($v_{\rm abs}=-0.035c$), ${\rm XSTAR}_{\rm 3,~X-N}$ probably still corresponds to a \hbox{small-scale} disk wind component that is within the BLR.
\end{enumerate}

\begin{figure*}
\centering
\includegraphics[scale=0.40]{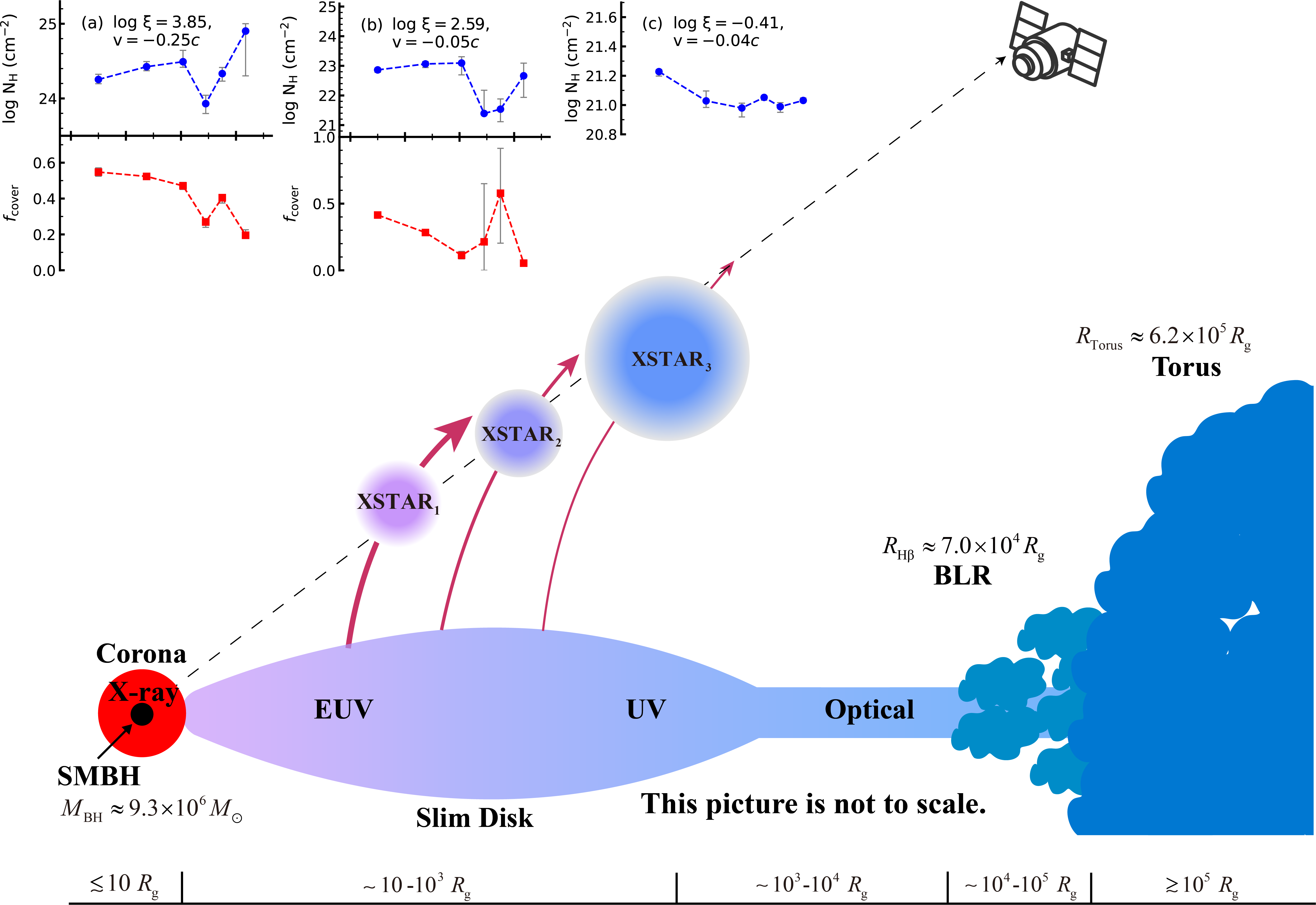}
\caption{Cartoon showing the locations of the three \xray\ absorbers in I~Zw~1.
The three absorbers, ordered by increasing distance from the central SMBH, are ${\rm XSTAR}_1$, ${\rm XSTAR}_2$, and ${\rm XSTAR}_3$.
We present the temporal evolution during the first 2020 \hbox{XMM-Newton} observation (Obs ID: 0851990101) of the column density and covering factor for each absorber.
The red arrowed curves represent the three \hbox{disk-wind} components associated with the corresponding clumpy absorbers; thicker curves indicate higher gas velocities.
}
\label{fig:cartoon}
\end{figure*}

We also examined if the parameter variations between the different time segments provide any useful constraints on the transverse velocities.
For ${\rm XSTAR}_{1,~\rm X-N}$, the covering factor varied by $\approx 0.20$ within a timescale of $\approx 10{\rm ~ks}$.
Assuming a typical coronal size of 10$R_{\rm g}$ \citep[e.g.,][]{Dai2010, Shemmer2014, Fabian2015}, the corresponding minimum transverse travel distance of the clumpy absorber is thus $2R_{\rm g}$.
This yields a lower limit on the transverse velocity of $|v_{\rm abs}|>0.009c$, which is too loose to provide any useful constraints on the location of ${\rm XSTAR}_{1,~\rm X-N}$.
For ${\rm XSTAR}_{2,~\rm X-N}$, the lower limit on the transverse velocity is $|v_{\rm abs}|>0.03c$, which is consistent with the value of $v_{\rm abs}=-0.049c$ from the \hbox{best-fit} model.
Based on the estimation of the absorber locations, we draw a schematic picture of the wind obscuration scenario for I~Zw~1 in Figure~\ref{fig:cartoon}.
We also labeled the estimated locations of the H$\beta$ BLR and the torus in Figure~\ref{fig:cartoon}.


To identify the primary drivers of the observed flux variations in the 2020 observations, we performed Spearman \hbox{rank-order} correlation tests on the observed \xray\ fluxes in the \hbox{XMM-Newton} \hbox{0.3--10~keV} and NuSTAR \hbox{10--24~keV} bands against the column densities and covering factors of the individual absorbers.
While no significant correlations are found for the \hbox{0.3--10~keV} flux, the \hbox{10--24~keV} flux exhibits a significant negative correlation exclusively with the covering factor of ${\rm XSTAR}_{\rm 1,~X-N}$ ($r_{\rm Spearman} = -0.700$, $p = 0.016$).
This demonstrates that the hard \xray\ flux variability is mainly driven by the varying covering factor of the innermost absorber, while the \hbox{0.3--10~keV} flux is modulated by the joint effects of the three absorbers, which blurs any \hbox{single-parameter} correlations.

Our analysis identifies three absorbers which are consistent in number with the findings of \cite{Rogantini2022}.
They jointly fit the \hbox{time-averaged}, 2020 \hbox{XMM-Newton} RGS and pn spectra, and identified two warm absorbers (${\rm log}~\xi\sim1.7$, $v_{\rm abs}\sim-0.007c$, and $N_{\rm H}\sim1\times10^{20}~{\rm cm}^{-2}$; ${\rm log}~\xi\sim-1$, $v_{\rm abs}\sim-0.006c$, and $N_{\rm H}\sim9\times10^{20}~{\rm cm}^{-2}$) and one UFO (${\rm log}~\xi\sim3.8$, $v_{\rm abs}\sim-0.26c$, and $N_{\rm H}\sim2\times10^{22}~{\rm cm}^{-2}$).
The ${\rm XSTAR}_{2,~\rm X-N}$ and ${\rm XSTAR}_{3,~\rm X-N}$ absorbers that we identified from the pn data exhibit larger ${\rm log}~\xi$, $v_{\rm abs}$, and $N_{\rm H}$ values compared to those of the two warm absorbers in \cite{Rogantini2022}, likely due to the different analysis approaches (including energy bandpass, time resolution, and absorption modeling).
While the ionization parameter and velocity of our ${\rm XSTAR}_{1,~\rm X-N}$ absorber (${\rm log}~\xi=3.85$, $v_{\rm abs}=-0.245c$) align notably well with those of their reported UFO, the column density of ${\rm XSTAR}_{1,~\rm X-N}$ is much higher ($N_{\rm H}\sim1\times10^{24}~{\rm cm}^{-2}$).
\cite{Rogantini2022} reported no Fe absorption lines which might be associated with the UFO.
We examined both the \hbox{time-averaged} and \hbox{time-resolved} pn spectra above \hbox{5~keV} by fitting a simple \hbox{power-law} model, and we found no apparent absorption features either.

\cite{Juranova2024} detected four distinct ionized outflows in I~Zw~1 from the 2015 HST Cosmic Origins Spectrograph UV spectrum, at \hbox{rest-frame} velocities of $-60$, $-280$, $-1950$, and $-2900~{\rm km}~{\rm s}^{-1}$ ($\approx0.01c$).
The column densities of the absorbers were constrained to \hbox{$\sim 5\times10^{17}$--$1\times10^{20}~{\rm cm}^{-2}$}, and they were considered to lie at distances comparable to the BLR.
A $-1870~{\rm km}~{\rm s}^{-1}$ UV absorber was also reported based on the 1997 HST Faint Object Spectrograph spectrum \citep{Laor1997}.
This $-1950/-1870~{\rm km}~{\rm s}^{-1}$ UV absorber is probably associated with the $-1870\pm70~{\rm km}~{\rm s}^{-1}$ \xray\ absorber identified in the 2015 RGS spectrum \citep{Silva2018}.
In the 2020 RGS spectrum, a similar \xray\ warm absorber (one of the warm absorbers mentioned above) with $v_{\rm abs}=-1750\pm100~{\rm km}~{\rm s}^{-1}$ was detected \citep{Rogantini2022}.
The column densities of these UV absorbers are too low to affect significantly the \hbox{0.3--24~keV} continuum we studied here, and thus we were not able to identify any of them.
On the other hand, our three \xray\ absorbers exhibit comparable ionization parameters to the UV absorbers, and their column densities and velocities are much higher.
If any of the \xray\ absorbers shields a large portion of the UV emission along the \hbox{line-of-sight} (i.e., a large covering factor), it should produce strong UV absorption lines with large blueshifts.
We inspected the HST spectrum from \cite{Juranova2024} and found no significant additional \ion{C}{4} $+$ \ion{N}{5} absorption doublet systems.
Therefore, our \xray\ absorbers likely do not cover a large fraction of the \hbox{UV-emitting} region (covering factors $\lesssim0.1$ considering the  \hbox{signal-to-noise} ratio of the HST spectrum).
The 1500~\AA\ continuum emission radius is $\sim1700R_{\rm g}$ under the standard thin accretion disk model \citep[e.g.,][]{Shakura1973, Frank2002}.
The three absorbers are probably all situated within this radius, and thus would not be expected to imprint strongly on the UV spectrum. 
Their much higher densities and outflow velocities than those of the UV absorbers also support a \hbox{small-scale} origin.

For the three NICER absorbers, we performed similar estimates.
The results are summarized in Table~\ref{tbl:wind_physics}.
Overall, the three absorbers overlap in location with the corresponding \hbox{XMM-Newton} $+$ NuSTAR absorbers.
As discussed in Section~\ref{sec:nicer_spec_analyses}, differences in the absorber properties from the two datasets are expected.
${\rm XSTAR}_{1,~\rm NI}$, ${\rm XSTAR}_{2,~\rm NI}$, and ${\rm XSTAR}_{3,~\rm NI}$ all likely originate from \hbox{small-scale}, \hbox{dust-free} disk winds.
We also performed Spearman \hbox{rank-order} correlation tests on the observed NICER \hbox{0.34--9~keV} flux versus the column density or covering factor of the individual absorbers.
We found significant correlations between the flux and both the covering factor ($r_{\rm Spearman} = -0.879$, $p = 0.001$) and column density ($r_{\rm Spearman} = -0.673$, $p = 0.033$) of ${\rm XSTAR}_{1,~\rm NI}$. 
This suggests that on a longer timescale ($\approx 100$~days), the properties of the innermost clumpy \hbox{disk-wind} absorber govern the observed \xray\ variability.

\subsection{Clumpy Disk-Wind Obscuration Scenario}\label{sec:clumpy_winds_in_AGNs}
Motivated by our previous investigations of the extreme \xray\ variability of \hbox{super-Eddington} accreting quasars \citep[e.g.,][]{Liu2019, Liu2021, Liu2022, Ni2020, Wang2024}, we performed the \hbox{time-resolved} spectral analyses of the \hbox{XMM-Newton}, NuSTAR, and NICER observations of I~Zw~1, to interpret its strong \xray\ variability within a pure obscuration scenario and to further examine properties and evolution of the \xray\ absorbers.
We found that the variable \hbox{time-resolved} spectra of I~Zw~1 can be well described using a simple \hbox{power-law} model modified by three \hbox{partial-covering} ionized absorbers, with the \hbox{power-law} normalization and photon index tied between different epochs.
Alternatively, relativistic disk reflection models have been invoked to interpret the strong \xray\ variability of NLS1s in general \citep[e.g.,][]{Fabian2012, Parker2014, Jiang2018, Wilkins2022, Ding2022}, where the normalization of \hbox{power-law} continuum is allowed to vary.
It is generally difficult to distinguish between a relativistic disk reflection model and a \hbox{partial-covering} absorption model with \xray\ data alone \citep[e.g.,][]{Tanaka2004, Miller2009, Marinucci2014, Brenneman2025}.
However, strongly \xray\ variable NLS1s generally lack corresponding strong optical/UV/IR variability, and at least for I~Zw~1, the apparent \xray\ flares still do not exceed the expectation from its optical/UV emission (Figure~\ref{fig:lc_xmm_nustar_om}a).
Therefore, the obscuration scenario appears to be the more natural explanation, where stable coronal \xray\ emission is modified by variable absorption along the line of sight.
This interpretation for objects such as I~Zw~1 also unifies the remarkable \xray\ properties of NLS1s and \hbox{super-Eddington} accreting quasars under the same physical framework, where \hbox{super-Eddington} accretion launches powerful winds from the inner accretion disk and the \hbox{dust-free} winds modify the observed \xray\ emission.
We note that the obscuration scenario can also accommodate an intrinsic continuum with a minor contribution from a disk reflection component, which is subsequently modified by wind obscuration.

A broad Gaussian Fe emission line was detected in addition to the \hbox{absorption-modified} continuum (Section~\ref{sec:xmm_spec_analyses}).
In disk reflection models, an asymmetric broad Fe K emission line is the canonical signature of relativistic reflection from the inner accretion disk.
It is plausible that the broad Fe line of I~Zw~1 includes some contribution from disk reflection, which does not contradict our overall model provided that the associated continuum component is not significant.
However, a broad Gaussian line could also arise from outflowing disk winds \citep[e.g.,][]{Reeves2019}.
For I~Zw~1, ${\rm XSTAR}_{\rm 3,~X-N}$ represents a relatively \hbox{large-scale} wind component that may subtend a large solid angle to the \xray\ corona for distant reflection, and its outflow velocity ($\approx -0.035c$) is of the right order of magnitude to match the observed $\approx0.5~{\rm keV}$ line width.
The line is unlikely to vary on short timescales, and the $\approx6.7~{\rm keV}$ centroid energy corresponds to the resonance line of \hbox{He-like} iron (\ion{Fe}{25}).
A few analogous examples include the broad Fe emission lines of PG~1211+143 \citep[e.g.,][]{Pounds2003} and PDS~456 \citep[e.g.,][]{Nardini2015, Reeves2020}; both quasars are also considered to be undergoing \hbox{super-Eddington} accretion.
It has also been reported that broad Fe lines are detected more frequently in NLS1s than in \hbox{broad-line} Seyfert 1 galaxies \citep[e.g.,][]{Liu2016, Waddell2022}.
Such lines have not been detected to date in our previous investigations of the extreme \xray\ variability of higher mass/luminosity \hbox{super-Eddington} accreting quasars \citep[e.g.,][]{Liu2019, Ni2020, Liu2022, Huang2023, Wang2024}.
In some objects (e.g., SDSS J0814+5325; \citealt{Huang2023}), the lack of line detection might be due to the limited spectral quality.
However, a few WLQs (e.g., PHL 1811; \citealt{Wang2022}) are affected by \hbox{Compton-thick} obscuration, and the $\sim6$--7~keV continuum is $\sim100$ times weaker that expected.
The lack of a strong Fe line is likely intrinsic, probably caused by a large \hbox{covering-factor} absorber that blocks the view to the far side of the absorber/reflector.
The presence of a broad Gaussian Fe emission line is nevertheless consistent with the \hbox{disk-wind} obscuration scenario.

\begin{figure*}
\centering
\includegraphics[scale=0.55]{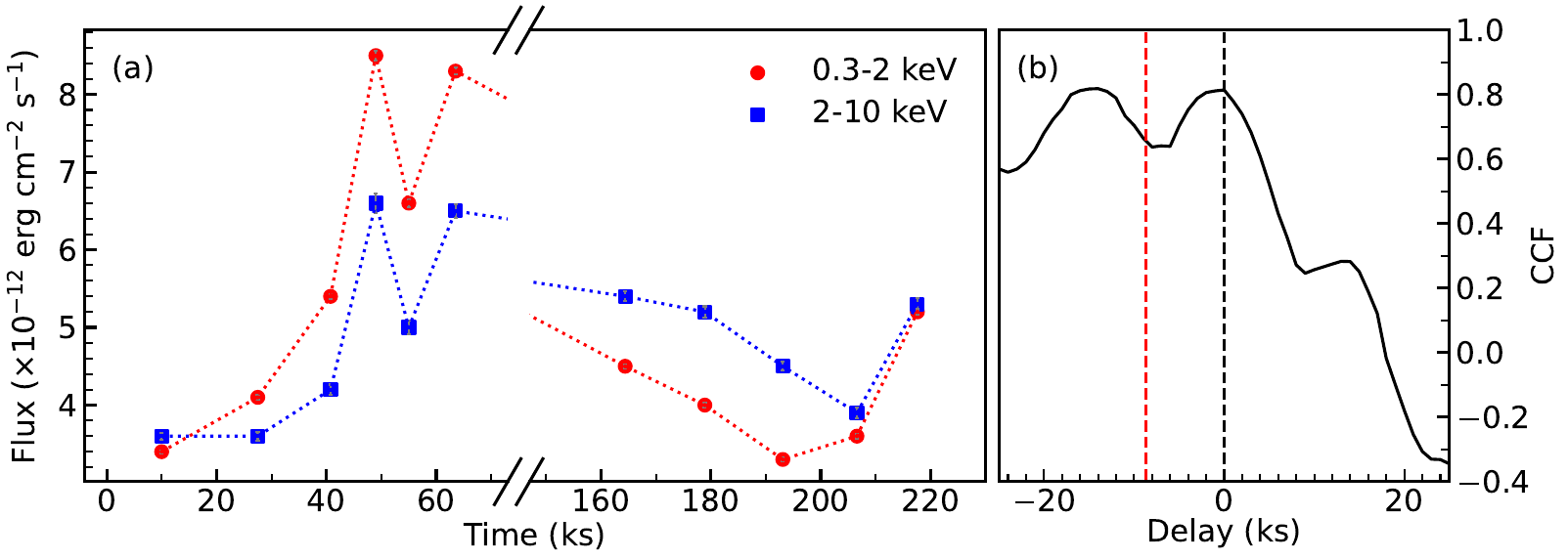}
\caption{(a) Soft \xray\ (\hbox{0.3--2~keV}; red circles) and hard \xray\ (\hbox{2--10~keV}; blue squares) flux light curves for the 11 segments of the 2020 \hbox{XMM-Newton} observations determined from the \hbox{best-fit} \hbox{partial-covering} absorption model.
Uncoordinated \hbox{soft--hard} \xray\ variability is present within the \hbox{10--30~ks} and \hbox{190--210~ks} time intervals.
(b) The CCF for the two flux light curves.
The peak with a negative time lag suggests that the \hbox{0.3--2~keV} variability precedes the \hbox{2--10~keV} variability.
The red dashed line represents the centroid time lag of $\tau_{\rm cent}=-8.7^{+5.9}_{-10.3}~{\rm ks}$.
}
\label{fig:lag}
\end{figure*}

Soft--hard \xray\ (e.g., \hbox{0.3--1~keV} vs. \hbox{1.2--4~keV} bands) time lags have sometimes been detected in the \xray\ light curves of NLS1s, which have been considered to support the relativistic disk reflection models where the lags reflect \hbox{light-travel-time} delays due to \xray\ reverberation from the inner region of the accretion disk \citep[e.g.,][]{Fabian2009, Uttley2014, Wilkins2017, Wilkins2021, Cackett2021}.
In some cases, a reverberation lag was additionally detected between the broad iron K band (\hbox{4--7~keV}) and the continuum band (\hbox{1.2--4~keV}), providing further support for the reflection models (e.g., \citealt{Wilkins2023}).
These reverberation time lags are often \hbox{time-variable} within individual observations; in disk reflection models, the intrinsic coronal emission (driving the \hbox{continuum-band} variability) as well as the coronal scale height (causing the varying time lags) are both considered to be variable.
For the 2020 \hbox{XMM-Newton} observations of I~Zw~1, soft lags of the order of 500~s were reported (e.g., Figure~7 of \citealt{Wilkins2023}).
In our pure obscuration scenario discussed here, neither the coronal emission nor the coronal height is significantly variable.
Since our spectral model can explain the \hbox{time-resolved} \xray\ spectra (Section~\ref{sec:xmm_spec_analyses}), it should also be able to reproduce any time lags determined from the \xray\ data within the time resolution of our study.
Varying \hbox{partial-covering} absorption can indeed produce varying \hbox{soft--hard} \xray\ time lags.
For example, with appropriate parameter setups, a decrease in the column density with an increase in the covering factor of the absorber would result in a smaller soft \xray\ flux but a larger hard \xray\ flux; the presence of such uncoordinated \hbox{soft--hard} \xray\ variability could effectively lead to ``time’’ lags between light curves that are not related to \hbox{light-travel-time} delays.
A number of such discrete lag events randomly distributed within a given observation could then produce, not a universal time lag, but multiple lags that vary significantly in either the temporal or frequency domain.
As a simple demonstration, we plot in Figure~\ref{fig:lag} the soft \xray\ (\hbox{0.3--2~keV}) and hard \xray\ (\hbox{2--10~keV}) flux light curves for the 11 segments of the 2020 \hbox{XMM-Newton} observations determined from the \hbox{best-fit} \hbox{partial-covering} absorption model. Uncoordinated \hbox{soft–hard} \xray\ variability is present within the \hbox{10--30~ks} and \hbox{190--210~ks} time intervals.
An analysis of the light curves with the ICCF even revealed a negative centroid time lag (hard lag) of $\tau_{\rm cent}=-8.7^{+5.9}_{-10.3}~{\rm ks}$.
Our time resolution (i.e., 11 segments) is insufficient to allow for direct comparisons with the $\sim500~{\rm s}$ soft lags found in \cite{Wilkins2023}; \hbox{partial-covering} absorption modeling of the spectra with finer time bins would be interesting but is beyond the scope of the current paper.

Within the \xray\ weak/variable \hbox{super-Eddington} accreting AGNs, different objects apparently exhibit different strengths of \xray\ weakness or variability.
Luminous quasars like PHL~1092 may have $f_{\rm weak}$ and $f_{\rm var}$ values of up to a few hundred \citep[e.g.,][]{Miniutti2012}, whereas for I~Zw~1, these factors are of the order of a few.
The \xray\ weakness and variability factors are likely governed by the wind strength (defined by parameters like density, velocity, solid angle, clumpiness), which is considered to be fundamentally linked to the normalized mass accretion rate and SMBH mass \citep[e.g.,][]{Giustini2019}.
I~Zw~1 has a relatively lower accretion rate and a smaller SMBH mass, and thus its wind is less dense (smaller $N_{\rm H}$) and/or more clumpy (smaller $f_{\rm cov}$), resulting in \hbox{less-significant} \xray\ absorption.

Another important observational parameter is the occurrence rate of strong \xray\ weakness for a given object.
A stronger wind (e.g., denser, larger solid angle, and less clumpy) likely results in a higher likelihood of observing \xray\ weak states and a lower likelihood of observing \xray\ \hbox{nominal-strength} states.
For a sample of these objects, this occurrence rate translates to the fraction of \xray\ weak sources within the population.
For example, WLQs exhibit a $\sim50\%$ \xray\ weak fraction \citep[e.g.,][]{Luo2015}, higher than the \hbox{$\sim20\%$--$30\%$} fractions found for other samples of \hbox{super-Eddington} accreting AGNs \citep[e.g.,][]{Liu2019, Nardini2019, Laurenti2022}.
Moreover, only a few \xray\ weak WLQs have been found to ever recover to \xray\ \hbox{nominal-strength} states \citep[e.g.,][]{Miniutti2012, Ni2020, Wang2024}.
These findings support the notion that WLQs are extremely \hbox{super-Eddington} accreting which was proposed to explain also their exceptional UV and optical \hbox{emission-line} properties \citep[e.g.,][]{Luo2015, Ni2022, Chen2024}.
A subpopulation of WLQs, the PHL 1811 analogs that were selected to have additional \ion{C}{4} blueshifts and strong \ion{Fe}{2} and \ion{Fe}{3} emission, appear to possess even more powerful winds as the \xray\ weak fraction is $\approx94\%$ and their average $f_{\rm weak}$ value is more than double the average value for typical WLQs \citep{Luo2015}.

The LRDs recently discovered by JWST have opened a new window to study \hbox{super-Eddington} accretion.
These \hbox{high-redshift} ($z\gtrsim4$) sources are identified by their \hbox{V-shaped} \hbox{UV--optical} continua (with spectral turnovers around the Balmer limit) and compact morphology.
A substantial fraction of LRDs display broad emission lines, indicating the presence of AGN activity \citep[e.g.,][]{Kocevski2023, Greene2024, Ji2025, Kocevski2025}.
LRDs share key features with local \hbox{super-Eddington} accreting AGNs, including minimal optical/UV variability \citep[e.g.,][]{Zhang2024, Tee2025} and extreme \xray\ weakness; some LRDs also show weak UV \ion{C}{4} and optical [\ion{O}{3}] emission lines.
LRDs are almost universally \xray\ weak \citep[e.g.,][]{Maiolino2024, Yue2024, Maiolino2025}, similar to the PHL 1811 analogs.
Recent studies have proposed that LRDs are AGNs embedded in ``cocoons'' of dense ionized/excited \hbox{dust-poor} gas \citep[e.g.,][]{D'Eugenio2025, de-Graaff2025, Inayoshi2025b, Kokorev2025, Torralba2025}.
Given our interpretation of the \xray\ weakness/variability of \hbox{super-Eddington} accreting AGNs, it is natural to consider that LRDs are extreme versions of WLQs, where powerful disk winds driven by extremely high accretion rates enshroud the entire nucleus.
Surveys of \xray\ weakness factors, \xray\ weakness fraction, and \xray\ variability in LRDs will help validate and refine their wind obscuration scenario.

\begin{figure*}
\centering
\includegraphics[scale=0.45]{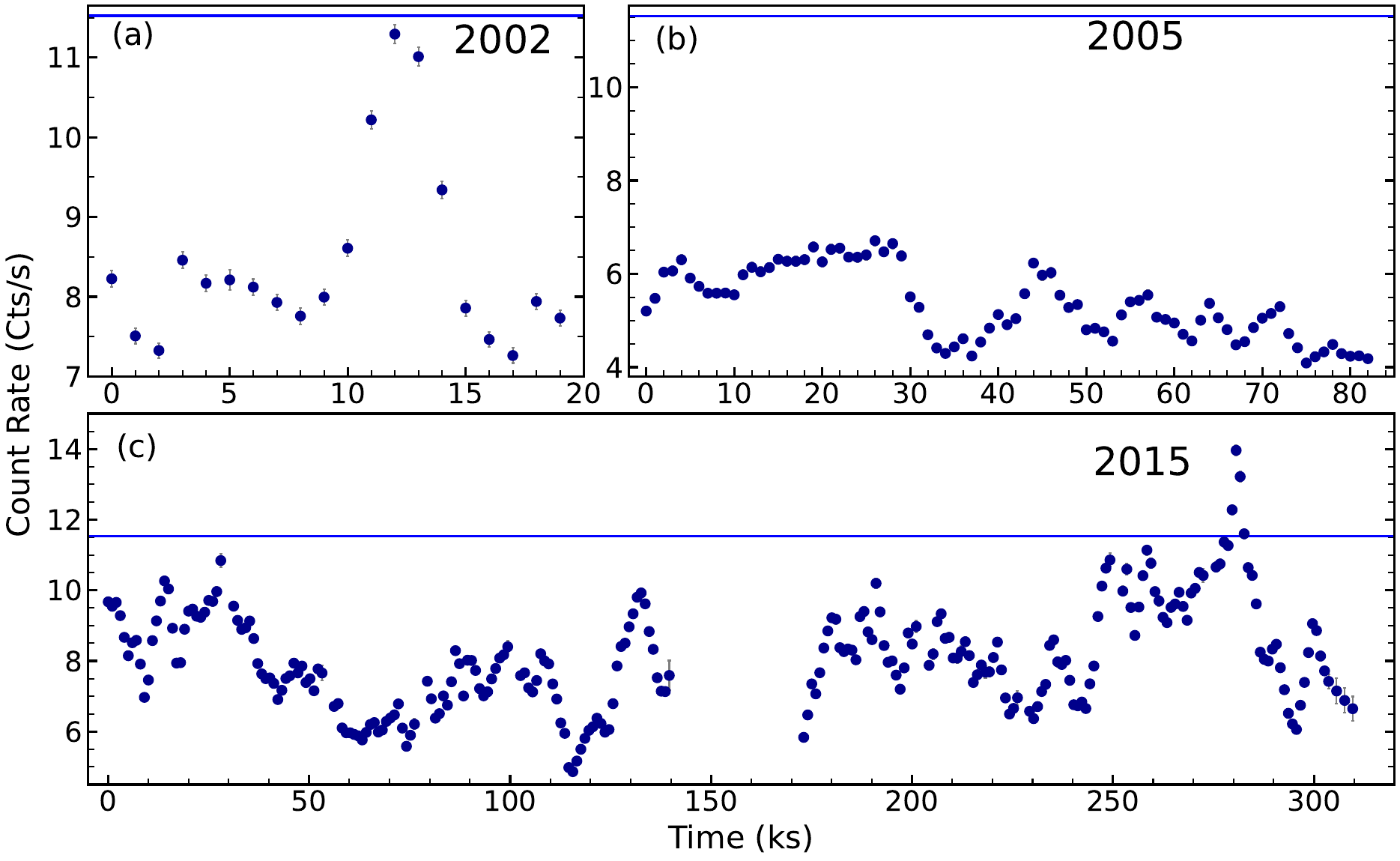}
\caption{\hbox{EPIC-pn} light curves in the \hbox{0.3--10~keV} band of I Zw 1 from the (a) 2002, (b) 2005, and (c) 2015 \hbox{XMM-Newton} observations, with a bin size of 1 ks.
The blue line in each panel represents the expected intrinsic \hbox{0.3--10~keV} count rate determined from the \hbox{best-fit} \hbox{partial-covering} absorption model in Section~\ref{sec:xmm_spec_analyses}.
}
\label{fig:previous_xmm}
\end{figure*}

This is our first attempt to apply the \hbox{wind-obscuration} scenario to local \hbox{super-Eddington} accreting AGNs.
The results appear promising.
Besides the 2020 \hbox{XMM-Newton} observations, I~Zw~1 also has a few earlier \hbox{XMM-Newton} observations with significant exposures ($\sim$20--140~ks).
A uniform analysis of all these observations is beyond the scope of the current paper, but we nevertheless briefly examined whether these data significantly violate our pure obscuration model.
We extracted \hbox{0.3--10~keV} pn \hbox{count-rate} light curves for the 2002, 2005, and 2015 observations, which are presented in Figure~\ref{fig:previous_xmm}.
They all exhibit strong \xray\ variability, with a maximum variability amplitude of $f_{\rm var}\approx3$.
For comparison, we include in these light curves the expected intrinsic \hbox{0.3--10~keV} count rate determined from the \hbox{best-fit} \hbox{partial-covering} absorption model in Section~\ref{sec:xmm_spec_analyses}.
Most of the observed count rates are below the expectation, consistent with an obscuration interpretation.
We further fitted the 2002 pn spectrum with our \hbox{partial-covering} absorption model (Equation~\ref{eq:model}).\footnote{The 2002 spectrum has a total exposure of $\sim$20~ks, comparable to the exposure times of the individual segments for the 2020 observations.
Due to the variation of the absorber parameters, longer observations are more difficult to explain without breaking them into segments.}
The \hbox{best-fit} model describes the spectrum well, with ${\chi}^2/{\rm d.o.f.}=155.5/143$ ($p_{\rm null}=0.22$).
The derived nominal level of intrinsic \xray\ emission ($\Delta\alpha_{\rm OX,~corr}=0.08$) is also consistent with our \hbox{best-fit} model in Section~\ref{sec:xmm_spec_analyses}.
Therefore, the previous \hbox{XMM-Newton} data do not appear to violate our obscuration model.

It will be useful to adopt the same approach to study other local \xray\ variable NLS1s or \hbox{super-Eddington} accreting AGNs with \hbox{high-quality} \xray\ spectra.
We list in Table~\ref{tbl:wind_agns} a sample of such AGNs compiled from the literature.
They are all characterized by strong \xray\ variability ($f_{\rm var}\gtrsim5$), and their \hbox{low-state} \xray\ spectra appear to have sufficient statistics for detailed analyses.
They have relatively large Eddington ratios ($\lambda_{\rm Edd} \gtrsim 0.2$), and we verified that they exhibit only mild optical/UV/IR variability on long timescales.
A few of these objects have been investigated previously using \hbox{partial-covering} absorption models \citep[e.g,][]{Tanaka2004, Boller2021, Midooka2023}.
However, these studies generally allowed the \hbox{power-law} normalization to vary freely between epochs and did not evaluate the $\Delta\alpha_{\rm OX}$ or $\Delta\alpha_{\rm OX,~corr}$ parameters.
A sample study will help validate our unification of the \xray\ weakness/variability in NLS1s and \hbox{super-Eddington} accreting quasars under the obscuration scenario, reveal the dependence of \xray\ weakness factor or \xray\ weakness fraction on accretion parameters, and provide additional insights into \hbox{accretion-disk} wind properties.

\begin{deluxetable*}{llccll}
\tablewidth{0pt} 
\tablecaption{NLS1s and Quasars with Strong \xray\ Variability ($f_{\rm var}\gtrsim5$)}
\tablehead{
\colhead{Object}  &
\colhead{$z$}  &
\colhead{${\rm log}~{L}_{\rm bol}$} &
\colhead{${\rm log}~M_{\rm BH}$} &
\colhead{$\lambda_{\rm Edd}$}  &
\colhead{References} \\
\colhead{ }  &
\colhead{ }  &
\colhead{$({\rm erg~s^{-1}})$} &
\colhead{$(M_{\odot})$} &
\colhead{ }  &
\colhead{ } \\
\colhead{(1)} &
\colhead{(2)} &
\colhead{(3)} &
\colhead{(4)} &
\colhead{(5)} &
\colhead{(6)}
}
\startdata
MCG-6-30-15 & $0.008$ & 43.6 & 6.2 & 0.2 & \cite{Wang2004}, \cite{Marinucci2014} \\
Mrk 335 & $0.026$ & 45.1 & 6.9 & 1.3 & \cite{Grupe2010} \\
Mrk 382 & $0.034$ & 44.4 & 6.5 & 0.6 & \cite{Liu2021} \\
1H 0707-495 & $0.040$ & 44.5 & 6.6 & $0.6$ & \cite{Done2016}, \cite{Boller2021}\\
Mrk 1048 & $0.042$ & 45.3 & 8.0 & 0.1 & \cite{Grupe2010} \\
Mrk 142 & $0.045$ & 44.3 & 6.2 & 0.9 & \cite{Cackett2020}\\
I Zwicky 1 & $0.061$ & 45.5 & 7.0 & $2.7$ & \cite{Rogantini2022}, this work\\
PG 1126-041 & $0.062$ & 45.3 & 7.7 & $0.3$ & \cite{Giustini2011}, \cite{Giustini2023}\\
PG 0844+349 & $0.064$ & 45.4 & 7.2 & 1.2 & \cite{Gallo2011}, \cite{Liu2021} \\
PG 1448+273 & $0.064$ & 45.5 & 7.0 & $2.4$ & \cite{Reeves2024}\\
IRAS 13224-3809 & $0.066$ & 44.6 & 6-7 & $1$--$3$ & \cite{Alston2019}, \cite{Midooka2023}\\
PG 1211+143 & $0.081$ & 45.7 & 7.9 & $0.5$ & \cite{Reeves2018}\\
SDSS J081456.10$+$532533.5 & $0.120$ & 45.0 & 7.4 & $0.3$ & \cite{Huang2023}\\
PDS~456 & $0.184$ & 47.4 & 9.3 & $0.9$ & \cite{Reeves2009}, \cite{Reeves2020},\\
 & & & & & \cite{XRISM2025}\\
RX J0134.2-4258 & $0.237$ & 46.0 & 7.2 & $4.6$ & \cite{Jin2022}\\
PHL~1092 & $0.396$ & 46.6 & 8.5 & $1.1$ & \cite{Miniutti2012} \\
\enddata
\tablecomments{Column (1): object name.
Column (2): redshift.
Column (3): bolometric luminosity.
Column (4): SMBH mass.
Column (5): Eddington ratio.
Column (6): references.
The objects are arranged in order of increasing redshift.}
\label{tbl:wind_agns}
\end{deluxetable*}


\section{Conclusion and Future Work}\label{sec:conclusion_future_work}

We have analyzed the strong and complex \xray\ variability in I~Zw~1 through time-resolved spectroscopy, using data from simultaneous \hbox{XMM-Newton} and NuSTAR observations in 2020 and NICER monitoring observations in 2022.
The key points are summarized
as follows.
\medskip
\begin{enumerate}[nosep]
    \item
    We reprocessed the data from the 2020 \hbox{XMM-Newton} and NuSTAR observations of I~Zw~1.
    Our analysis reveals strong ($f_{\rm var}\sim3$) and rapid \xray\ variability including two \hbox{flare-like} features in the \hbox{XMM-Newton} light curve.
    There is no coordinated UV variability during the 2020 \hbox{XMM-Newton} observations, and the NuSTAR \hbox{hard-band} light curve shows milder variability ($f_{\rm var}\sim2$).
    The \hbox{XMM-Newton} and NuSTAR count rates are always below the expectations derived from the optical/UV radiation and the \hbox{\aox--\ltkf} relation, suggestive of \xray\ absorption.
    See Section~\ref{sec:xmm_obs} and Section~\ref{sec:nustar_obs}.
    \item
    We have monitored I~Zw~1 with NICER over a \hbox{100-day} period from 14 September to 23 December 2022.
    The NICER \xray\ light curve reveals strong and rapid \xray\ variability, with a maximum amplitude of \hbox{$f_{\rm var}\sim4$}.
    The NICER fluxes are also below the expectation from the \hbox{\aox--\ltkf} relation.
    See Section~\ref{sec:nicer_obs}.
    \item 
    We found that the \hbox{long-term} optical/IR variability amplitudes of I~Zw~1 are mild ($\approx30\%$).
    We constructed an intrinsic SED of I~Zw~1 and it is generally consistent with the composite SED of typical quasars.
    See Section~\ref{sec:multiwave_var} and Section~\ref{sec:uv_ext_sed}.
    \item
    We explained the \hbox{time-resolved} \xray\ spectra utilizing a \hbox{partial-covering} absorption model with a stable corona and three varying ionized absorbers; an additional broad Gaussian Fe emission line was identified, which can reasonably be accommodated within our framework.
    The strong \xray\ variability is ascribed to changes in the column densities and covering factors of the absorbers.
    See Section~\ref{sec:xmm_spec_analyses} and Section~\ref{sec:nicer_spec_analyses}.
    \item
    The three \xray\ absorbers are probably associated with fast disk winds launched from the inner accretion disk.
    We also discussed unifying the remarkable \xray\ properties of NLS1s and \hbox{super-Eddington} accreting quasars under the same physical framework, where \hbox{super-Eddington} accretion launches powerful winds from the inner accretion disk and the \hbox{dust-free} winds modify the observed \xray\ emission.
    See Section~\ref{sec:properties_of_winds} and Section~\ref{sec:clumpy_winds_in_AGNs}.
\end{enumerate}

Our results demonstrate that \hbox{disk-wind} obscuration can explain the observed strong \xray\ variability of I~Zw~1 on the \hbox{$\sim$2-day} and \hbox{$\sim$100-day} timescales.
The \hbox{accretion-disk} optical/UV emission and the coronal \xray\ emission remained stable during the observation periods.
We also constrain the basic properties and locations of the three \hbox{disk-wind} absorbers.
For future work, besides the sample study proposed in Section~\ref{sec:clumpy_winds_in_AGNs}, it might also be useful to obtain XRISM observations of I~Zw~1.
Its superior spectral resolution will likely allow better characterization of the disk winds via spectral analyses.
\\

We thank the anonymous referee for the helpful comments that improved the 
presentation of this paper.
We thank Edward M. Cackett, Pu Du, Qiusheng Gu, Chen Hu, Zhiyuan Li, Yan-Rong Li, Jian-Min Wang, and Zhiyu Zhang for helpful discussions.
J.H. and B.L. acknowledge financial support from the National Natural Science Foundation of China grant 12573016.
L.C.H. was supported by the National Natural Science Foundation of China (12233001) and the China Manned Space Program (CMS-CSST-2025-A09).

\software{{\sc astropy} \citep{astropy:2013, astropy:2018, astropy:2022},
          {\sc dustmaps} \citep{2018JOSS....3..695M},
          {\sc iccf} \citep{Du2025}.}

\bibliographystyle{aasjournal}
\bibliography{ms_IZW1_Huang}

@ARTICLE{HI4PI2016,
       author = {{HI4PI Collaboration} and {Ben Bekhti}, N. and {Fl{\"o}er}, L. and {Keller}, R. and {Kerp}, J. and {Lenz}, D. and {Winkel}, B. and {Bailin}, J. and {Calabretta}, M.~R. and {Dedes}, L. and {Ford}, H.~A. and {Gibson}, B.~K. and {Haud}, U. and {Janowiecki}, S. and {Kalberla}, P.~M.~W. and {Lockman}, F.~J. and {McClure-Griffiths}, N.~M. and {Murphy}, T. and {Nakanishi}, H. and {Pisano}, D.~J. and {Staveley-Smith}, L.},
        title = "{HI4PI: A full-sky H I survey based on EBHIS and GASS}",
      journal = {\aap},
         year = 2016,
        month = oct,
       volume = {594},
          eid = {A116},
        pages = {A116},
          doi = {10.1051/0004-6361/201629178},
archivePrefix = {arXiv},
       eprint = {1610.06175},
 primaryClass = {astro-ph.GA},
       adsurl = {https://ui.adsabs.harvard.edu/abs/2016A&A...594A.116H}
}

@ARTICLE{Shakura1973,
       author = {{Shakura}, N.~I. and {Sunyaev}, R.~A.},
        title = "{Black holes in binary systems. Observational appearance.}",
      journal = {\aap},
         year = 1973,
        month = jan,
       volume = {24},
        pages = {337-355},
       adsurl = {https://ui.adsabs.harvard.edu/abs/1973A&A....24..337S}
}

@ARTICLE{Sunyaev1980,
       author = {{Sunyaev}, R.~A. and {Titarchuk}, L.~G.},
        title = "{Comptonization of X-rays in plasma clouds. Typical radiation spectra.}",
      journal = {\aap},
         year = 1980,
        month = jun,
       volume = {500},
        pages = {167-184},
       adsurl = {https://ui.adsabs.harvard.edu/abs/1980A&A....86..121S}
}

@ARTICLE{Haardt1993,
       author = {{Haardt}, Francesco and {Maraschi}, Laura},
        title = "{X-Ray Spectra from Two-Phase Accretion Disks}",
      journal = {\apj},
         year = 1993,
        month = aug,
       volume = {413},
        pages = {507},
          doi = {10.1086/173020},
       adsurl = {https://ui.adsabs.harvard.edu/abs/1993ApJ...413..507H}
}

@ARTICLE{Steffen2006,
       author = {{Steffen}, A.~T. and {Strateva}, I. and {Brandt}, W.~N. and {Alexander}, D.~M. and {Koekemoer}, A.~M. and {Lehmer}, B.~D. and {Schneider}, D.~P. and {Vignali}, C.},
        title = "{The X-Ray-to-Optical Properties of Optically Selected Active Galaxies over Wide Luminosity and Redshift Ranges}",
      journal = {\aj},
         year = 2006,
        month = jun,
       volume = {131},
       number = {6},
        pages = {2826-2842},
          doi = {10.1086/503627},
archivePrefix = {arXiv},
       eprint = {astro-ph/0602407},
 primaryClass = {astro-ph},
       adsurl = {https://ui.adsabs.harvard.edu/abs/2006AJ....131.2826S}
}

@ARTICLE{Just2007,
       author = {{Just}, D.~W. and {Brandt}, W.~N. and {Shemmer}, O. and {Steffen}, A.~T. and {Schneider}, D.~P. and {Chartas}, G. and {Garmire}, G.~P.},
        title = "{The X-Ray Properties of the Most Luminous Quasars from the Sloan Digital Sky Survey}",
      journal = {\apj},
         year = 2007,
        month = aug,
       volume = {665},
       number = {2},
        pages = {1004-1022},
          doi = {10.1086/519990},
archivePrefix = {arXiv},
       eprint = {0705.3059},
 primaryClass = {astro-ph},
       adsurl = {https://ui.adsabs.harvard.edu/abs/2007ApJ...665.1004J}
}

@ARTICLE{Pu2020,
       author = {{Pu}, Xingting and {Luo}, B. and {Brandt}, W.~N. and {Timlin}, John D. and {Liu}, Hezhen and {Ni}, Q. and {Wu}, Jianfeng},
        title = "{On the Fraction of X-Ray-weak Quasars from the Sloan Digital Sky Survey}",
      journal = {\apj},
         year = 2020,
        month = sep,
       volume = {900},
       number = {2},
          eid = {141},
        pages = {141},
          doi = {10.3847/1538-4357/abacc5},
archivePrefix = {arXiv},
       eprint = {2008.02277},
 primaryClass = {astro-ph.GA},
       adsurl = {https://ui.adsabs.harvard.edu/abs/2020ApJ...900..141P}
}

@ARTICLE{Risaliti&Lusso2019,
       author = {{Risaliti}, G. and {Lusso}, E.},
        title = "{Cosmological Constraints from the Hubble Diagram of Quasars at High Redshifts}",
      journal = {Nature Astronomy},
         year = 2019,
        month = jan,
       volume = {3},
        pages = {272-277},
          doi = {10.1038/s41550-018-0657-z},
archivePrefix = {arXiv},
       eprint = {1811.02590},
 primaryClass = {astro-ph.CO},
       adsurl = {https://ui.adsabs.harvard.edu/abs/2019NatAs...3..272R}
}

@ARTICLE{Luo2015,
       author = {{Luo}, B. and {Brandt}, W.~N. and {Hall}, P.~B. and {Wu}, Jianfeng and {Anderson}, S.~F. and {Garmire}, G.~P. and {Gibson}, R.~R. and {Plotkin}, R.~M. and {Richards}, G.~T. and {Schneider}, D.~P. and {Shemmer}, O. and {Shen}, Yue},
        title = "{X-ray Insights into the Nature of PHL 1811 Analogs and Weak Emission-line Quasars: Unification with a Geometrically Thick Accretion Disk?}",
      journal = {\apj},
         year = 2015,
        month = jun,
       volume = {805},
       number = {2},
          eid = {122},
        pages = {122},
          doi = {10.1088/0004-637X/805/2/122},
archivePrefix = {arXiv},
       eprint = {1503.02085},
 primaryClass = {astro-ph.GA},
       adsurl = {https://ui.adsabs.harvard.edu/abs/2015ApJ...805..122L}
}

@ARTICLE{Nardini2019,
       author = {{Nardini}, E. and {Lusso}, E. and {Risaliti}, G. and {Bisogni}, S. and {Civano}, F. and {Elvis}, M. and {Fabbiano}, G. and {Gilli}, R. and {Marconi}, A. and {Salvestrini}, F. and {Vignali}, C.},
        title = "{The most luminous blue quasars at 3.0 < z < 3.3. I. A tale of two X-ray populations}",
      journal = {\aap},
         year = 2019,
        month = dec,
       volume = {632},
          eid = {A109},
        pages = {A109},
          doi = {10.1051/0004-6361/201936911},
archivePrefix = {arXiv},
       eprint = {1910.04604},
 primaryClass = {astro-ph.GA},
       adsurl = {https://ui.adsabs.harvard.edu/abs/2019A&A...632A.109N}
}

@ARTICLE{Timlin2020,
       author = {{Timlin}, John D., III and {Brandt}, W.~N. and {Zhu}, S. and {Liu}, H. and {Luo}, B. and {Ni}, Q.},
        title = "{The frequency of extreme X-ray variability for radio-quiet quasars}",
      journal = {\mnras},
         year = 2020,
        month = nov,
       volume = {498},
       number = {3},
        pages = {4033-4050},
          doi = {10.1093/mnras/staa2661},
archivePrefix = {arXiv},
       eprint = {2008.12778},
 primaryClass = {astro-ph.HE},
       adsurl = {https://ui.adsabs.harvard.edu/abs/2020MNRAS.498.4033T}
}

@ARTICLE{Laurenti2022,
       author = {{Laurenti}, M. and {Piconcelli}, E. and {Zappacosta}, L. and {Tombesi}, F. and {Vignali}, C. and {Bianchi}, S. and {Marziani}, P. and {Vagnetti}, F. and {Bongiorno}, A. and {Bischetti}, M. and {del Olmo}, A. and {Lanzuisi}, G. and {Luminari}, A. and {Middei}, R. and {Perri}, M. and {Ricci}, C. and {Vietri}, G.},
        title = "{X-ray spectroscopic survey of highly accreting AGN}",
      journal = {\aap},
         year = 2022,
        month = jan,
       volume = {657},
          eid = {A57},
        pages = {A57},
          doi = {10.1051/0004-6361/202141829},
archivePrefix = {arXiv},
       eprint = {2110.06939},
 primaryClass = {astro-ph.GA},
       adsurl = {https://ui.adsabs.harvard.edu/abs/2022A&A...657A..57L}
}

@ARTICLE{Ni2022,
       author = {{Ni}, Q. and {Brandt}, W.~N. and {Luo}, B. and {Garmire}, G.~P. and {Hall}, P.~B. and {Plotkin}, R.~M. and {Shemmer}, O. and {Timlin}, J.~D. and {Vito}, F. and {Wu}, J. and {Yi}, W.},
        title = "{Sensitive Chandra coverage of a representative sample of weak-line quasars: revealing the full range of X-ray properties}",
      journal = {\mnras},
         year = 2022,
        month = apr,
       volume = {511},
       number = {4},
        pages = {5251-5264},
          doi = {10.1093/mnras/stac394},
archivePrefix = {arXiv},
       eprint = {2202.05279},
 primaryClass = {astro-ph.GA},
       adsurl = {https://ui.adsabs.harvard.edu/abs/2022MNRAS.511.5251N}
}

@ARTICLE{Leighly2007b,
       author = {{Leighly}, Karen M. and {Halpern}, Jules P. and {Jenkins}, Edward B. and {Casebeer}, Darrin},
        title = "{The Intrinsically X-Ray-weak Quasar PHL 1811. II. Optical and UV Spectra and Analysis}",
      journal = {\apjs},
         year = 2007,
        month = nov,
       volume = {173},
       number = {1},
        pages = {1-36},
          doi = {10.1086/519768},
archivePrefix = {arXiv},
       eprint = {0705.0940},
 primaryClass = {astro-ph},
       adsurl = {https://ui.adsabs.harvard.edu/abs/2007ApJS..173....1L}
}

@ARTICLE{Leighly2007a,
       author = {{Leighly}, Karen M. and {Halpern}, Jules P. and {Jenkins}, Edward B. and {Grupe}, Dirk and {Choi}, Jiehae and {Prescott}, Kimberly B.},
        title = "{The Intrinsically X-Ray Weak Quasar PHL 1811. I. X-Ray Observations and Spectral Energy Distribution}",
      journal = {\apj},
         year = 2007,
        month = jul,
       volume = {663},
       number = {1},
        pages = {103-117},
          doi = {10.1086/518017},
archivePrefix = {arXiv},
       eprint = {astro-ph/0611349},
 primaryClass = {astro-ph},
       adsurl = {https://ui.adsabs.harvard.edu/abs/2007ApJ...663..103L}
}

@ARTICLE{Wang2022,
       author = {{Wang}, Chaojun and {Luo}, B. and {Brandt}, W.~N. and {Alexander}, D.~M. and {Bauer}, F.~E. and {Gallagher}, S.~C. and {Huang}, Jian and {Liu}, Hezhen and {Stern}, D.},
        title = "{NuSTAR Observations of Intrinsically X-Ray Weak Quasar Candidates: An Obscuration-only Scenario}",
      journal = {\apj},
         year = 2022,
        month = sep,
       volume = {936},
       number = {2},
          eid = {95},
        pages = {95},
          doi = {10.3847/1538-4357/ac886e},
archivePrefix = {arXiv},
       eprint = {2208.04961},
 primaryClass = {astro-ph.HE},
       adsurl = {https://ui.adsabs.harvard.edu/abs/2022ApJ...936...95W}
}

@ARTICLE{McHardy2006,
       author = {{McHardy}, I.~M. and {Koerding}, E. and {Knigge}, C. and {Uttley}, P. and {Fender}, R.~P.},
        title = "{Active galactic nuclei as scaled-up Galactic black holes}",
      journal = {\nat},
         year = 2006,
        month = dec,
       volume = {444},
       number = {7120},
        pages = {730-732},
          doi = {10.1038/nature05389},
archivePrefix = {arXiv},
       eprint = {astro-ph/0612273},
 primaryClass = {astro-ph},
       adsurl = {https://ui.adsabs.harvard.edu/abs/2006Natur.444..730M}
}

@ARTICLE{Yang2016,
       author = {{Yang}, G. and {Brandt}, W.~N. and {Luo}, B. and {Xue}, Y.~Q. and {Bauer}, F.~E. and {Sun}, M.~Y. and {Kim}, S. and {Schulze}, S. and {Zheng}, X.~C. and {Paolillo}, M. and {Shemmer}, O. and {Liu}, T. and {Schneider}, D.~P. and {Vignali}, C. and {Vito}, F. and {Wang}, J. -X.},
        title = "{Long-term X-Ray Variability of Typical Active Galactic Nuclei in the Distant Universe}",
      journal = {\apj},
         year = 2016,
        month = nov,
       volume = {831},
       number = {2},
          eid = {145},
        pages = {145},
          doi = {10.3847/0004-637X/831/2/145},
archivePrefix = {arXiv},
       eprint = {1608.08224},
 primaryClass = {astro-ph.HE},
       adsurl = {https://ui.adsabs.harvard.edu/abs/2016ApJ...831..145Y}
}

@ARTICLE{Zheng2017,
       author = {{Zheng}, X.~C. and {Xue}, Y.~Q. and {Brandt}, W.~N. and {Li}, J.~Y. and {Paolillo}, M. and {Yang}, G. and {Zhu}, S.~F. and {Luo}, B. and {Sun}, M.~Y. and {Hughes}, T.~M. and {Bauer}, F.~E. and {Vito}, F. and {Wang}, J.~X. and {Liu}, T. and {Vignali}, C. and {Shu}, X.~W.},
        title = "{Deepest View of AGN X-Ray Variability with the 7 Ms Chandra Deep Field-South Survey}",
      journal = {\apj},
         year = 2017,
        month = nov,
       volume = {849},
       number = {2},
          eid = {127},
        pages = {127},
          doi = {10.3847/1538-4357/aa9378},
archivePrefix = {arXiv},
       eprint = {1710.04358},
 primaryClass = {astro-ph.GA},
       adsurl = {https://ui.adsabs.harvard.edu/abs/2017ApJ...849..127Z}
}

@ARTICLE{Gibson2012,
       author = {{Gibson}, Robert R. and {Brandt}, W.~N.},
        title = "{The X-Ray Variability of a Large, Serendipitous Sample of Spectroscopic Quasars}",
      journal = {\apj},
         year = 2012,
        month = feb,
       volume = {746},
       number = {1},
          eid = {54},
        pages = {54},
          doi = {10.1088/0004-637X/746/1/54},
archivePrefix = {arXiv},
       eprint = {1110.5341},
 primaryClass = {astro-ph.CO},
       adsurl = {https://ui.adsabs.harvard.edu/abs/2012ApJ...746...54G}
}

@ARTICLE{Miniutti2012,
       author = {{Miniutti}, G. and {Brandt}, W.~N. and {Schneider}, D.~P. and {Fabian}, A.~C. and {Gallo}, L.~C. and {Boller}, Th.},
        title = "{Insights on the X-ray weak quasar phenomenon from XMM-Newton monitoring of PHL 1092}",
      journal = {\mnras},
         year = 2012,
        month = sep,
       volume = {425},
       number = {3},
        pages = {1718-1737},
          doi = {10.1111/j.1365-2966.2012.21648.x},
archivePrefix = {arXiv},
       eprint = {1207.0694},
 primaryClass = {astro-ph.HE},
       adsurl = {https://ui.adsabs.harvard.edu/abs/2012MNRAS.425.1718M}
}

@ARTICLE{Liu2019,
       author = {{Liu}, Hezhen and {Luo}, B. and {Brandt}, W.~N. and {Brotherton}, Michael S. and {Du}, Pu and {Gallagher}, S.~C. and {Hu}, Chen and {Shemmer}, Ohad and {Wang}, Jian-Min},
        title = "{SDSS J075101.42+291419.1: A Super-Eddington Accreting Quasar with Extreme X-Ray Variability}",
      journal = {\apj},
         year = 2019,
        month = jun,
       volume = {878},
       number = {2},
          eid = {79},
        pages = {79},
          doi = {10.3847/1538-4357/ab1d5b},
archivePrefix = {arXiv},
       eprint = {1904.12876},
 primaryClass = {astro-ph.GA},
       adsurl = {https://ui.adsabs.harvard.edu/abs/2019ApJ...878...79L}
}

@ARTICLE{Ni2020,
       author = {{Ni}, Q. and {Brandt}, W.~N. and {Yi}, W. and {Luo}, B. and {Timlin}, J.~D., III and {Hall}, P.~B. and {Liu}, Hezhen and {Plotkin}, R.~M. and {Shemmer}, O. and {Vito}, F. and {Wu}, Jianfeng},
        title = "{An Extreme X-Ray Variability Event of a Weak-line Quasar}",
      journal = {\apjl},
         year = 2020,
        month = feb,
       volume = {889},
       number = {2},
          eid = {L37},
        pages = {L37},
          doi = {10.3847/2041-8213/ab6d78},
archivePrefix = {arXiv},
       eprint = {2001.08216},
 primaryClass = {astro-ph.GA},
       adsurl = {https://ui.adsabs.harvard.edu/abs/2020ApJ...889L..37N}
}

@ARTICLE{Liu2022,
       author = {{Liu}, Hezhen and {Luo}, B. and {Brandt}, W.~N. and {Huang}, Jian and {Pu}, Xingting and {Yi}, Weimin and {Yu}, Li-Ming},
        title = "{A Rapid and Large-amplitude X-Ray Dimming Event in a z {\ensuremath{\approx}} 2.6 Radio-quiet Quasar}",
      journal = {\apj},
         year = 2022,
        month = may,
       volume = {930},
       number = {1},
          eid = {53},
        pages = {53},
          doi = {10.3847/1538-4357/ac6265},
archivePrefix = {arXiv},
       eprint = {2203.15824},
 primaryClass = {astro-ph.HE},
       adsurl = {https://ui.adsabs.harvard.edu/abs/2022ApJ...930...53L}
}

@ARTICLE{Huang2023,
       author = {{Huang}, Jian and {Luo}, Bin and {Brandt}, W.~N. and {Du}, Pu and {Garmire}, Gordon P. and {Hu}, Chen and {Liu}, Hezhen and {Ni}, Qingling and {Wang}, Jian-Min},
        title = "{Strong and Rapid X-Ray Variability of the Super-Eddington Accreting Quasar SDSS J081456.10+532533.5}",
      journal = {\apj},
         year = 2023,
        month = jun,
       volume = {950},
       number = {1},
          eid = {18},
        pages = {18},
          doi = {10.3847/1538-4357/accd64},
archivePrefix = {arXiv},
       eprint = {2304.07323},
 primaryClass = {astro-ph.HE},
       adsurl = {https://ui.adsabs.harvard.edu/abs/2023ApJ...950...18H}
}

@ARTICLE{Jiang2014,
       author = {{Jiang}, Yan-Fei and {Stone}, James M. and {Davis}, Shane W.},
        title = "{Radiation Magnetohydrodynamic Simulations of the Formation of Hot Accretion Disk Coronae}",
      journal = {\apj},
         year = 2014,
        month = apr,
       volume = {784},
       number = {2},
          eid = {169},
        pages = {169},
          doi = {10.1088/0004-637X/784/2/169},
archivePrefix = {arXiv},
       eprint = {1402.2979},
 primaryClass = {astro-ph.HE},
       adsurl = {https://ui.adsabs.harvard.edu/abs/2014ApJ...784..169J}
}

@ARTICLE{Sadowski2014,
       author = {{Sadowski}, Aleksander and {Narayan}, Ramesh and {McKinney}, Jonathan C. and {Tchekhovskoy}, Alexander},
        title = "{Numerical simulations of super-critical black hole accretion flows in general relativity}",
      journal = {\mnras},
         year = 2014,
        month = mar,
       volume = {439},
       number = {1},
        pages = {503-520},
          doi = {10.1093/mnras/stt2479},
archivePrefix = {arXiv},
       eprint = {1311.5900},
 primaryClass = {astro-ph.HE},
       adsurl = {https://ui.adsabs.harvard.edu/abs/2014MNRAS.439..503S}
}

@ARTICLE{Jiang2019,
       author = {{Jiang}, Yan-Fei and {Stone}, James M. and {Davis}, Shane W.},
        title = "{Super-Eddington Accretion Disks around Supermassive Black Holes}",
      journal = {\apj},
         year = 2019,
        month = aug,
       volume = {880},
       number = {2},
          eid = {67},
        pages = {67},
          doi = {10.3847/1538-4357/ab29ff},
archivePrefix = {arXiv},
       eprint = {1709.02845},
 primaryClass = {astro-ph.HE},
       adsurl = {https://ui.adsabs.harvard.edu/abs/2019ApJ...880...67J}
}

@ARTICLE{Reeves2019,
       author = {{Reeves}, J.~N. and {Braito}, V.},
        title = "{A Momentum-conserving Accretion Disk Wind in the Narrow-line Seyfert 1 I Zwicky 1}",
      journal = {\apj},
         year = 2019,
        month = oct,
       volume = {884},
       number = {1},
          eid = {80},
        pages = {80},
          doi = {10.3847/1538-4357/ab41f9},
archivePrefix = {arXiv},
       eprint = {1909.05039},
 primaryClass = {astro-ph.GA},
       adsurl = {https://ui.adsabs.harvard.edu/abs/2019ApJ...884...80R}
}

@ARTICLE{Boller2021,
       author = {{Boller}, Th. and {Liu}, T. and {Weber}, P. and {Arcodia}, R. and {Dauser}, T. and {Wilms}, J. and {Nandra}, K. and {Buchner}, J. and {Merloni}, A. and {Freyberg}, M.~J. and {Krumpe}, M. and {Waddell}, S.~G.~H.},
        title = "{Extreme ultra-soft X-ray variability in an eROSITA observation of the narrow-line Seyfert 1 galaxy 1H 0707-495}",
      journal = {\aap},
         year = 2021,
        month = mar,
       volume = {647},
          eid = {A6},
        pages = {A6},
          doi = {10.1051/0004-6361/202039316},
archivePrefix = {arXiv},
       eprint = {2011.03307},
 primaryClass = {astro-ph.HE},
       adsurl = {https://ui.adsabs.harvard.edu/abs/2021A&A...647A...6B}
}

@ARTICLE{Parker2021,
       author = {{Parker}, M.~L. and {Alston}, W.~N. and {H{\"a}rer}, L. and {Igo}, Z. and {Joyce}, A. and {Buisson}, D.~J.~K. and {Chainakun}, P. and {Fabian}, A.~C. and {Jiang}, J. and {Kosec}, P. and {Matzeu}, G.~A. and {Pinto}, C. and {Xu}, Y. and {Zaidouni}, F.},
        title = "{The nature of the extreme X-ray variability in the NLS1 1H 0707-495}",
      journal = {\mnras},
         year = 2021,
        month = dec,
       volume = {508},
       number = {2},
        pages = {1798-1816},
          doi = {10.1093/mnras/stab2434},
archivePrefix = {arXiv},
       eprint = {2108.10167},
 primaryClass = {astro-ph.HE},
       adsurl = {https://ui.adsabs.harvard.edu/abs/2021MNRAS.508.1798P}
}

@ARTICLE{Jin2022,
       author = {{Jin}, Chichuan and {Done}, Chris and {Ward}, Martin and {Panessa}, Francesca and {Liu}, Bo and {Liu}, Heyang},
        title = "{Multiwavelength campaign on the Super-Eddington NLS1 RX J0134.2-4258 - I. Peculiar X-ray spectra and variability}",
      journal = {\mnras},
         year = 2022,
        month = jun,
       volume = {512},
       number = {4},
        pages = {5642-5656},
          doi = {10.1093/mnras/stac827},
archivePrefix = {arXiv},
       eprint = {2203.13419},
 primaryClass = {astro-ph.HE},
       adsurl = {https://ui.adsabs.harvard.edu/abs/2022MNRAS.512.5642J}
}

@ARTICLE{Boroson1992,
       author = {{Boroson}, Todd A. and {Green}, Richard F.},
        title = "{The Emission-Line Properties of Low-Redshift Quasi-stellar Objects}",
      journal = {\apjs},
         year = 1992,
        month = may,
       volume = {80},
        pages = {109},
          doi = {10.1086/191661},
       adsurl = {https://ui.adsabs.harvard.edu/abs/1992ApJS...80..109B}
}

@ARTICLE{Huang2019,
       author = {{Huang}, Ying-Ke and {Hu}, Chen and {Zhao}, Yu-Lin and {Zhang}, Zhi-Xiang and {Lu}, Kai-Xing and {Wang}, Kai and {Zhang}, Yue and {Du}, Pu and {Li}, Yan-Rong and {Bai}, Jin-Ming and {Ho}, Luis C. and {Bian}, Wei-Hao and {Yuan}, Ye-Fei and {Wang}, Jian-Min},
        title = "{Reverberation Mapping of the Narrow-line Seyfert 1 Galaxy I Zwicky 1: Black Hole Mass}",
      journal = {\apj},
         year = 2019,
        month = may,
       volume = {876},
       number = {2},
          eid = {102},
        pages = {102},
          doi = {10.3847/1538-4357/ab16ef},
archivePrefix = {arXiv},
       eprint = {1904.06146},
 primaryClass = {astro-ph.GA},
       adsurl = {https://ui.adsabs.harvard.edu/abs/2019ApJ...876..102H}
}

@ARTICLE{Kaastra2016,
       author = {{Kaastra}, J.~S. and {Bleeker}, J.~A.~M.},
        title = "{Optimal binning of X-ray spectra and response matrix design}",
      journal = {\aap},
         year = 2016,
        month = mar,
       volume = {587},
          eid = {A151},
        pages = {A151},
          doi = {10.1051/0004-6361/201527395},
archivePrefix = {arXiv},
       eprint = {1601.05309},
 primaryClass = {astro-ph.IM},
       adsurl = {https://ui.adsabs.harvard.edu/abs/2016A&A...587A.151K}
}

@ARTICLE{Jansen2001,
       author = {{Jansen}, F. and {Lumb}, D. and {Altieri}, B. and {Clavel}, J. and {Ehle}, M. and {Erd}, C. and {Gabriel}, C. and {Guainazzi}, M. and {Gondoin}, P. and {Much}, R. and {Munoz}, R. and {Santos}, M. and {Schartel}, N. and {Texier}, D. and {Vacanti}, G.},
        title = "{XMM-Newton observatory. I. The spacecraft and operations}",
      journal = {\aap},
         year = 2001,
        month = jan,
       volume = {365},
        pages = {L1-L6},
          doi = {10.1051/0004-6361:20000036},
       adsurl = {https://ui.adsabs.harvard.edu/abs/2001A&A...365L...1J}
}

@ARTICLE{Harrison2013,
       author = {{Harrison}, Fiona A. and {Craig}, William W. and {Christensen}, Finn E. and {Hailey}, Charles J. and {Zhang}, William W. and {Boggs}, Steven E. and {Stern}, Daniel and {Cook}, W. Rick and {Forster}, Karl and {Giommi}, Paolo and {Grefenstette}, Brian W. and {Kim}, Yunjin and {Kitaguchi}, Takao and {Koglin}, Jason E. and {Madsen}, Kristin K. and {Mao}, Peter H. and {Miyasaka}, Hiromasa and {Mori}, Kaya and {Perri}, Matteo and {Pivovaroff}, Michael J. and {Puccetti}, Simonetta and {Rana}, Vikram R. and {Westergaard}, Niels J. and {Willis}, Jason and {Zoglauer}, Andreas and {An}, Hongjun and {Bachetti}, Matteo and {Barri{\`e}re}, Nicolas M. and {Bellm}, Eric C. and {Bhalerao}, Varun and {Brejnholt}, Nicolai F. and {Fuerst}, Felix and {Liebe}, Carl C. and {Markwardt}, Craig B. and {Nynka}, Melania and {Vogel}, Julia K. and {Walton}, Dominic J. and {Wik}, Daniel R. and {Alexander}, David M. and {Cominsky}, Lynn R. and {Hornschemeier}, Ann E. and {Hornstrup}, Allan and {Kaspi}, Victoria M. and {Madejski}, Greg M. and {Matt}, Giorgio and {Molendi}, Silvano and {Smith}, David M. and {Tomsick}, John A. and {Ajello}, Marco and {Ballantyne}, David R. and {Balokovi{\'c}}, Mislav and {Barret}, Didier and {Bauer}, Franz E. and {Blandford}, Roger D. and {Brandt}, W. Niel and {Brenneman}, Laura W. and {Chiang}, James and {Chakrabarty}, Deepto and {Chenevez}, Jerome and {Comastri}, Andrea and {Dufour}, Francois and {Elvis}, Martin and {Fabian}, Andrew C. and {Farrah}, Duncan and {Fryer}, Chris L. and {Gotthelf}, Eric V. and {Grindlay}, Jonathan E. and {Helfand}, David J. and {Krivonos}, Roman and {Meier}, David L. and {Miller}, Jon M. and {Natalucci}, Lorenzo and {Ogle}, Patrick and {Ofek}, Eran O. and {Ptak}, Andrew and {Reynolds}, Stephen P. and {Rigby}, Jane R. and {Tagliaferri}, Gianpiero and {Thorsett}, Stephen E. and {Treister}, Ezequiel and {Urry}, C. Megan},
        title = "{The Nuclear Spectroscopic Telescope Array (NuSTAR) High-energy X-Ray Mission}",
      journal = {\apj},
         year = 2013,
        month = jun,
       volume = {770},
       number = {2},
          eid = {103},
        pages = {103},
          doi = {10.1088/0004-637X/770/2/103},
archivePrefix = {arXiv},
       eprint = {1301.7307},
 primaryClass = {astro-ph.IM},
       adsurl = {https://ui.adsabs.harvard.edu/abs/2013ApJ...770..103H}
}

@ARTICLE{Ding2022,
       author = {{Ding}, Yuanze and {Li}, Ruancun and {Ho}, Luis C. and {Ricci}, Claudio},
        title = "{Accretion Disk Outflow during the X-Ray Flare of the Super-Eddington Active Nucleus of I Zwicky 1}",
      journal = {\apj},
         year = 2022,
        month = jun,
       volume = {931},
       number = {2},
          eid = {77},
        pages = {77},
          doi = {10.3847/1538-4357/ac6955},
       adsurl = {https://ui.adsabs.harvard.edu/abs/2022ApJ...931...77D}
}

@ARTICLE{Rogantini2022,
       author = {{Rogantini}, D. and {Costantini}, E. and {Gallo}, L.~C. and {Wilkins}, D.~R. and {Brandt}, W.~N. and {Mehdipour}, M.},
        title = "{The multi-epoch X-ray tale of I Zwicky 1 outflows}",
      journal = {\mnras},
         year = 2022,
        month = nov,
       volume = {516},
       number = {4},
        pages = {5171-5186},
          doi = {10.1093/mnras/stac2552},
archivePrefix = {arXiv},
       eprint = {2209.02747},
 primaryClass = {astro-ph.HE},
       adsurl = {https://ui.adsabs.harvard.edu/abs/2022MNRAS.516.5171R}
}

@INPROCEEDINGS{Gendreau2016,
       author = {{Gendreau}, Keith C. and {Arzoumanian}, Zaven and {Adkins}, Phillip W. and {Albert}, Cheryl L. and {Anders}, John F. and {Aylward}, Andrew T. and {Baker}, Charles L. and {Balsamo}, Erin R. and {Bamford}, William A. and {Benegalrao}, Suyog S. and {Berry}, Daniel L. and {Bhalwani}, Shiraz and {Black}, J. Kevin and {Blaurock}, Carl and {Bronke}, Ginger M. and {Brown}, Gary L. and {Budinoff}, Jason G. and {Cantwell}, Jeffrey D. and {Cazeau}, Thoniel and {Chen}, Philip T. and {Clement}, Thomas G. and {Colangelo}, Andrew T. and {Coleman}, Jerry S. and {Coopersmith}, Jonathan D. and {Dehaven}, William E. and {Doty}, John P. and {Egan}, Mark D. and {Enoto}, Teruaki and {Fan}, Terry W. and {Ferro}, Deneen M. and {Foster}, Richard and {Galassi}, Nicholas M. and {Gallo}, Luis D. and {Green}, Chris M. and {Grosh}, Dave and {Ha}, Kong Q. and {Hasouneh}, Monther A. and {Heefner}, Kristofer B. and {Hestnes}, Phyllis and {Hoge}, Lisa J. and {Jacobs}, Tawanda M. and {J{\o}rgensen}, John L. and {Kaiser}, Michael A. and {Kellogg}, James W. and {Kenyon}, Steven J. and {Koenecke}, Richard G. and {Kozon}, Robert P. and {LaMarr}, Beverly and {Lambertson}, Mike D. and {Larson}, Anne M. and {Lentine}, Steven and {Lewis}, Jesse H. and {Lilly}, Michael G. and {Liu}, Kuochia Alice and {Malonis}, Andrew and {Manthripragada}, Sridhar S. and {Markwardt}, Craig B. and {Matonak}, Bryan D. and {Mcginnis}, Isaac E. and {Miller}, Roger L. and {Mitchell}, Alissa L. and {Mitchell}, Jason W. and {Mohammed}, Jelila S. and {Monroe}, Charles A. and {Montt de Garcia}, Kristina M. and {Mul{\'e}}, Peter D. and {Nagao}, Louis T. and {Ngo}, Son N. and {Norris}, Eric D. and {Norwood}, Dwight A. and {Novotka}, Joseph and {Okajima}, Takashi and {Olsen}, Lawrence G. and {Onyeachu}, Chimaobi O. and {Orosco}, Henry Y. and {Peterson}, Jacqualine R. and {Pevear}, Kristina N. and {Pham}, Karen K. and {Pollard}, Sue E. and {Pope}, John S. and {Powers}, Daniel F. and {Powers}, Charles E. and {Price}, Samuel R. and {Prigozhin}, Gregory Y. and {Ramirez}, Julian B. and {Reid}, Winston J. and {Remillard}, Ronald A. and {Rogstad}, Eric M. and {Rosecrans}, Glenn P. and {Rowe}, John N. and {Sager}, Jennifer A. and {Sanders}, Claude A. and {Savadkin}, Bruce and {Saylor}, Maxine R. and {Schaeffer}, Alexander F. and {Schweiss}, Nancy S. and {Semper}, Sean R. and {Serlemitsos}, Peter J. and {Shackelford}, Larry V. and {Soong}, Yang and {Struebel}, Jonathan and {Vezie}, Michael L. and {Villasenor}, Joel S. and {Winternitz}, Luke B. and {Wofford}, George I. and {Wright}, Michael R. and {Yang}, Mike Y. and {Yu}, Wayne H.},
        title = "{The Neutron star Interior Composition Explorer (NICER): design and development}",
    booktitle = {Space Telescopes and Instrumentation 2016: Ultraviolet to Gamma Ray},
         year = 2016,
       editor = {{den Herder}, Jan-Willem A. and {Takahashi}, Tadayuki and {Bautz}, Marshall},
       series = {Society of Photo-Optical Instrumentation Engineers (SPIE) Conference Series},
       volume = {9905},
        month = jul,
          eid = {99051H},
        pages = {99051H},
          doi = {10.1117/12.2231304},
       adsurl = {https://ui.adsabs.harvard.edu/abs/2016SPIE.9905E..1HG}
}

@INPROCEEDINGS{Arnaud1996,
       author = {{Arnaud}, K.~A.},
        title = "{XSPEC: The First Ten Years}",
    booktitle = {Astronomical Data Analysis Software and Systems V},
         year = 1996,
       editor = {{Jacoby}, George H. and {Barnes}, Jeannette},
       series = {Astronomical Society of the Pacific Conference Series},
       volume = {101},
        month = jan,
        pages = {17},
       adsurl = {https://ui.adsabs.harvard.edu/abs/1996ASPC..101...17A}
}

@ARTICLE{Wilkins2021,
       author = {{Wilkins}, D.~R. and {Gallo}, L.~C. and {Costantini}, E. and {Brandt}, W.~N. and {Blandford}, R.~D.},
        title = "{Light bending and X-ray echoes from behind a supermassive black hole}",
      journal = {\nat},
         year = 2021,
        month = jul,
       volume = {595},
       number = {7869},
        pages = {657-660},
          doi = {10.1038/s41586-021-03667-0},
archivePrefix = {arXiv},
       eprint = {2107.13555},
 primaryClass = {astro-ph.HE},
       adsurl = {https://ui.adsabs.harvard.edu/abs/2021Natur.595..657W}
}

@ARTICLE{Wilkins2022,
       author = {{Wilkins}, D.~R. and {Gallo}, L.~C. and {Costantini}, E. and {Brandt}, W.~N. and {Blandford}, R.~D.},
        title = "{Acceleration and cooling of the corona during X-ray flares from the Seyfert galaxy I Zw 1}",
      journal = {\mnras},
         year = 2022,
        month = may,
       volume = {512},
       number = {1},
        pages = {761-775},
          doi = {10.1093/mnras/stac416},
archivePrefix = {arXiv},
       eprint = {2202.06958},
 primaryClass = {astro-ph.HE},
       adsurl = {https://ui.adsabs.harvard.edu/abs/2022MNRAS.512..761W}
}

@ARTICLE{Planck2020,
       author = {{Planck Collaboration} and {Aghanim}, N. and {Akrami}, Y. and {Ashdown}, M. and {Aumont}, J. and {Baccigalupi}, C. and {Ballardini}, M. and {Banday}, A.~J. and {Barreiro}, R.~B. and {Bartolo}, N. and {Basak}, S. and {Battye}, R. and {Benabed}, K. and {Bernard}, J. -P. and {Bersanelli}, M. and {Bielewicz}, P. and {Bock}, J.~J. and {Bond}, J.~R. and {Borrill}, J. and {Bouchet}, F.~R. and {Boulanger}, F. and {Bucher}, M. and {Burigana}, C. and {Butler}, R.~C. and {Calabrese}, E. and {Cardoso}, J. -F. and {Carron}, J. and {Challinor}, A. and {Chiang}, H.~C. and {Chluba}, J. and {Colombo}, L.~P.~L. and {Combet}, C. and {Contreras}, D. and {Crill}, B.~P. and {Cuttaia}, F. and {de Bernardis}, P. and {de Zotti}, G. and {Delabrouille}, J. and {Delouis}, J. -M. and {Di Valentino}, E. and {Diego}, J.~M. and {Dor{\'e}}, O. and {Douspis}, M. and {Ducout}, A. and {Dupac}, X. and {Dusini}, S. and {Efstathiou}, G. and {Elsner}, F. and {En{\ss}lin}, T.~A. and {Eriksen}, H.~K. and {Fantaye}, Y. and {Farhang}, M. and {Fergusson}, J. and {Fernandez-Cobos}, R. and {Finelli}, F. and {Forastieri}, F. and {Frailis}, M. and {Fraisse}, A.~A. and {Franceschi}, E. and {Frolov}, A. and {Galeotta}, S. and {Galli}, S. and {Ganga}, K. and {G{\'e}nova-Santos}, R.~T. and {Gerbino}, M. and {Ghosh}, T. and {Gonz{\'a}lez-Nuevo}, J. and {G{\'o}rski}, K.~M. and {Gratton}, S. and {Gruppuso}, A. and {Gudmundsson}, J.~E. and {Hamann}, J. and {Handley}, W. and {Hansen}, F.~K. and {Herranz}, D. and {Hildebrandt}, S.~R. and {Hivon}, E. and {Huang}, Z. and {Jaffe}, A.~H. and {Jones}, W.~C. and {Karakci}, A. and {Keih{\"a}nen}, E. and {Keskitalo}, R. and {Kiiveri}, K. and {Kim}, J. and {Kisner}, T.~S. and {Knox}, L. and {Krachmalnicoff}, N. and {Kunz}, M. and {Kurki-Suonio}, H. and {Lagache}, G. and {Lamarre}, J. -M. and {Lasenby}, A. and {Lattanzi}, M. and {Lawrence}, C.~R. and {Le Jeune}, M. and {Lemos}, P. and {Lesgourgues}, J. and {Levrier}, F. and {Lewis}, A. and {Liguori}, M. and {Lilje}, P.~B. and {Lilley}, M. and {Lindholm}, V. and {L{\'o}pez-Caniego}, M. and {Lubin}, P.~M. and {Ma}, Y. -Z. and {Mac{\'\i}as-P{\'e}rez}, J.~F. and {Maggio}, G. and {Maino}, D. and {Mandolesi}, N. and {Mangilli}, A. and {Marcos-Caballero}, A. and {Maris}, M. and {Martin}, P.~G. and {Martinelli}, M. and {Mart{\'\i}nez-Gonz{\'a}lez}, E. and {Matarrese}, S. and {Mauri}, N. and {McEwen}, J.~D. and {Meinhold}, P.~R. and {Melchiorri}, A. and {Mennella}, A. and {Migliaccio}, M. and {Millea}, M. and {Mitra}, S. and {Miville-Desch{\^e}nes}, M. -A. and {Molinari}, D. and {Montier}, L. and {Morgante}, G. and {Moss}, A. and {Natoli}, P. and {N{\o}rgaard-Nielsen}, H.~U. and {Pagano}, L. and {Paoletti}, D. and {Partridge}, B. and {Patanchon}, G. and {Peiris}, H.~V. and {Perrotta}, F. and {Pettorino}, V. and {Piacentini}, F. and {Polastri}, L. and {Polenta}, G. and {Puget}, J. -L. and {Rachen}, J.~P. and {Reinecke}, M. and {Remazeilles}, M. and {Renzi}, A. and {Rocha}, G. and {Rosset}, C. and {Roudier}, G. and {Rubi{\~n}o-Mart{\'\i}n}, J.~A. and {Ruiz-Granados}, B. and {Salvati}, L. and {Sandri}, M. and {Savelainen}, M. and {Scott}, D. and {Shellard}, E.~P.~S. and {Sirignano}, C. and {Sirri}, G. and {Spencer}, L.~D. and {Sunyaev}, R. and {Suur-Uski}, A. -S. and {Tauber}, J.~A. and {Tavagnacco}, D. and {Tenti}, M. and {Toffolatti}, L. and {Tomasi}, M. and {Trombetti}, T. and {Valenziano}, L. and {Valiviita}, J. and {Van Tent}, B. and {Vibert}, L. and {Vielva}, P. and {Villa}, F. and {Vittorio}, N. and {Wandelt}, B.~D. and {Wehus}, I.~K. and {White}, M. and {White}, S.~D.~M. and {Zacchei}, A. and {Zonca}, A.},
        title = "{Planck 2018 results. VI. Cosmological parameters}",
      journal = {\aap},
         year = 2020,
        month = sep,
       volume = {641},
          eid = {A6},
        pages = {A6},
          doi = {10.1051/0004-6361/201833910},
archivePrefix = {arXiv},
       eprint = {1807.06209},
 primaryClass = {astro-ph.CO},
       adsurl = {https://ui.adsabs.harvard.edu/abs/2020A&A...641A...6P}
}

@ARTICLE{Struder2001,
       author = {{Str{\"u}der}, L. and {Briel}, U. and {Dennerl}, K. and {Hartmann}, R. and {Kendziorra}, E. and {Meidinger}, N. and {Pfeffermann}, E. and {Reppin}, C. and {Aschenbach}, B. and {Bornemann}, W. and {Br{\"a}uninger}, H. and {Burkert}, W. and {Elender}, M. and {Freyberg}, M. and {Haberl}, F. and {Hartner}, G. and {Heuschmann}, F. and {Hippmann}, H. and {Kastelic}, E. and {Kemmer}, S. and {Kettenring}, G. and {Kink}, W. and {Krause}, N. and {M{\"u}ller}, S. and {Oppitz}, A. and {Pietsch}, W. and {Popp}, M. and {Predehl}, P. and {Read}, A. and {Stephan}, K.~H. and {St{\"o}tter}, D. and {Tr{\"u}mper}, J. and {Holl}, P. and {Kemmer}, J. and {Soltau}, H. and {St{\"o}tter}, R. and {Weber}, U. and {Weichert}, U. and {von Zanthier}, C. and {Carathanassis}, D. and {Lutz}, G. and {Richter}, R.~H. and {Solc}, P. and {B{\"o}ttcher}, H. and {Kuster}, M. and {Staubert}, R. and {Abbey}, A. and {Holland}, A. and {Turner}, M. and {Balasini}, M. and {Bignami}, G.~F. and {La Palombara}, N. and {Villa}, G. and {Buttler}, W. and {Gianini}, F. and {Lain{\'e}}, R. and {Lumb}, D. and {Dhez}, P.},
        title = "{The European Photon Imaging Camera on XMM-Newton: The pn-CCD camera}",
      journal = {\aap},
         year = 2001,
        month = jan,
       volume = {365},
        pages = {L18-L26},
          doi = {10.1051/0004-6361:20000066},
       adsurl = {https://ui.adsabs.harvard.edu/abs/2001A&A...365L..18S}
}

@ARTICLE{Mason2001,
       author = {{Mason}, K.~O. and {Breeveld}, A. and {Much}, R. and {Carter}, M. and {Cordova}, F.~A. and {Cropper}, M.~S. and {Fordham}, J. and {Huckle}, H. and {Ho}, C. and {Kawakami}, H. and {Kennea}, J. and {Kennedy}, T. and {Mittaz}, J. and {Pandel}, D. and {Priedhorsky}, W.~C. and {Sasseen}, T. and {Shirey}, R. and {Smith}, P. and {Vreux}, J. -M.},
        title = "{The XMM-Newton optical/UV monitor telescope}",
      journal = {\aap},
         year = 2001,
        month = jan,
       volume = {365},
        pages = {L36-L44},
          doi = {10.1051/0004-6361:20000044},
archivePrefix = {arXiv},
       eprint = {astro-ph/0011216},
 primaryClass = {astro-ph},
       adsurl = {https://ui.adsabs.harvard.edu/abs/2001A&A...365L..36M}
}

@ARTICLE{Fitzpatrick2019,
       author = {{Fitzpatrick}, E.~L. and {Massa}, Derck and {Gordon}, Karl D. and {Bohlin}, Ralph and {Clayton}, Geoffrey C.},
        title = "{An Analysis of the Shapes of Interstellar Extinction Curves. VII. Milky Way Spectrophotometric Optical-through-ultraviolet Extinction and Its R-dependence}",
      journal = {\apj},
         year = 2019,
        month = dec,
       volume = {886},
       number = {2},
          eid = {108},
        pages = {108},
          doi = {10.3847/1538-4357/ab4c3a},
archivePrefix = {arXiv},
       eprint = {1910.08852},
 primaryClass = {astro-ph.GA},
       adsurl = {https://ui.adsabs.harvard.edu/abs/2019ApJ...886..108F}
}

@ARTICLE{Schlegel1998,
       author = {{Schlegel}, David J. and {Finkbeiner}, Douglas P. and {Davis}, Marc},
        title = "{Maps of Dust Infrared Emission for Use in Estimation of Reddening and Cosmic Microwave Background Radiation Foregrounds}",
      journal = {\apj},
         year = 1998,
        month = jun,
       volume = {500},
       number = {2},
        pages = {525-553},
          doi = {10.1086/305772},
archivePrefix = {arXiv},
       eprint = {astro-ph/9710327},
 primaryClass = {astro-ph},
       adsurl = {https://ui.adsabs.harvard.edu/abs/1998ApJ...500..525S}
}

@INPROCEEDINGS{Blackburn1995,
       author = {{Blackburn}, J.~K.},
        title = "{FTOOLS: A FITS Data Processing and Analysis Software Package}",
    booktitle = {Astronomical Data Analysis Software and Systems IV},
         year = 1995,
       editor = {{Shaw}, R.~A. and {Payne}, H.~E. and {Hayes}, J.~J.~E.},
       series = {Astronomical Society of the Pacific Conference Series},
       volume = {77},
        month = jan,
        pages = {367},
       adsurl = {https://ui.adsabs.harvard.edu/abs/1995ASPC...77..367B}
}

@ARTICLE{Freeman2002,
       author = {{Freeman}, P.~E. and {Kashyap}, V. and {Rosner}, R. and {Lamb}, D.~Q.},
        title = "{A Wavelet-Based Algorithm for the Spatial Analysis of Poisson Data}",
      journal = {\apjs},
         year = 2002,
        month = jan,
       volume = {138},
       number = {1},
        pages = {185-218},
          doi = {10.1086/324017},
archivePrefix = {arXiv},
       eprint = {astro-ph/0108429},
 primaryClass = {astro-ph},
       adsurl = {https://ui.adsabs.harvard.edu/abs/2002ApJS..138..185F}
}

@ARTICLE{Mainzer2011,
       author = {{Mainzer}, A. and {Bauer}, J. and {Grav}, T. and {Masiero}, J. and {Cutri}, R.~M. and {Dailey}, J. and {Eisenhardt}, P. and {McMillan}, R.~S. and {Wright}, E. and {Walker}, R. and {Jedicke}, R. and {Spahr}, T. and {Tholen}, D. and {Alles}, R. and {Beck}, R. and {Brandenburg}, H. and {Conrow}, T. and {Evans}, T. and {Fowler}, J. and {Jarrett}, T. and {Marsh}, K. and {Masci}, F. and {McCallon}, H. and {Wheelock}, S. and {Wittman}, M. and {Wyatt}, P. and {DeBaun}, E. and {Elliott}, G. and {Elsbury}, D. and {Gautier}, T., IV and {Gomillion}, S. and {Leisawitz}, D. and {Maleszewski}, C. and {Micheli}, M. and {Wilkins}, A.},
        title = "{Preliminary Results from NEOWISE: An Enhancement to the Wide-field Infrared Survey Explorer for Solar System Science}",
      journal = {\apj},
         year = 2011,
        month = apr,
       volume = {731},
       number = {1},
          eid = {53},
        pages = {53},
          doi = {10.1088/0004-637X/731/1/53},
archivePrefix = {arXiv},
       eprint = {1102.1996},
 primaryClass = {astro-ph.EP},
       adsurl = {https://ui.adsabs.harvard.edu/abs/2011ApJ...731...53M}
}

@ARTICLE{Masci2019,
       author = {{Masci}, Frank J. and {Laher}, Russ R. and {Rusholme}, Ben and {Shupe}, David L. and {Groom}, Steven and {Surace}, Jason and {Jackson}, Edward and {Monkewitz}, Serge and {Beck}, Ron and {Flynn}, David and {Terek}, Scott and {Landry}, Walter and {Hacopians}, Eugean and {Desai}, Vandana and {Howell}, Justin and {Brooke}, Tim and {Imel}, David and {Wachter}, Stefanie and {Ye}, Quan-Zhi and {Lin}, Hsing-Wen and {Cenko}, S. Bradley and {Cunningham}, Virginia and {Rebbapragada}, Umaa and {Bue}, Brian and {Miller}, Adam A. and {Mahabal}, Ashish and {Bellm}, Eric C. and {Patterson}, Maria T. and {Juri{\'c}}, Mario and {Golkhou}, V. Zach and {Ofek}, Eran O. and {Walters}, Richard and {Graham}, Matthew and {Kasliwal}, Mansi M. and {Dekany}, Richard G. and {Kupfer}, Thomas and {Burdge}, Kevin and {Cannella}, Christopher B. and {Barlow}, Tom and {Van Sistine}, Angela and {Giomi}, Matteo and {Fremling}, Christoffer and {Blagorodnova}, Nadejda and {Levitan}, David and {Riddle}, Reed and {Smith}, Roger M. and {Helou}, George and {Prince}, Thomas A. and {Kulkarni}, Shrinivas R.},
        title = "{The Zwicky Transient Facility: Data Processing, Products, and Archive}",
      journal = {\pasp},
         year = 2019,
        month = jan,
       volume = {131},
       number = {995},
        pages = {018003},
          doi = {10.1088/1538-3873/aae8ac},
archivePrefix = {arXiv},
       eprint = {1902.01872},
 primaryClass = {astro-ph.IM},
       adsurl = {https://ui.adsabs.harvard.edu/abs/2019PASP..131a8003M}
}

@ARTICLE{Wright2010,
       author = {{Wright}, Edward L. and {Eisenhardt}, Peter R.~M. and {Mainzer}, Amy K. and {Ressler}, Michael E. and {Cutri}, Roc M. and {Jarrett}, Thomas and {Kirkpatrick}, J. Davy and {Padgett}, Deborah and {McMillan}, Robert S. and {Skrutskie}, Michael and {Stanford}, S.~A. and {Cohen}, Martin and {Walker}, Russell G. and {Mather}, John C. and {Leisawitz}, David and {Gautier}, Thomas N., III and {McLean}, Ian and {Benford}, Dominic and {Lonsdale}, Carol J. and {Blain}, Andrew and {Mendez}, Bryan and {Irace}, William R. and {Duval}, Valerie and {Liu}, Fengchuan and {Royer}, Don and {Heinrichsen}, Ingolf and {Howard}, Joan and {Shannon}, Mark and {Kendall}, Martha and {Walsh}, Amy L. and {Larsen}, Mark and {Cardon}, Joel G. and {Schick}, Scott and {Schwalm}, Mark and {Abid}, Mohamed and {Fabinsky}, Beth and {Naes}, Larry and {Tsai}, Chao-Wei},
        title = "{The Wide-field Infrared Survey Explorer (WISE): Mission Description and Initial On-orbit Performance}",
      journal = {\aj},
         year = 2010,
        month = dec,
       volume = {140},
       number = {6},
        pages = {1868-1881},
          doi = {10.1088/0004-6256/140/6/1868},
archivePrefix = {arXiv},
       eprint = {1008.0031},
 primaryClass = {astro-ph.IM},
       adsurl = {https://ui.adsabs.harvard.edu/abs/2010AJ....140.1868W}
}

@ARTICLE{Skrutskie2006,
       author = {{Skrutskie}, M.~F. and {Cutri}, R.~M. and {Stiening}, R. and {Weinberg}, M.~D. and {Schneider}, S. and {Carpenter}, J.~M. and {Beichman}, C. and {Capps}, R. and {Chester}, T. and {Elias}, J. and {Huchra}, J. and {Liebert}, J. and {Lonsdale}, C. and {Monet}, D.~G. and {Price}, S. and {Seitzer}, P. and {Jarrett}, T. and {Kirkpatrick}, J.~D. and {Gizis}, J.~E. and {Howard}, E. and {Evans}, T. and {Fowler}, J. and {Fullmer}, L. and {Hurt}, R. and {Light}, R. and {Kopan}, E.~L. and {Marsh}, K.~A. and {McCallon}, H.~L. and {Tam}, R. and {Van Dyk}, S. and {Wheelock}, S.},
        title = "{The Two Micron All Sky Survey (2MASS)}",
      journal = {\aj},
         year = 2006,
        month = feb,
       volume = {131},
       number = {2},
        pages = {1163-1183},
          doi = {10.1086/498708},
       adsurl = {https://ui.adsabs.harvard.edu/abs/2006AJ....131.1163S}
}

@ARTICLE{Gaskell1986,
       author = {{Gaskell}, C.~M. and {Sparke}, L.~S.},
        title = "{Line Variations in Quasars and Seyfert Galaxies}",
      journal = {\apj},
         year = 1986,
        month = jun,
       volume = {305},
        pages = {175},
          doi = {10.1086/164238},
       adsurl = {https://ui.adsabs.harvard.edu/abs/1986ApJ...305..175G}
}

@ARTICLE{Krawczyk2013,
       author = {{Krawczyk}, Coleman M. and {Richards}, Gordon T. and {Mehta}, Sajjan S. and {Vogeley}, Michael S. and {Gallagher}, S.~C. and {Leighly}, Karen M. and {Ross}, Nicholas P. and {Schneider}, Donald P.},
        title = "{Mean Spectral Energy Distributions and Bolometric Corrections for Luminous Quasars}",
      journal = {\apjs},
         year = 2013,
        month = may,
       volume = {206},
       number = {1},
          eid = {4},
        pages = {4},
          doi = {10.1088/0067-0049/206/1/4},
archivePrefix = {arXiv},
       eprint = {1304.5573},
 primaryClass = {astro-ph.CO},
       adsurl = {https://ui.adsabs.harvard.edu/abs/2013ApJS..206....4K}
}

@ARTICLE{Gordon2003,
       author = {{Gordon}, Karl D. and {Clayton}, Geoffrey C. and {Misselt}, K.~A. and {Landolt}, Arlo U. and {Wolff}, Michael J.},
        title = "{A Quantitative Comparison of the Small Magellanic Cloud, Large Magellanic Cloud, and Milky Way Ultraviolet to Near-Infrared Extinction Curves}",
      journal = {\apj},
         year = 2003,
        month = sep,
       volume = {594},
       number = {1},
        pages = {279-293},
          doi = {10.1086/376774},
archivePrefix = {arXiv},
       eprint = {astro-ph/0305257},
 primaryClass = {astro-ph},
       adsurl = {https://ui.adsabs.harvard.edu/abs/2003ApJ...594..279G}
}

@ARTICLE{Glikman2012,
       author = {{Glikman}, Eilat and {Urrutia}, Tanya and {Lacy}, Mark and {Djorgovski}, S. George and {Mahabal}, Ashish and {Myers}, Adam D. and {Ross}, Nicholas P. and {Petitjean}, Patrick and {Ge}, Jian and {Schneider}, Donald P. and {York}, Donald G.},
        title = "{FIRST-2MASS Red Quasars: Transitional Objects Emerging from the Dust}",
      journal = {\apj},
         year = 2012,
        month = sep,
       volume = {757},
       number = {1},
          eid = {51},
        pages = {51},
          doi = {10.1088/0004-637X/757/1/51},
archivePrefix = {arXiv},
       eprint = {1207.2175},
 primaryClass = {astro-ph.CO},
       adsurl = {https://ui.adsabs.harvard.edu/abs/2012ApJ...757...51G}
}

@ARTICLE{Hopkins2004,
       author = {{Hopkins}, Philip F. and {Strauss}, Michael A. and {Hall}, Patrick B. and {Richards}, Gordon T. and {Cooper}, Ariana S. and {Schneider}, Donald P. and {Vanden Berk}, Daniel E. and {Jester}, Sebastian and {Brinkmann}, J. and {Szokoly}, Gyula P.},
        title = "{Dust Reddening in Sloan Digital Sky Survey Quasars}",
      journal = {\aj},
         year = 2004,
        month = sep,
       volume = {128},
       number = {3},
        pages = {1112-1123},
          doi = {10.1086/423291},
archivePrefix = {arXiv},
       eprint = {astro-ph/0406293},
 primaryClass = {astro-ph},
       adsurl = {https://ui.adsabs.harvard.edu/abs/2004AJ....128.1112H}
}

@ARTICLE{Rudy2000,
       author = {{Rudy}, Richard J. and {Mazuk}, S. and {Puetter}, R.~C. and {Hamann}, F.},
        title = "{The 1 Micron Fe II Lines of the Seyfert Galaxy I Zw 1}",
      journal = {\apj},
         year = 2000,
        month = aug,
       volume = {539},
       number = {1},
        pages = {166-171},
          doi = {10.1086/309222},
       adsurl = {https://ui.adsabs.harvard.edu/abs/2000ApJ...539..166R}
}

@ARTICLE{Kallman2004,
       author = {{Kallman}, T.~R. and {Palmeri}, P. and {Bautista}, M.~A. and {Mendoza}, C. and {Krolik}, J.~H.},
        title = "{Photoionization Modeling and the K Lines of Iron}",
      journal = {\apjs},
         year = 2004,
        month = dec,
       volume = {155},
       number = {2},
        pages = {675-701},
          doi = {10.1086/424039},
archivePrefix = {arXiv},
       eprint = {astro-ph/0405210},
 primaryClass = {astro-ph},
       adsurl = {https://ui.adsabs.harvard.edu/abs/2004ApJS..155..675K}
}

@ARTICLE{Reeves2008,
       author = {{Reeves}, James and {Done}, Chris and {Pounds}, Ken and {Terashima}, Yuichi and {Hayashida}, Kiyoshi and {Anabuki}, Naohisa and {Uchino}, Masahiro and {Turner}, Martin},
        title = "{On why the iron K-shell absorption in AGN is not a signature of the local warm/hot intergalactic medium}",
      journal = {\mnras},
         year = 2008,
        month = mar,
       volume = {385},
       number = {1},
        pages = {L108-L112},
          doi = {10.1111/j.1745-3933.2008.00443.x},
archivePrefix = {arXiv},
       eprint = {0801.1587},
 primaryClass = {astro-ph},
       adsurl = {https://ui.adsabs.harvard.edu/abs/2008MNRAS.385L.108R}
}

@ARTICLE{Grevesse&Sauval1998,
       author = {{Grevesse}, N. and {Sauval}, A.~J.},
        title = "{Standard Solar Composition}",
      journal = {\ssr},
         year = 1998,
        month = may,
       volume = {85},
        pages = {161-174},
          doi = {10.1023/A:1005161325181},
       adsurl = {https://ui.adsabs.harvard.edu/abs/1998SSRv...85..161G}
}

@ARTICLE{Kocevski2023,
       author = {{Kocevski}, Dale D. and {Onoue}, Masafusa and {Inayoshi}, Kohei and {Trump}, Jonathan R. and {Arrabal Haro}, Pablo and {Grazian}, Andrea and {Dickinson}, Mark and {Finkelstein}, Steven L. and {Kartaltepe}, Jeyhan S. and {Hirschmann}, Michaela and {Aird}, James and {Holwerda}, Benne W. and {Fujimoto}, Seiji and {Juneau}, St{\'e}phanie and {Amor{\'\i}n}, Ricardo O. and {Backhaus}, Bren E. and {Bagley}, Micaela B. and {Barro}, Guillermo and {Bell}, Eric F. and {Bisigello}, Laura and {Calabr{\`o}}, Antonello and {Cleri}, Nikko J. and {Cooper}, M.~C. and {Ding}, Xuheng and {Grogin}, Norman A. and {Ho}, Luis C. and {Hutchison}, Taylor A. and {Inoue}, Akio K. and {Jiang}, Linhua and {Jones}, Brenda and {Koekemoer}, Anton M. and {Li}, Wenxiu and {Li}, Zhengrong and {McGrath}, Elizabeth J. and {Molina}, Juan and {Papovich}, Casey and {P{\'e}rez-Gonz{\'a}lez}, Pablo G. and {Pirzkal}, Nor and {Wilkins}, Stephen M. and {Yang}, Guang and {Yung}, L.~Y. Aaron},
        title = "{Hidden Little Monsters: Spectroscopic Identification of Low-mass, Broad-line AGNs at z > 5 with CEERS}",
      journal = {\apjl},
         year = 2023,
        month = sep,
       volume = {954},
       number = {1},
          eid = {L4},
        pages = {L4},
          doi = {10.3847/2041-8213/ace5a0},
archivePrefix = {arXiv},
       eprint = {2302.00012},
 primaryClass = {astro-ph.GA},
       adsurl = {https://ui.adsabs.harvard.edu/abs/2023ApJ...954L...4K}
}

@ARTICLE{Greene2024,
       author = {{Greene}, Jenny E. and {Labbe}, Ivo and {Goulding}, Andy D. and {Furtak}, Lukas J. and {Chemerynska}, Iryna and {Kokorev}, Vasily and {Dayal}, Pratika and {Volonteri}, Marta and {Williams}, Christina C. and {Wang}, Bingjie and {Setton}, David J. and {Burgasser}, Adam J. and {Bezanson}, Rachel and {Atek}, Hakim and {Brammer}, Gabriel and {Cutler}, Sam E. and {Feldmann}, Robert and {Fujimoto}, Seiji and {Glazebrook}, Karl and {de Graaff}, Anna and {Khullar}, Gourav and {Leja}, Joel and {Marchesini}, Danilo and {Maseda}, Michael V. and {Matthee}, Jorryt and {Miller}, Tim B. and {Naidu}, Rohan P. and {Nanayakkara}, Themiya and {Oesch}, Pascal A. and {Pan}, Richard and {Papovich}, Casey and {Price}, Sedona H. and {van Dokkum}, Pieter and {Weaver}, John R. and {Whitaker}, Katherine E. and {Zitrin}, Adi},
        title = "{UNCOVER Spectroscopy Confirms the Surprising Ubiquity of Active Galactic Nuclei in Red Sources at z > 5}",
      journal = {\apj},
         year = 2024,
        month = mar,
       volume = {964},
       number = {1},
          eid = {39},
        pages = {39},
          doi = {10.3847/1538-4357/ad1e5f},
archivePrefix = {arXiv},
       eprint = {2309.05714},
 primaryClass = {astro-ph.GA},
       adsurl = {https://ui.adsabs.harvard.edu/abs/2024ApJ...964...39G}
}

@ARTICLE{Ji2025,
       author = {{Ji}, Xihan and {Maiolino}, Roberto and {{\"U}bler}, Hannah and {Scholtz}, Jan and {D'Eugenio}, Francesco and {Sun}, Fengwu and {Perna}, Michele and {Turner}, Hannah and {Carniani}, Stefano and {Arribas}, Santiago and {Bennett}, Jake S. and {Bunker}, Andrew and {Charlot}, St{\'e}phane and {Cresci}, Giovanni and {Curti}, Mirko and {Egami}, Eiichi and {Fabian}, Andy and {Inayoshi}, Kohei and {Isobe}, Yuki and {Jones}, Gareth and {Juod{\v{z}}balis}, Ignas and {Kumari}, Nimisha and {Lyu}, Jianwei and {Mazzolari}, Giovanni and {Parlanti}, Eleonora and {Robertson}, Brant and {Rodr{\'\i}guez Del Pino}, Bruno and {Schneider}, Raffaella and {Sijacki}, Debora and {Tacchella}, Sandro and {Trinca}, Alessandro and {Valiante}, Rosa and {Venturi}, Giacomo and {Volonteri}, Marta and {Willott}, Chris and {Witten}, Callum and {Witstok}, Joris},
        title = "{BlackTHUNDER ─ A non-stellar Balmer break in a black hole-dominated little red dot at z = 7.04}",
      journal = {\mnras},
         year = 2025,
        month = dec,
       volume = {544},
       number = {4},
        pages = {3900-3935},
          doi = {10.1093/mnras/staf1867},
archivePrefix = {arXiv},
       eprint = {2501.13082},
 primaryClass = {astro-ph.GA},
       adsurl = {https://ui.adsabs.harvard.edu/abs/2025MNRAS.544.3900J}
}

@ARTICLE{Kocevski2025,
       author = {{Kocevski}, Dale D. and {Finkelstein}, Steven L. and {Barro}, Guillermo and {Taylor}, Anthony J. and {Calabr{\`o}}, Antonello and {Laloux}, Brivael and {Buchner}, Johannes and {Trump}, Jonathan R. and {Leung}, Gene C.~K. and {Yang}, Guang and {Dickinson}, Mark and {P{\'e}rez-Gonz{\'a}lez}, Pablo G. and {Pacucci}, Fabio and {Inayoshi}, Kohei and {Somerville}, Rachel S. and {McGrath}, Elizabeth J. and {Akins}, Hollis B. and {Bagley}, Micaela B. and {Bowler}, Rebecca A.~A. and {Bisigello}, Laura and {Carnall}, Adam and {Casey}, Caitlin M. and {Cheng}, Yingjie and {Cleri}, Nikko J. and {Costantin}, Luca and {Cullen}, Fergus and {Davis}, Kelcey and {Donnan}, Callum T. and {Dunlop}, James S. and {Ellis}, Richard S. and {Ferguson}, Henry C. and {Fujimoto}, Seiji and {Fontana}, Adriano and {Giavalisco}, Mauro and {Grazian}, Andrea and {Grogin}, Norman A. and {Hathi}, Nimish P. and {Hirschmann}, Michaela and {Huertas-Company}, Marc and {Holwerda}, Benne W. and {Illingworth}, Garth and {Juneau}, St{\'e}phanie and {Kartaltepe}, Jeyhan S. and {Koekemoer}, Anton M. and {Li}, Wenxiu and {Lucas}, Ray A. and {Magee}, Dan and {Mason}, Charlotte and {McLeod}, Derek J. and {McLure}, Ross J. and {Napolitano}, Lorenzo and {Papovich}, Casey and {Pirzkal}, Nor and {Rodighiero}, Giulia and {Santini}, Paola and {Wilkins}, Stephen M. and {Yung}, L.~Y. Aaron},
        title = "{The Rise of Faint, Red Active Galactic Nuclei at z > 4: A Sample of Little Red Dots in the JWST Extragalactic Legacy Fields}",
      journal = {\apj},
         year = 2025,
        month = jun,
       volume = {986},
       number = {2},
          eid = {126},
        pages = {126},
          doi = {10.3847/1538-4357/adbc7d},
archivePrefix = {arXiv},
       eprint = {2404.03576},
 primaryClass = {astro-ph.GA},
       adsurl = {https://ui.adsabs.harvard.edu/abs/2025ApJ...986..126K}
}

@ARTICLE{Zhang2024,
       author = {{Zhang}, Zijian and {Jiang}, Linhua and {Liu}, Weiyang and {Ho}, Luis C.},
        title = "{Analysis of Multi-epoch JWST Images of {\ensuremath{\sim}}300 Little Red Dots: Tentative Detection of Variability in a Minority of Sources}",
      journal = {\apj},
         year = 2025,
        month = may,
       volume = {985},
       number = {1},
          eid = {119},
        pages = {119},
          doi = {10.3847/1538-4357/adcb3e},
archivePrefix = {arXiv},
       eprint = {2411.02729},
 primaryClass = {astro-ph.GA},
       adsurl = {https://ui.adsabs.harvard.edu/abs/2025ApJ...985..119Z}
}

@ARTICLE{Maiolino2024,
       author = {{Maiolino}, Roberto and {Scholtz}, Jan and {Curtis-Lake}, Emma and {Carniani}, Stefano and {Baker}, William and {de Graaff}, Anna and {Tacchella}, Sandro and {{\"U}bler}, Hannah and {D'Eugenio}, Francesco and {Witstok}, Joris and {Curti}, Mirko and {Arribas}, Santiago and {Bunker}, Andrew J. and {Charlot}, St{\'e}phane and {Chevallard}, Jacopo and {Eisenstein}, Daniel J. and {Egami}, Eiichi and {Ji}, Zhiyuan and {Jones}, Gareth C. and {Lyu}, Jianwei and {Rawle}, Tim and {Robertson}, Brant and {Rujopakarn}, Wiphu and {Perna}, Michele and {Sun}, Fengwu and {Venturi}, Giacomo and {Williams}, Christina C. and {Willott}, Chris},
        title = "{JADES: The diverse population of infant black holes at 4 < z < 11: Merging, tiny, poor, but mighty}",
      journal = {\aap},
         year = 2024,
        month = nov,
       volume = {691},
          eid = {A145},
        pages = {A145},
          doi = {10.1051/0004-6361/202347640},
archivePrefix = {arXiv},
       eprint = {2308.01230},
 primaryClass = {astro-ph.GA},
       adsurl = {https://ui.adsabs.harvard.edu/abs/2024A&A...691A.145M}
}

@ARTICLE{Maiolino2025,
       author = {{Maiolino}, Roberto and {Risaliti}, Guido and {Signorini}, Matilde and {Trefoloni}, Bartolomeo and {Juod{\v{z}}balis}, Ignas and {Scholtz}, Jan and {{\"U}bler}, Hannah and {D'Eugenio}, Francesco and {Carniani}, Stefano and {Fabian}, Andy and {Ji}, Xihan and {Mazzolari}, Giovanni and {Bertola}, Elena and {Brusa}, Marcella and {Bunker}, Andrew J. and {Charlot}, Stephane and {Comastri}, Andrea and {Cresci}, Giovanni and {DeCoursey}, Christa Noel and {Egami}, Eiichi and {Fiore}, Fabrizio and {Gilli}, Roberto and {Perna}, Michele and {Tacchella}, Sandro and {Venturi}, Giacomo},
        title = "{JWST meets Chandra: a large population of Compton thick, feedback-free, and intrinsically X-ray weak AGN, with a sprinkle of SNe}",
      journal = {\mnras},
         year = 2025,
        month = apr,
       volume = {538},
       number = {3},
        pages = {1921-1943},
          doi = {10.1093/mnras/staf359},
archivePrefix = {arXiv},
       eprint = {2405.00504},
 primaryClass = {astro-ph.GA},
       adsurl = {https://ui.adsabs.harvard.edu/abs/2025MNRAS.538.1921M}
}

@ARTICLE{Blustin2005,
       author = {{Blustin}, A.~J. and {Page}, M.~J. and {Fuerst}, S.~V. and {Branduardi-Raymont}, G. and {Ashton}, C.~E.},
        title = "{The nature and origin of Seyfert warm absorbers}",
      journal = {\aap},
         year = 2005,
        month = feb,
       volume = {431},
        pages = {111-125},
          doi = {10.1051/0004-6361:20041775},
archivePrefix = {arXiv},
       eprint = {astro-ph/0411297},
 primaryClass = {astro-ph},
       adsurl = {https://ui.adsabs.harvard.edu/abs/2005A&A...431..111B}
}

@ARTICLE{Gofford2015,
       author = {{Gofford}, J. and {Reeves}, J.~N. and {McLaughlin}, D.~E. and {Braito}, V. and {Turner}, T.~J. and {Tombesi}, F. and {Cappi}, M.},
        title = "{The Suzaku view of highly ionized outflows in AGN - II. Location, energetics and scalings with bolometric luminosity}",
      journal = {\mnras},
         year = 2015,
        month = aug,
       volume = {451},
       number = {4},
        pages = {4169-4182},
          doi = {10.1093/mnras/stv1207},
archivePrefix = {arXiv},
       eprint = {1506.00614},
 primaryClass = {astro-ph.HE},
       adsurl = {https://ui.adsabs.harvard.edu/abs/2015MNRAS.451.4169G}
}

@ARTICLE{King2016,
       author = {{King}, Andrew and {Muldrew}, Stuart I.},
        title = "{Black hole winds II: Hyper-Eddington winds and feedback}",
      journal = {\mnras},
         year = 2016,
        month = jan,
       volume = {455},
       number = {2},
        pages = {1211-1217},
          doi = {10.1093/mnras/stv2347},
archivePrefix = {arXiv},
       eprint = {1510.01736},
 primaryClass = {astro-ph.HE},
       adsurl = {https://ui.adsabs.harvard.edu/abs/2016MNRAS.455.1211K}
}

@ARTICLE{Chen2023,
       author = {{Chen}, Yong-Jie and {Liu}, Jun-Rong and {Zhai}, Shuo and {Yao}, Zhu-Heng and {Li}, Yan-Rong and {Du}, Pu and {Hu}, Chen and {Guo}, Wei-Jian and {Xiao}, Ming and {Songsheng}, Yu-Yang and {Wang}, Jian-Min},
        title = "{Mid-infrared dusty torus sizes in active galactic nuclei with H{\ensuremath{\beta}} reverberation mapping}",
      journal = {\mnras},
         year = 2023,
        month = jul,
       volume = {522},
       number = {3},
        pages = {3439-3457},
          doi = {10.1093/mnras/stad1136},
       adsurl = {https://ui.adsabs.harvard.edu/abs/2023MNRAS.522.3439C}
}

@ARTICLE{Reeves2009,
       author = {{Reeves}, J.~N. and {O'Brien}, P.~T. and {Braito}, V. and {Behar}, E. and {Miller}, L. and {Turner}, T.~J. and {Fabian}, A.~C. and {Kaspi}, S. and {Mushotzky}, R. and {Ward}, M.},
        title = "{A Compton-thick Wind in the High-luminosity Quasar, PDS 456}",
      journal = {\apj},
         year = 2009,
        month = aug,
       volume = {701},
       number = {1},
        pages = {493-507},
          doi = {10.1088/0004-637X/701/1/493},
archivePrefix = {arXiv},
       eprint = {0906.0312},
 primaryClass = {astro-ph.HE},
       adsurl = {https://ui.adsabs.harvard.edu/abs/2009ApJ...701..493R}
}

@ARTICLE{Nardini2015,
       author = {{Nardini}, E. and {Reeves}, J.~N. and {Gofford}, J. and {Harrison}, F.~A. and {Risaliti}, G. and {Braito}, V. and {Costa}, M.~T. and {Matzeu}, G.~A. and {Walton}, D.~J. and {Behar}, E. and {Boggs}, S.~E. and {Christensen}, F.~E. and {Craig}, W.~W. and {Hailey}, C.~J. and {Matt}, G. and {Miller}, J.~M. and {O'Brien}, P.~T. and {Stern}, D. and {Turner}, T.~J. and {Ward}, M.~J.},
        title = "{Black hole feedback in the luminous quasar PDS 456}",
      journal = {Science},
         year = 2015,
        month = feb,
       volume = {347},
       number = {6224},
        pages = {860-863},
          doi = {10.1126/science.1259202},
archivePrefix = {arXiv},
       eprint = {1502.06636},
 primaryClass = {astro-ph.HE},
       adsurl = {https://ui.adsabs.harvard.edu/abs/2015Sci...347..860N}
}

@ARTICLE{Reeves2020,
       author = {{Reeves}, J.~N. and {Braito}, V. and {Chartas}, G. and {Hamann}, F. and {Laha}, S. and {Nardini}, E.},
        title = "{Resolving the Soft X-Ray Ultrafast Outflow in PDS 456}",
      journal = {\apj},
         year = 2020,
        month = may,
       volume = {895},
       number = {1},
          eid = {37},
        pages = {37},
          doi = {10.3847/1538-4357/ab8cc4},
archivePrefix = {arXiv},
       eprint = {2004.12439},
 primaryClass = {astro-ph.HE},
       adsurl = {https://ui.adsabs.harvard.edu/abs/2020ApJ...895...37R}
}

@ARTICLE{XRISM2025,
       author = {{XRISM Collaboration} and {Audard}, Marc and {Awaki}, Hisamitsu and {Ballhausen}, Ralf and {Bamba}, Aya and {Behar}, Ehud and {Boissay-Malaquin}, Rozenn and {Brenneman}, Laura and {Brown}, Gregory V. and {Corrales}, Lia and {Costantini}, Elisa and {Cumbee}, Renata and {Trigo}, Mar{\'\i}a D{\'\i}az and {Done}, Chris and {Dotani}, Tadayasu and {Ebisawa}, Ken and {Eckart}, Megan and {Eckert}, Dominique and {Enoto}, Teruaki and {Eguchi}, Satoshi and {Ezoe}, Yuichiro and {Foster}, Adam and {Fujimoto}, Ryuichi and {Fujita}, Yutaka and {Fukazawa}, Yasushi and {Fukushima}, Kotaro and {Furuzawa}, Akihiro and {Gallo}, Luigi and {Garc{\'\i}a}, Javier A. and {Gu}, Liyi and {Guainazzi}, Matteo and {Hagino}, Kouichi and {Hamaguchi}, Kenji and {Hatsukade}, Isamu and {Hayashi}, Katsuhiro and {Hayashi}, Takayuki and {Hell}, Natalie and {Hodges-Kluck}, Edmund and {Hornschemeier}, Ann and {Ichinohe}, Yuto and {Ishida}, Manabu and {Ishikawa}, Kumi and {Ishisaki}, Yoshitaka and {Kaastra}, Jelle and {Kallman}, Timothy and {Kara}, Erin and {Katsuda}, Satoru and {Kanemaru}, Yoshiaki and {Kelley}, Richard and {Kilbourne}, Caroline and {Kitamoto}, Shunji and {Kobayashi}, Shogo and {Kohmura}, Takayoshi and {Kubota}, Aya and {Leutenegger}, Maurice and {Loewenstein}, Michael and {Maeda}, Yoshitomo and {Markevitch}, Maxim and {Matsumoto}, Hironori and {Matsushita}, Kyoko and {McCammon}, Dan and {McNamara}, Brian and {Mernier}, Fran{\c{c}}ois and {Miller}, Eric D. and {Miller}, Jon M. and {Mitsuishi}, Ikuyuki and {Mizumoto}, Misaki and {Mizuno}, Tsunefumi and {Mori}, Koji and {Mukai}, Koji and {Murakami}, Hiroshi and {Mushotzky}, Richard and {Nakajima}, Hiroshi and {Nakazawa}, Kazuhiro and {Ness}, Jan-Uwe and {Nobukawa}, Kumiko and {Nobukawa}, Masayoshi and {Noda}, Hirofumi and {Odaka}, Hirokazu and {Ogawa}, Shoji and {Ogorzalek}, Anna and {Okajima}, Takashi and {Ota}, Naomi and {Paltani}, Stephane and {Petre}, Robert and {Plucinsky}, Paul and {Porter}, Frederick Scott and {Pottschmidt}, Katja and {Sato}, Kosuke and {Sato}, Toshiki and {Sawada}, Makoto and {Seta}, Hiromi and {Shidatsu}, Megumi and {Simionescu}, Aurora and {Smith}, Randall and {Suzuki}, Hiromasa and {Szymkowiak}, Andrew and {Takahashi}, Hiromitsu and {Takeo}, Mai and {Tamagawa}, Toru and {Tamura}, Keisuke and {Tanaka}, Takaaki and {Tanimoto}, Atsushi and {Tashiro}, Makoto and {Terada}, Yukikatsu and {Terashima}, Yuichi and {Tsuboi}, Yohko and {Tsujimoto}, Masahiro and {Tsunemi}, Hiroshi and {Tsuru}, Takeshi G. and {Uchida}, Hiroyuki and {Uchida}, Nagomi and {Uchida}, Yuusuke and {Uchiyama}, Hideki and {Ueda}, Yoshihiro and {Uno}, Shinichiro and {Vink}, Jacco and {Watanabe}, Shin and {Williams}, Brian J. and {Yamada}, Satoshi and {Yamada}, Shinya and {Yamaguchi}, Hiroya and {Yamaoka}, Kazutaka and {Yamasaki}, Noriko and {Yamauchi}, Makoto and {Yamauchi}, Shigeo and {Yaqoob}, Tahir and {Yoneyama}, Tomokage and {Yoshida}, Tessei and {Yukita}, Mihoko and {Zhuravleva}, Irina and {Braito}, Valentina and {Cond{\`o}}, Pierpaolo and {Fukumura}, Keigo and {Gonzalez}, Adam and {Luminari}, Alfredo and {Miyamoto}, Aiko and {Mizukawa}, Ryuki and {Reeves}, James and {Sato}, Riki and {Tombesi}, Francesco and {Xu}, Yerong},
        title = "{Structured ionized winds shooting out from a quasar at relativistic speeds}",
      journal = {\nat},
         year = 2025,
        month = may,
       volume = {641},
       number = {8065},
        pages = {1132-1136},
          doi = {10.1038/s41586-025-08968-2},
archivePrefix = {arXiv},
       eprint = {2505.09171},
 primaryClass = {astro-ph.HE},
       adsurl = {https://ui.adsabs.harvard.edu/abs/2025Natur.641.1132X}
}

@ARTICLE{Yue2024,
       author = {{Yue}, Minghao and {Eilers}, Anna-Christina and {Ananna}, Tonima Tasnim and {Panagiotou}, Christos and {Kara}, Erin and {Miyaji}, Takamitsu},
        title = "{Stacking X-Ray Observations of ``Little Red Dots'': Implications for Their Active Galactic Nucleus Properties}",
      journal = {\apjl},
         year = 2024,
        month = oct,
       volume = {974},
       number = {2},
          eid = {L26},
        pages = {L26},
          doi = {10.3847/2041-8213/ad7eba},
archivePrefix = {arXiv},
       eprint = {2404.13290},
 primaryClass = {astro-ph.GA},
       adsurl = {https://ui.adsabs.harvard.edu/abs/2024ApJ...974L..26Y}
}

@ARTICLE{Done2016,
       author = {{Done}, Chris and {Jin}, Chichuan},
        title = "{The mass and spin of the extreme Narrow Line Seyfert 1 Galaxy 1H 0707-495 and its implications for the trigger for relativistic jets}",
      journal = {\mnras},
         year = 2016,
        month = aug,
       volume = {460},
       number = {2},
        pages = {1716-1724},
          doi = {10.1093/mnras/stw1070},
archivePrefix = {arXiv},
       eprint = {1506.04547},
 primaryClass = {astro-ph.HE},
       adsurl = {https://ui.adsabs.harvard.edu/abs/2016MNRAS.460.1716D}
}

@ARTICLE{Giustini2011,
       author = {{Giustini}, M. and {Cappi}, M. and {Chartas}, G. and {Dadina}, M. and {Eracleous}, M. and {Ponti}, G. and {Proga}, D. and {Tombesi}, F. and {Vignali}, C. and {Palumbo}, G.~G.~C.},
        title = "{Variable X-ray absorption in the mini-BAL QSO PG 1126-041}",
      journal = {\aap},
         year = 2011,
        month = dec,
       volume = {536},
          eid = {A49},
        pages = {A49},
          doi = {10.1051/0004-6361/201117732},
archivePrefix = {arXiv},
       eprint = {1109.6026},
 primaryClass = {astro-ph.CO},
       adsurl = {https://ui.adsabs.harvard.edu/abs/2011A&A...536A..49G}
}

@ARTICLE{Giustini2023,
       author = {{Giustini}, M. and {Rodr{\'\i}guez Hidalgo}, P. and {Reeves}, J.~N. and {Matzeu}, G. and {Braito}, V. and {Eracleous}, M. and {Chartas}, G. and {Schartel}, N. and {Vignali}, C. and {Hall}, P.~B. and {Waters}, T. and {Ponti}, G. and {Proga}, D. and {Dadina}, M. and {Cappi}, M. and {Miniutti}, G. and {de Vries}, L.},
        title = "{Coordinated X-ray and UV absorption within the accretion disk wind of the active galactic nucleus PG 1126-041}",
      journal = {\aap},
         year = 2023,
        month = nov,
       volume = {679},
          eid = {A73},
        pages = {A73},
          doi = {10.1051/0004-6361/202244270},
archivePrefix = {arXiv},
       eprint = {2306.05469},
 primaryClass = {astro-ph.HE},
       adsurl = {https://ui.adsabs.harvard.edu/abs/2023A&A...679A..73G}
}

@ARTICLE{Reeves2024,
       author = {{Reeves}, J.~N. and {Braito}, V. and {Luminari}, A. and {Porquet}, D. and {Laurenti}, M. and {Matzeu}, G. and {Lobban}, A. and {Hagen}, S.},
        title = "{An Eddington-limited Accretion Disk Wind in the Narrow-line Seyfert 1 PG 1448+273}",
      journal = {\apj},
         year = 2024,
        month = oct,
       volume = {974},
       number = {1},
          eid = {58},
        pages = {58},
          doi = {10.3847/1538-4357/ad6b95},
archivePrefix = {arXiv},
       eprint = {2408.15095},
 primaryClass = {astro-ph.HE},
       adsurl = {https://ui.adsabs.harvard.edu/abs/2024ApJ...974...58R}
}

@ARTICLE{Alston2019,
       author = {{Alston}, W.~N. and {Fabian}, A.~C. and {Buisson}, D.~J.~K. and {Kara}, E. and {Parker}, M.~L. and {Lohfink}, A.~M. and {Uttley}, P. and {Wilkins}, D.~R. and {Pinto}, C. and {De Marco}, B. and {Cackett}, E.~M. and {Middleton}, M.~J. and {Walton}, D.~J. and {Reynolds}, C.~S. and {Jiang}, J. and {Gallo}, L.~C. and {Zogbhi}, A. and {Miniutti}, G. and {Dovciak}, M. and {Young}, A.~J.},
        title = "{The remarkable X-ray variability of IRAS 13224-3809 - I. The variability process}",
      journal = {\mnras},
         year = 2019,
        month = jan,
       volume = {482},
       number = {2},
        pages = {2088-2106},
          doi = {10.1093/mnras/sty2527},
archivePrefix = {arXiv},
       eprint = {1803.10444},
 primaryClass = {astro-ph.HE},
       adsurl = {https://ui.adsabs.harvard.edu/abs/2019MNRAS.482.2088A}
}

@ARTICLE{Midooka2023,
       author = {{Midooka}, Takuya and {Mizumoto}, Misaki and {Ebisawa}, Ken},
        title = "{Radiatively Driven Clumpy X-Ray Absorbers in the NLS1 Galaxy IRAS 13224-3809}",
      journal = {\apj},
         year = 2023,
        month = sep,
       volume = {954},
       number = {1},
          eid = {47},
        pages = {47},
          doi = {10.3847/1538-4357/ace71a},
archivePrefix = {arXiv},
       eprint = {2307.12023},
 primaryClass = {astro-ph.HE},
       adsurl = {https://ui.adsabs.harvard.edu/abs/2023ApJ...954...47M}
}

@ARTICLE{Reeves2018,
       author = {{Reeves}, J.~N. and {Lobban}, A. and {Pounds}, K.~A.},
        title = "{The Variable Fast Soft X-Ray Wind in PG 1211+143}",
      journal = {\apj},
         year = 2018,
        month = feb,
       volume = {854},
       number = {1},
          eid = {28},
        pages = {28},
          doi = {10.3847/1538-4357/aaa776},
archivePrefix = {arXiv},
       eprint = {1801.03784},
 primaryClass = {astro-ph.HE},
       adsurl = {https://ui.adsabs.harvard.edu/abs/2018ApJ...854...28R}
}

@ARTICLE{Wu2022,
       author = {{Wu}, Qiaoya and {Shen}, Yue},
        title = "{A Catalog of Quasar Properties from Sloan Digital Sky Survey Data Release 16}",
      journal = {\apjs},
         year = 2022,
        month = dec,
       volume = {263},
       number = {2},
          eid = {42},
        pages = {42},
          doi = {10.3847/1538-4365/ac9ead},
archivePrefix = {arXiv},
       eprint = {2209.03987},
 primaryClass = {astro-ph.GA},
       adsurl = {https://ui.adsabs.harvard.edu/abs/2022ApJS..263...42W}
}

@ARTICLE{Wang2024,
       author = {{Wang}, Shouyi and {Brandt}, W.~N. and {Luo}, Bin and {Yu}, Zhibo and {Zou}, Fan and {Huang}, Jian and {Ni}, Qingling and {Vito}, Fabio},
        title = "{The Remarkable X-Ray Spectra and Variability of the Ultraluminous Weak-line Quasar SDSS J1521+5202}",
      journal = {\apj},
         year = 2024,
        month = oct,
       volume = {974},
       number = {1},
          eid = {2},
        pages = {2},
          doi = {10.3847/1538-4357/ad7589},
archivePrefix = {arXiv},
       eprint = {2408.16060},
 primaryClass = {astro-ph.HE},
       adsurl = {https://ui.adsabs.harvard.edu/abs/2024ApJ...974....2W}
}

@ARTICLE{Huang2020,
       author = {{Huang}, Jian and {Luo}, Bin and {Du}, Pu and {Hu}, Chen and {Wang}, Jian-Min and {Li}, Yi-Jia},
        title = "{On the Relation between the Hard X-Ray Photon Index and Accretion Rate for Super-Eddington Accreting Quasars}",
      journal = {\apj},
         year = 2020,
        month = jun,
       volume = {895},
       number = {2},
          eid = {114},
        pages = {114},
          doi = {10.3847/1538-4357/ab9019},
archivePrefix = {arXiv},
       eprint = {2005.01749},
 primaryClass = {astro-ph.HE},
       adsurl = {https://ui.adsabs.harvard.edu/abs/2020ApJ...895..114H}
}

@INPROCEEDINGS{Shappee2014,
       author = {{Shappee}, Benjamin and {Prieto}, J. and {Stanek}, K.~Z. and {Kochanek}, C.~S. and {Holoien}, T. and {Jencson}, J. and {Basu}, U. and {Beacom}, J.~F. and {Szczygiel}, D. and {Pojmanski}, G. and {Brimacombe}, J. and {Dubberley}, M. and {Elphick}, M. and {Foale}, S. and {Hawkins}, E. and {Mullins}, D. and {Rosing}, W. and {Ross}, R. and {Walker}, Z.},
        title = "{All Sky Automated Survey for SuperNovae (ASAS-SN or ``Assassin'')}",
    booktitle = {American Astronomical Society Meeting Abstracts \#223},
         year = 2014,
       series = {American Astronomical Society Meeting Abstracts},
       volume = {223},
        month = jan,
          eid = {236.03},
        pages = {236.03},
       adsurl = {https://ui.adsabs.harvard.edu/abs/2014AAS...22323603S}
}

@ARTICLE{Laor1997,
       author = {{Laor}, Ari and {Jannuzi}, Buell T. and {Green}, Richard F. and {Boroson}, Todd A.},
        title = "{The Ultraviolet Properties of the Narrow-Line Quasar I Zw 1}",
      journal = {\apj},
         year = 1997,
        month = nov,
       volume = {489},
       number = {2},
        pages = {656-671},
          doi = {10.1086/304816},
archivePrefix = {arXiv},
       eprint = {astro-ph/9706264},
 primaryClass = {astro-ph},
       adsurl = {https://ui.adsabs.harvard.edu/abs/1997ApJ...489..656L}
}

@ARTICLE{Huang2025,
       author = {{Huang}, Jian and {Luo}, Bin and {Brandt}, W.~N. and {Chen}, Ying and {Ni}, Qingling and {Xue}, Yongquan and {Zhang}, Zijian},
        title = "{Photometric Selection of Type 1 Quasars in the XMM-LSS Field with Machine Learning and the Disk{\textendash}Corona Connection}",
      journal = {\apj},
         year = 2025,
        month = feb,
       volume = {979},
       number = {2},
          eid = {107},
        pages = {107},
          doi = {10.3847/1538-4357/ad9baf},
archivePrefix = {arXiv},
       eprint = {2412.06923},
 primaryClass = {astro-ph.GA},
       adsurl = {https://ui.adsabs.harvard.edu/abs/2025ApJ...979..107H}
}

@ARTICLE{Chen2024,
       author = {{Chen}, Y. and {Luo}, B. and {Brandt}, W.~N. and {Zuo}, Wenwen and {Dix}, Cooper and {Ha}, Trung and {Matthews}, Brandon and {Paul}, Jeremiah D. and {Plotkin}, Richard M. and {Shemmer}, Ohad},
        title = "{Rest-frame Optical Spectroscopy of Ten z {\ensuremath{\sim}} 2 Weak Emission-line Quasars}",
      journal = {\apj},
         year = 2024,
        month = sep,
       volume = {972},
       number = {2},
          eid = {191},
        pages = {191},
          doi = {10.3847/1538-4357/ad5f89},
archivePrefix = {arXiv},
       eprint = {2407.03422},
 primaryClass = {astro-ph.GA},
       adsurl = {https://ui.adsabs.harvard.edu/abs/2024ApJ...972..191C}
}

@ARTICLE{King2015,
       author = {{King}, Andrew and {Pounds}, Ken},
        title = "{Powerful Outflows and Feedback from Active Galactic Nuclei}",
      journal = {\araa},
         year = 2015,
        month = aug,
       volume = {53},
        pages = {115-154},
          doi = {10.1146/annurev-astro-082214-122316},
archivePrefix = {arXiv},
       eprint = {1503.05206},
 primaryClass = {astro-ph.GA},
       adsurl = {https://ui.adsabs.harvard.edu/abs/2015ARA&A..53..115K}
}

@ARTICLE{Grupe2010,
       author = {{Grupe}, Dirk and {Komossa}, Stefanie and {Leighly}, Karen M. and {Page}, Kim L.},
        title = "{The Simultaneous Optical-to-X-Ray Spectral Energy Distribution of Soft X-Ray Selected Active Galactic Nuclei Observed by Swift}",
      journal = {\apjs},
         year = 2010,
        month = mar,
       volume = {187},
       number = {1},
        pages = {64-106},
          doi = {10.1088/0067-0049/187/1/64},
archivePrefix = {arXiv},
       eprint = {1001.3140},
 primaryClass = {astro-ph.CO},
       adsurl = {https://ui.adsabs.harvard.edu/abs/2010ApJS..187...64G}
}

@ARTICLE{Gallo2011,
       author = {{Gallo}, L.~C. and {Grupe}, D. and {Schartel}, N. and {Komossa}, S. and {Miniutti}, G. and {Fabian}, A.~C. and {Santos-Lleo}, M.},
        title = "{The quasar PG 0844+349 in an X-ray weak state}",
      journal = {\mnras},
         year = 2011,
        month = mar,
       volume = {412},
       number = {1},
        pages = {161-170},
          doi = {10.1111/j.1365-2966.2010.17894.x},
archivePrefix = {arXiv},
       eprint = {1010.4453},
 primaryClass = {astro-ph.HE},
       adsurl = {https://ui.adsabs.harvard.edu/abs/2011MNRAS.412..161G}
}

@ARTICLE{Liu2021,
       author = {{Liu}, Hezhen and {Luo}, B. and {Brandt}, W.~N. and {Brotherton}, Michael S. and {Gallagher}, S.~C. and {Ni}, Q. and {Shemmer}, Ohad and {Timlin}, III, J.~D.},
        title = "{On the Observational Difference between the Accretion Disk-Corona Connections among Super- and Sub-Eddington Accreting Active Galactic Nuclei}",
      journal = {\apj},
         year = 2021,
        month = apr,
       volume = {910},
       number = {2},
          eid = {103},
        pages = {103},
          doi = {10.3847/1538-4357/abe37f},
archivePrefix = {arXiv},
       eprint = {2102.02832},
 primaryClass = {astro-ph.GA},
       adsurl = {https://ui.adsabs.harvard.edu/abs/2021ApJ...910..103L}
}

@ARTICLE{Cackett2020,
       author = {{Cackett}, Edward M. and {Gelbord}, Jonathan and {Li}, Yan-Rong and {Horne}, Keith and {Wang}, Jian-Min and {Barth}, Aaron J. and {Bai}, Jin-Ming and {Bian}, Wei-Hao and {Carroll}, Russell W. and {Du}, Pu and {Edelson}, Rick and {Goad}, Michael R. and {Ho}, Luis C. and {Hu}, Chen and {Khatu}, Viraja C. and {Luo}, Bin and {Miller}, Jake and {Yuan}, Ye-Fei},
        title = "{Supermassive Black Holes with High Accretion Rates in Active Galactic Nuclei. XI. Accretion Disk Reverberation Mapping of Mrk 142}",
      journal = {\apj},
         year = 2020,
        month = jun,
       volume = {896},
       number = {1},
          eid = {1},
        pages = {1},
          doi = {10.3847/1538-4357/ab91b5},
archivePrefix = {arXiv},
       eprint = {2005.03685},
 primaryClass = {astro-ph.HE},
       adsurl = {https://ui.adsabs.harvard.edu/abs/2020ApJ...896....1C}
}

@ARTICLE{Inayoshi2025,
       author = {{Inayoshi}, Kohei and {Kimura}, Shigeo S. and {Noda}, Hirofumi},
        title = "{Weakness of X-rays and variability in high-redshift active galactic nuclei with super-Eddington accretion Get access Arrow}",
      journal = {\pasj},
         year = 2025,
        month = jun,
       volume = {77},
       number = {4},
        pages = {811-822},
          doi = {10.1093/pasj/psaf050},
archivePrefix = {arXiv},
       eprint = {2412.03653},
 primaryClass = {astro-ph.HE},
       adsurl = {https://ui.adsabs.harvard.edu/abs/2025PASJ...77..811I}
}

@ARTICLE{Trefoloni2023,
       author = {{Trefoloni}, Bartolomeo and {Lusso}, Elisabeta and {Nardini}, Emanuele and {Risaliti}, Guido and {Bargiacchi}, Giada and {Bisogni}, Susanna and {Civano}, Francesca M. and {Elvis}, Martin and {Fabbiano}, Giuseppina and {Gilli}, Roberto and {Marconi}, Alessandro and {Richards}, Gordon T. and {Sacchi}, Andrea and {Salvestrini}, Francesco and {Signorini}, Matilde and {Vignali}, Cristian},
        title = "{The most luminous blue quasars at 3.0 < z < 3.3. III. LBT spectra and accretion parameters}",
      journal = {\aap},
         year = 2023,
        month = sep,
       volume = {677},
          eid = {A111},
        pages = {A111},
          doi = {10.1051/0004-6361/202346024},
archivePrefix = {arXiv},
       eprint = {2305.07699},
 primaryClass = {astro-ph.GA},
       adsurl = {https://ui.adsabs.harvard.edu/abs/2023A&A...677A.111T}
}

@ARTICLE{Gallo2007,
       author = {{Gallo}, L.~C. and {Brandt}, W.~N. and {Costantini}, E. and {Fabian}, A.~C. and {Iwasawa}, K. and {Papadakis}, I.~E.},
        title = "{A longer XMM-Newton look at I Zwicky 1: variability of the X-ray continuum, absorption and iron K{\ensuremath{\alpha}} line}",
      journal = {\mnras},
         year = 2007,
        month = may,
       volume = {377},
       number = {1},
        pages = {391-401},
          doi = {10.1111/j.1365-2966.2007.11601.x},
archivePrefix = {arXiv},
       eprint = {astro-ph/0610283},
 primaryClass = {astro-ph},
       adsurl = {https://ui.adsabs.harvard.edu/abs/2007MNRAS.377..391G}
}

@ARTICLE{Kang2023,
       author = {{Kang}, Jia-Lai and {Wang}, Jun-Xian},
        title = "{On joint analysing XMM-NuSTAR spectra of active galactic nuclei}",
      journal = {arXiv e-prints},
         year = 2023,
        month = nov,
          eid = {arXiv:2311.15499},
        pages = {arXiv:2311.15499},
          doi = {10.48550/arXiv.2311.15499},
archivePrefix = {arXiv},
       eprint = {2311.15499},
 primaryClass = {astro-ph.HE},
       adsurl = {https://ui.adsabs.harvard.edu/abs/2023arXiv231115499K}
}

@ARTICLE{Tanaka2004,
       author = {{Tanaka}, Yasuo and {Boller}, Thomas and {Gallo}, Luigi and {Keil}, Ralph and {Ueda}, Yoshihiro},
        title = "{Partial Covering Interpretation of the X-Ray Spectrum of the NLS1 1H 0707-495}",
      journal = {\pasj},
         year = 2004,
        month = jun,
       volume = {56},
        pages = {L9-L13},
          doi = {10.1093/pasj/56.3.L9},
archivePrefix = {arXiv},
       eprint = {astro-ph/0405158},
 primaryClass = {astro-ph},
       adsurl = {https://ui.adsabs.harvard.edu/abs/2004PASJ...56L...9T}
}

@ARTICLE{Turner2009,
       author = {{Turner}, T.~J. and {Miller}, L. and {Kraemer}, S.~B. and {Reeves}, J.~N. and {Pounds}, K.~A.},
        title = "{Suzaku Observation of a Hard Excess in 1H 0419 - 577: Detection of a Compton-Thick Partial-Covering Absorber}",
      journal = {\apj},
         year = 2009,
        month = jun,
       volume = {698},
       number = {1},
        pages = {99-105},
          doi = {10.1088/0004-637X/698/1/99},
archivePrefix = {arXiv},
       eprint = {0903.4347},
 primaryClass = {astro-ph.CO},
       adsurl = {https://ui.adsabs.harvard.edu/abs/2009ApJ...698...99T}
}

@ARTICLE{Ni2018,
       author = {{Ni}, Q. and {Brandt}, W.~N. and {Luo}, B. and {Hall}, P.~B. and {Shen}, Yue and {Anderson}, S.~F. and {Plotkin}, R.~M. and {Richards}, Gordon T. and {Schneider}, D.~P. and {Shemmer}, O. and {Wu}, Jianfeng},
        title = "{Connecting the X-ray properties of weak-line and typical quasars: testing for a geometrically thick accretion disk}",
      journal = {\mnras},
         year = 2018,
        month = nov,
       volume = {480},
       number = {4},
        pages = {5184-5202},
          doi = {10.1093/mnras/sty1989},
archivePrefix = {arXiv},
       eprint = {1807.08757},
 primaryClass = {astro-ph.GA},
       adsurl = {https://ui.adsabs.harvard.edu/abs/2018MNRAS.480.5184N}
}

@ARTICLE{Fabian2015,
       author = {{Fabian}, A.~C. and {Lohfink}, A. and {Kara}, E. and {Parker}, M.~L. and {Vasudevan}, R. and {Reynolds}, C.~S.},
        title = "{Properties of AGN coronae in the NuSTAR era}",
      journal = {\mnras},
         year = 2015,
        month = aug,
       volume = {451},
       number = {4},
        pages = {4375-4383},
          doi = {10.1093/mnras/stv1218},
archivePrefix = {arXiv},
       eprint = {1505.07603},
 primaryClass = {astro-ph.HE},
       adsurl = {https://ui.adsabs.harvard.edu/abs/2015MNRAS.451.4375F}
}

@ARTICLE{Shemmer2014,
       author = {{Shemmer}, Ohad and {Brandt}, W.~N. and {Paolillo}, Maurizio and {Kaspi}, Shai and {Vignali}, Cristian and {Stein}, Matthew S. and {Lira}, Paulina and {Schneider}, Donald P. and {Gibson}, Robert R.},
        title = "{Exploratory X-Ray Monitoring of Luminous Radio-quiet Quasars at High Redshift: Initial Results}",
      journal = {\apj},
         year = 2014,
        month = mar,
       volume = {783},
       number = {2},
          eid = {116},
        pages = {116},
          doi = {10.1088/0004-637X/783/2/116},
archivePrefix = {arXiv},
       eprint = {1401.5496},
 primaryClass = {astro-ph.CO},
       adsurl = {https://ui.adsabs.harvard.edu/abs/2014ApJ...783..116S}
}

@ARTICLE{Dai2010,
       author = {{Dai}, X. and {Kochanek}, C.~S. and {Chartas}, G. and {Koz{\l}owski}, S. and {Morgan}, C.~W. and {Garmire}, G. and {Agol}, E.},
        title = "{The Sizes of the X-ray and Optical Emission Regions of RXJ 1131-1231}",
      journal = {\apj},
         year = 2010,
        month = jan,
       volume = {709},
       number = {1},
        pages = {278-285},
          doi = {10.1088/0004-637X/709/1/278},
archivePrefix = {arXiv},
       eprint = {0906.4342},
 primaryClass = {astro-ph.HE},
       adsurl = {https://ui.adsabs.harvard.edu/abs/2010ApJ...709..278D}
}

@misc{Hu2025,
      title={Clumpy Outflows from Super-Eddington Accreting Black Holes I: Radiation Hydrodynamics Simulations and Observational Implications}, 
      author={Haojie Hu and Yuta Asahina and Shogo Yoshioka and Hiroyuki R. Takahashi and Ken Ohsuga},
      year={2025},
      eprint={2510.17696},
      archivePrefix={arXiv},
      primaryClass={astro-ph.HE},
      url={https://arxiv.org/abs/2510.17696}, 
}

@ARTICLE{Brenneman2025,
       author = {{Brenneman}, Laura W. and {Wilkins}, Daniel R. and {Ogorza{\l}ek}, Anna and {Rogantini}, Daniele and {Fabian}, Andrew C. and {Garc{\'\i}a}, Javier A. and {Jur{\'a}{\v{n}}ov{\'a}}, Anna and {Mizumoto}, Misaki and {Noda}, Hirofumi and {Behar}, Ehud and {Boissay-Malaquin}, Rozenn and {Guainazzi}, Matteo and {Okajima}, Takashi and {Hoffman}, Erika and {Keshet}, Noa and {Kaastra}, Jelle and {Kara}, Erin and {Yamauchi}, Makoto},
        title = "{A Sharper View of the X-Ray Spectrum of MCG─6-30-15 with XRISM, XMM-Newton, and NuSTAR}",
      journal = {\apj},
         year = 2025,
        month = dec,
       volume = {995},
       number = {2},
          eid = {200},
        pages = {200},
          doi = {10.3847/1538-4357/ae1225},
archivePrefix = {arXiv},
       eprint = {2510.08926},
 primaryClass = {astro-ph.HE},
       adsurl = {https://ui.adsabs.harvard.edu/abs/2025ApJ...995..200B}
}

@article{Marinucci2014,
doi = {10.1088/0004-637X/787/1/83},
url = {https://doi.org/10.1088/0004-637X/787/1/83},
year = {2014},
month = {may},
publisher = {The American Astronomical Society},
volume = {787},
number = {1},
pages = {83},
author = {Marinucci, A. and Matt, G. and Miniutti, G. and Guainazzi, M. and Parker, M. L. and Brenneman, L. and Fabian, A. C. and Kara, E. and Arevalo, P. and Ballantyne, D. R. and Boggs, S. E. and Cappi, M. and Christensen, F. E. and Craig, W. W. and Elvis, M. and Hailey, C. J. and Harrison, F. A. and Reynolds, C. S. and Risaliti, G. and Stern, D. K. and Walton, D. J. and Zhang, W.},
title = {THE BROADBAND SPECTRAL VARIABILITY OF MCG−6-30-15 OBSERVED BY NUSTAR AND XMM-NEWTON},
journal = {The Astrophysical Journal}
}

@ARTICLE{Miller2009,
       author = {{Miller}, L. and {Turner}, T.~J. and {Reeves}, J.~N.},
        title = "{The absorption-dominated model for the X-ray spectra of typeI active galaxies: MCG-6-30-15}",
      journal = {\mnras},
         year = 2009,
        month = oct,
       volume = {399},
       number = {1},
        pages = {L69-L73},
          doi = {10.1111/j.1745-3933.2009.00726.x},
archivePrefix = {arXiv},
       eprint = {0907.3114},
 primaryClass = {astro-ph.HE},
       adsurl = {https://ui.adsabs.harvard.edu/abs/2009MNRAS.399L..69M}
}

@ARTICLE{Fabian2012,
       author = {{Fabian}, A.~C. and {Zoghbi}, A. and {Wilkins}, D. and {Dwelly}, T. and {Uttley}, P. and {Schartel}, N. and {Miniutti}, G. and {Gallo}, L. and {Grupe}, D. and {Komossa}, S. and {Santos-Lle{\'o}}, M.},
        title = "{1H 0707-495 in 2011: an X-ray source within a gravitational radius of the event horizon}",
      journal = {\mnras},
         year = 2012,
        month = jan,
       volume = {419},
       number = {1},
        pages = {116-123},
          doi = {10.1111/j.1365-2966.2011.19676.x},
archivePrefix = {arXiv},
       eprint = {1108.5988},
 primaryClass = {astro-ph.HE},
       adsurl = {https://ui.adsabs.harvard.edu/abs/2012MNRAS.419..116F}
}

@ARTICLE{Jiang2018,
       author = {{Jiang}, J. and {Parker}, M.~L. and {Fabian}, A.~C. and {Alston}, W.~N. and {Buisson}, D.~J.~K. and {Cackett}, E.~M. and {Chiang}, C.-Y. and {Dauser}, T. and {Gallo}, L.~C. and {Garc{\'\i}a}, J.~A. and {Harrison}, F.~A. and {Lohfink}, A.~M. and {De Marco}, B. and {Kara}, E. and {Miller}, J.~M. and {Miniutti}, G. and {Pinto}, C. and {Walton}, D.~J. and {Wilkins}, D.~R.},
        title = "{The 1.5 Ms observing campaign on IRAS 13224-3809 - I. X-ray spectral analysis}",
      journal = {\mnras},
         year = 2018,
        month = jul,
       volume = {477},
       number = {3},
        pages = {3711-3726},
          doi = {10.1093/mnras/sty836},
archivePrefix = {arXiv},
       eprint = {1804.00349},
 primaryClass = {astro-ph.HE},
       adsurl = {https://ui.adsabs.harvard.edu/abs/2018MNRAS.477.3711J}
}

@ARTICLE{Parker2014,
       author = {{Parker}, M.~L. and {Wilkins}, D.~R. and {Fabian}, A.~C. and {Grupe}, D. and {Dauser}, T. and {Matt}, G. and {Harrison}, F.~A. and {Brenneman}, L. and {Boggs}, S.~E. and {Christensen}, F.~E. and {Craig}, W.~W. and {Gallo}, L.~C. and {Hailey}, C.~J. and {Kara}, E. and {Komossa}, S. and {Marinucci}, A. and {Miller}, J.~M. and {Risaliti}, G. and {Stern}, D. and {Walton}, D.~J. and {Zhang}, W.~W.},
        title = "{The NuSTAR spectrum of Mrk 335: extreme relativistic effects within two gravitational radii of the event horizon?}",
      journal = {\mnras},
         year = 2014,
        month = sep,
       volume = {443},
       number = {2},
        pages = {1723-1732},
          doi = {10.1093/mnras/stu1246},
archivePrefix = {arXiv},
       eprint = {1407.8223},
 primaryClass = {astro-ph.HE},
       adsurl = {https://ui.adsabs.harvard.edu/abs/2014MNRAS.443.1723P}
}

@ARTICLE{Tee2025,
       author = {{Tee}, Wei Leong and {Fan}, Xiaohui and {Wang}, Feige and {Yang}, Jinyi},
        title = "{Lack of Rest-frame Ultraviolet Variability in Little Red Dots Based on HST and JWST Observations}",
      journal = {\apjl},
         year = 2025,
        month = apr,
       volume = {983},
       number = {1},
          eid = {L26},
        pages = {L26},
          doi = {10.3847/2041-8213/adc5e3},
archivePrefix = {arXiv},
       eprint = {2412.05242},
 primaryClass = {astro-ph.GA},
       adsurl = {https://ui.adsabs.harvard.edu/abs/2025ApJ...983L..26T}
}

@ARTICLE{Fan1999,
       author = {{Fan}, Xiaohui and {Strauss}, Michael A. and {Gunn}, James E. and {Lupton}, Robert H. and {Carilli}, C.~L. and {Rupen}, M.~P. and {Schmidt}, Gary D. and {Moustakas}, Leonidas A. and {Davis}, Marc and {Annis}, James and et al.},
        title = "{The Discovery of a High-Redshift Quasar without Emission Lines from Sloan Digital Sky Survey Commissioning Data}",
      journal = {\apjl},
         year = 1999,
        month = dec,
       volume = {526},
       number = {2},
        pages = {L57-L60},
          doi = {10.1086/312382},
archivePrefix = {arXiv},
       eprint = {astro-ph/9910001},
 primaryClass = {astro-ph},
       adsurl = {https://ui.adsabs.harvard.edu/abs/1999ApJ...526L..57F}
}

@ARTICLE{Diamond-Stanic2009,
       author = {{Diamond-Stanic}, Aleksandar M. and {Fan}, Xiaohui and {Brandt}, W.~N. and {Shemmer}, Ohad and {Strauss}, Michael A. and {Anderson}, Scott F. and {Carilli}, Christopher L. and {Gibson}, Robert R. and {Jiang}, Linhua and {Kim}, J. Serena and et al.},
        title = "{High-redshift SDSS Quasars with Weak Emission Lines}",
      journal = {\apj},
         year = 2009,
        month = jul,
       volume = {699},
       number = {1},
        pages = {782-799},
          doi = {10.1088/0004-637X/699/1/782},
archivePrefix = {arXiv},
       eprint = {0904.2181},
 primaryClass = {astro-ph.GA},
       adsurl = {https://ui.adsabs.harvard.edu/abs/2009ApJ...699..782D}
}

@ARTICLE{Shemmer2010,
       author = {{Shemmer}, Ohad and {Trakhtenbrot}, Benny and {Anderson}, Scott F. and {Brandt}, W.~N. and {Diamond-Stanic}, Aleksandar M. and {Fan}, Xiaohui and {Lira}, Paulina and {Netzer}, Hagai and {Plotkin}, Richard M. and {Richards}, Gordon T. and et al.},
        title = "{Weak Line Quasars at High Redshift: Extremely High Accretion Rates or Anemic Broad-line Regions?}",
      journal = {\apjl},
         year = 2010,
        month = oct,
       volume = {722},
       number = {2},
        pages = {L152-L156},
          doi = {10.1088/2041-8205/722/2/L152},
archivePrefix = {arXiv},
       eprint = {1009.2091},
 primaryClass = {astro-ph.CO},
       adsurl = {https://ui.adsabs.harvard.edu/abs/2010ApJ...722L.152S}
}

@ARTICLE{Plotkin2010,
       author = {{Plotkin}, Richard M. and {Anderson}, Scott F. and {Brandt}, W.~N. and {Diamond-Stanic}, Aleksandar M. and {Fan}, Xiaohui and {MacLeod}, Chelsea L. and {Schneider}, Donald P. and {Shemmer}, Ohad},
        title = "{Multiwavelength Observations of Radio-quiet Quasars with Weak Emission Lines}",
      journal = {\apj},
         year = 2010,
        month = sep,
       volume = {721},
       number = {1},
        pages = {562-575},
          doi = {10.1088/0004-637X/721/1/562},
archivePrefix = {arXiv},
       eprint = {1007.5058},
 primaryClass = {astro-ph.CO},
       adsurl = {https://ui.adsabs.harvard.edu/abs/2010ApJ...721..562P}
}

@ARTICLE{Giustini2019,
       author = {{Giustini}, Margherita and {Proga}, Daniel},
        title = "{A global view of the inner accretion and ejection flow around super massive black holes. Radiation-driven accretion disk winds in a physical context}",
      journal = {\aap},
         year = 2019,
        month = oct,
       volume = {630},
          eid = {A94},
        pages = {A94},
          doi = {10.1051/0004-6361/201833810},
archivePrefix = {arXiv},
       eprint = {1904.07341},
 primaryClass = {astro-ph.GA},
       adsurl = {https://ui.adsabs.harvard.edu/abs/2019A&A...630A..94G}
}

@ARTICLE{de-Graaff2025,
       author = {{de Graaff}, Anna and {Hviding}, Raphael E. and {Naidu}, Rohan P. and {Greene}, Jenny E. and {Miller}, Tim B. and {Leja}, Joel and {Matthee}, Jorryt and {Brammer}, Gabriel and {Katz}, Harley and {Bezanson}, Rachel and {Boogaard}, Leindert A. and {Bose}, Sownak and {Chisholm}, John and {Cleri}, Nikko J. and {Dayal}, Pratika and {Feldmann}, Robert and {Fudamoto}, Yoshinobu and {Fujimoto}, Seiji and {Furtak}, Lukas J. and {Glazebrook}, Karl and {Gottumukkala}, Rashmi and {Heintz}, Kasper E. and {Kokorev}, Vasily and {Labbe}, Ivo and {Maseda}, Michael V. and {McConachie}, Ian and {Nanayakkara}, Themiya and {Nelson}, Erica and {Nowaczyk}, Przemys{\l}aw and {Oesch}, Pascal A. and {Rix}, Hans-Walter and {Setton}, David J. and {Torralba}, Alberto and {Walter}, Fabian and {Wang}, Bingjie and {Weibel}, Andrea and {van der Wel}, Arjen},
        title = "{Little Red Dots host Black Hole Stars: A unified family of gas-reddened AGN revealed by JWST/NIRSpec spectroscopy}",
      journal = {arXiv e-prints},
         year = 2025,
        month = nov,
          eid = {arXiv:2511.21820},
        pages = {arXiv:2511.21820},
          doi = {10.48550/arXiv.2511.21820},
archivePrefix = {arXiv},
       eprint = {2511.21820},
 primaryClass = {astro-ph.GA},
       adsurl = {https://ui.adsabs.harvard.edu/abs/2025arXiv251121820D}
}

@ARTICLE{Torralba2025,
       author = {{Torralba}, Alberto and {Matthee}, Jorryt and {Pezzulli}, Gabriele and {Naidu}, Rohan P. and {Ishikawa}, Yuzo and {Brammer}, Gabriel B. and {Chang}, Seok-Jun and {Chisholm}, John and {de Graaff}, Anna and {D'Eugenio}, Francesco and {Di Cesare}, Claudia and {Eilers}, Anna-Christina and {Greene}, Jenny E. and {Gronke}, Max and {Iani}, Edoardo and {Kokorev}, Vasily and {Kotiwale}, Gauri and {Kramarenko}, Ivan and {Ma}, Yilun and {Mascia}, Sara and {Navarrete}, Benjam{\'\i}n and {Nelson}, Erica and {Oesch}, Pascal and {Simcoe}, Robert A. and {Wuyts}, Stijn},
        title = "{The warm outer layer of a Little Red Dot as the source of [Fe II] and collisional Balmer lines with scattering wings}",
      journal = {arXiv e-prints},
         year = 2025,
        month = sep,
          eid = {arXiv:2510.00103},
        pages = {arXiv:2510.00103},
          doi = {10.48550/arXiv.2510.00103},
archivePrefix = {arXiv},
       eprint = {2510.00103},
 primaryClass = {astro-ph.GA},
       adsurl = {https://ui.adsabs.harvard.edu/abs/2025arXiv251000103T}
}

@ARTICLE{Kokorev2025,
       author = {{Kokorev}, Vasily and {Chisholm}, John and {Naidu}, Rohan P. and {Fujimoto}, Seiji and {Atek}, Hakim and {Brammer}, Gabriel and {Finkelstein}, Steven L. and {Akins}, Hollis B. and {Berg}, Danielle A. and {Furtak}, Lukas J. and {Fei}, Qinyue and {Hsiao}, Tiger Yu-Yang and {Labb{\'e}}, Ivo and {Matthee}, Jorryt and {Mu{\~n}oz}, Julian B. and {Oesch}, Pascal A. and {Pan}, Richard and {Rinaldi}, Pierluigi and {Saldana-Lopez}, Alberto and {Schaerer}, Daniel and {Volonteri}, Marta and {Zitrin}, Adi},
        title = "{The Deepest GLIMPSE of a Dense Gas Cocoon Enshrouding a Little Red Dot}",
      journal = {arXiv e-prints},
         year = 2025,
        month = nov,
          eid = {arXiv:2511.07515},
        pages = {arXiv:2511.07515},
          doi = {10.48550/arXiv.2511.07515},
archivePrefix = {arXiv},
       eprint = {2511.07515},
 primaryClass = {astro-ph.GA},
       adsurl = {https://ui.adsabs.harvard.edu/abs/2025arXiv251107515K}
}

@ARTICLE{Inayoshi2025b,
       author = {{Inayoshi}, Kohei and {Ho}, Luis C.},
        title = "{A Critical Evaluation of the Physical Nature of the Little Red Dots}",
      journal = {arXiv e-prints},
         year = 2025,
        month = dec,
          eid = {arXiv:2512.03130},
        pages = {arXiv:2512.03130},
archivePrefix = {arXiv},
       eprint = {2512.03130},
 primaryClass = {astro-ph.GA},
       adsurl = {https://ui.adsabs.harvard.edu/abs/2025arXiv251203130I}
}

@ARTICLE{Proga2000,
       author = {{Proga}, Daniel and {Stone}, James M. and {Kallman}, Timothy R.},
        title = "{Dynamics of Line-driven Disk Winds in Active Galactic Nuclei}",
      journal = {\apj},
         year = 2000,
        month = nov,
       volume = {543},
       number = {2},
        pages = {686-696},
          doi = {10.1086/317154},
archivePrefix = {arXiv},
       eprint = {astro-ph/0005315},
 primaryClass = {astro-ph},
       adsurl = {https://ui.adsabs.harvard.edu/abs/2000ApJ...543..686P}
}

@ARTICLE{Nomura2020,
       author = {{Nomura}, Mariko and {Ohsuga}, Ken and {Done}, Chris},
        title = "{Line-driven disc wind in near-Eddington active galactic nuclei: decrease of mass accretion rate due to powerful outflow}",
      journal = {\mnras},
         year = 2020,
        month = may,
       volume = {494},
       number = {3},
        pages = {3616-3626},
          doi = {10.1093/mnras/staa948},
archivePrefix = {arXiv},
       eprint = {1811.01966},
 primaryClass = {astro-ph.HE},
       adsurl = {https://ui.adsabs.harvard.edu/abs/2020MNRAS.494.3616N}
}

@ARTICLE{Dong2012,
       author = {{Dong}, Ruobing and {Greene}, Jenny E. and {Ho}, Luis C.},
        title = "{X-Ray Properties of Intermediate-mass Black Holes in Active Galaxies. III. Spectral Energy Distribution and Possible Evidence for Intrinsically X-Ray-weak Active Galactic Nuclei}",
      journal = {\apj},
         year = 2012,
        month = dec,
       volume = {761},
       number = {1},
          eid = {73},
        pages = {73},
          doi = {10.1088/0004-637X/761/1/73},
archivePrefix = {arXiv},
       eprint = {1210.6653},
 primaryClass = {astro-ph.GA},
       adsurl = {https://ui.adsabs.harvard.edu/abs/2012ApJ...761...73D}
}

@ARTICLE{Wang2004,
       author = {{Wang}, Jian-Min and {Qu}, Jin-Lu and {Xue}, Sui-Jian},
        title = "{The Additional Line Component within the Iron K{\ensuremath{\alpha}} Profile in MCG -6-30-15: Evidence for Blob Ejection?}",
      journal = {\apj},
         year = 2004,
        month = jul,
       volume = {609},
       number = {1},
        pages = {107-112},
          doi = {10.1086/420892},
archivePrefix = {arXiv},
       eprint = {astro-ph/0407124},
 primaryClass = {astro-ph},
       adsurl = {https://ui.adsabs.harvard.edu/abs/2004ApJ...609..107W}
}

@ARTICLE{Drewes2026,
       author = {{Drewes}, Farin and {Vieliute}, Roberta and {Hern{\'a}ndez Santisteban}, Juan V. and {Horne}, Keith and {Barth}, Aaron J. and {Cackett}, Edward M. and {Romero Colmenero}, Encarni and {Goad}, Michael R. and {Kaspi}, Shai and {Landt}, Hermine and {Lira}, Paulina and {Netzer}, Hagai and {Vestergaard}, Marianne and {Winkler}, Hartmut},
        title = "{A Phenomenological Study of the Accretion Disk in the Super-Eddington AGN I Zw 1}",
      journal = {\mnras},
         year = 2026,
        month = jan,
          doi = {10.1093/mnras/stag067},
archivePrefix = {arXiv},
       eprint = {2601.05818},
 primaryClass = {astro-ph.GA},
       adsurl = {https://ui.adsabs.harvard.edu/abs/2026MNRAS.tmp...64D}
}

@ARTICLE{Boller1996,
       author = {{Boller}, T. and {Brandt}, W.~N. and {Fink}, H.},
        title = "{Soft X-ray properties of narrow-line Seyfert 1 galaxies.}",
      journal = {\aap},
         year = 1996,
        month = jan,
       volume = {305},
        pages = {53},
          doi = {10.48550/arXiv.astro-ph/9504093},
archivePrefix = {arXiv},
       eprint = {astro-ph/9504093},
 primaryClass = {astro-ph},
       adsurl = {https://ui.adsabs.harvard.edu/abs/1996A&A...305...53B}
}

@ARTICLE{Juranova2024,
       author = {{Jur{\'a}{\v{n}}ov{\'a}}, A. and {Costantini}, E. and {Kriss}, G.~A. and {Mehdipour}, M. and {Brandt}, W.~N. and {Di Gesu}, L. and {Fabian}, A.~C. and {Gallo}, L. and {Giustini}, M. and {Rogantini}, D. and et al.},
        title = "{The outflowing ionised gas of I Zw 1 observed by HST COS}",
      journal = {\aap},
         year = 2024,
        month = jun,
       volume = {686},
          eid = {A99},
        pages = {A99},
          doi = {10.1051/0004-6361/202449544},
archivePrefix = {arXiv},
       eprint = {2404.10060},
 primaryClass = {astro-ph.GA},
       adsurl = {https://ui.adsabs.harvard.edu/abs/2024A&A...686A..99J}
}

@ARTICLE{Vestergaard2001,
       author = {{Vestergaard}, M. and {Wilkes}, B.~J.},
        title = "{An Empirical Ultraviolet Template for Iron Emission in Quasars as Derived from I Zwicky 1}",
      journal = {\apjs},
         year = 2001,
        month = may,
       volume = {134},
       number = {1},
        pages = {1-33},
          doi = {10.1086/320357},
archivePrefix = {arXiv},
       eprint = {astro-ph/0104320},
 primaryClass = {astro-ph},
       adsurl = {https://ui.adsabs.harvard.edu/abs/2001ApJS..134....1V}
}

@ARTICLE{Peterson1998,
       author = {{Peterson}, Bradley M. and {Wanders}, Ignaz and {Horne}, Keith and {Collier}, Stefan and {Alexander}, Tal and {Kaspi}, Shai and {Maoz}, Dan},
        title = "{On Uncertainties in Cross-Correlation Lags and the Reality of Wavelength-dependent Continuum Lags in Active Galactic Nuclei}",
      journal = {\pasp},
         year = 1998,
        month = jun,
       volume = {110},
       number = {748},
        pages = {660-670},
          doi = {10.1086/316177},
archivePrefix = {arXiv},
       eprint = {astro-ph/9802103},
 primaryClass = {astro-ph},
       adsurl = {https://ui.adsabs.harvard.edu/abs/1998PASP..110..660P}
}

@ARTICLE{2018JOSS....3..695M,
       author = {{Green}, {Gregory M.}},
        title = "{dustmaps: A Python interface for maps of interstellar dust}",
      journal = {The Journal of Open Source Software},
         year = "2018",
        month = "Jun",
       volume = {3},
       number = {26},
        pages = {695},
          oridoi = {10.21105/joss.00695},
       adsurl = {https://ui.adsabs.harvard.edu/abs/2018JOSS....3..695G}
}

@ARTICLE{astropy:2018,
       author = {{Astropy Collaboration} and {Price-Whelan}, A.~M. and {Sip{\H{o}}cz}, B.~M. and {G{\"u}nther}, H.~M. and {Lim}, P.~L. and {Crawford}, S.~M. and {Conseil}, S. and {Shupe}, D.~L. and {Craig}, M.~W. and {Dencheva}, N. and {Ginsburg}, A. and {VanderPlas}, J.~T. and {Bradley}, L.~D. and {P{\'e}rez-Su{\'a}rez}, D. and {de Val-Borro}, M. and {Aldcroft}, T.~L. and {Cruz}, K.~L. and {Robitaille}, T.~P. and {Tollerud}, E.~J. and {Ardelean}, C. and {Babej}, T. and {Bach}, Y.~P. and {Bachetti}, M. and {Bakanov}, A.~V. and {Bamford}, S.~P. and {Barentsen}, G. and {Barmby}, P. and {Baumbach}, A. and {Berry}, K.~L. and {Biscani}, F. and {Boquien}, M. and {Bostroem}, K.~A. and {Bouma}, L.~G. and {Brammer}, G.~B. and {Bray}, E.~M. and {Breytenbach}, H. and {Buddelmeijer}, H. and {Burke}, D.~J. and {Calderone}, G. and {Cano Rodr{\'\i}guez}, J.~L. and {Cara}, M. and {Cardoso}, J.~V.~M. and {Cheedella}, S. and {Copin}, Y. and {Corrales}, L. and {Crichton}, D. and {D'Avella}, D. and {Deil}, C. and {Depagne}, {\'E}. and {Dietrich}, J.~P. and {Donath}, A. and {Droettboom}, M. and {Earl}, N. and {Erben}, T. and {Fabbro}, S. and {Ferreira}, L.~A. and {Finethy}, T. and {Fox}, R.~T. and {Garrison}, L.~H. and {Gibbons}, S.~L.~J. and {Goldstein}, D.~A. and {Gommers}, R. and {Greco}, J.~P. and {Greenfield}, P. and {Groener}, A.~M. and {Grollier}, F. and {Hagen}, A. and {Hirst}, P. and {Homeier}, D. and {Horton}, A.~J. and {Hosseinzadeh}, G. and {Hu}, L. and {Hunkeler}, J.~S. and {Ivezi{\'c}}, {\v{Z}}. and {Jain}, A. and {Jenness}, T. and {Kanarek}, G. and {Kendrew}, S. and {Kern}, N.~S. and {Kerzendorf}, W.~E. and {Khvalko}, A. and {King}, J. and {Kirkby}, D. and {Kulkarni}, A.~M. and {Kumar}, A. and {Lee}, A. and {Lenz}, D. and {Littlefair}, S.~P. and {Ma}, Z. and {Macleod}, D.~M. and {Mastropietro}, M. and {McCully}, C. and {Montagnac}, S. and {Morris}, B.~M. and {Mueller}, M. and {Mumford}, S.~J. and {Muna}, D. and {Murphy}, N.~A. and {Nelson}, S. and {Nguyen}, G.~H. and {Ninan}, J.~P. and {N{\"o}the}, M. and {Ogaz}, S. and {Oh}, S. and {Parejko}, J.~K. and {Parley}, N. and {Pascual}, S. and {Patil}, R. and {Patil}, A.~A. and {Plunkett}, A.~L. and {Prochaska}, J.~X. and {Rastogi}, T. and {Reddy Janga}, V. and {Sabater}, J. and {Sakurikar}, P. and {Seifert}, M. and {Sherbert}, L.~E. and {Sherwood-Taylor}, H. and {Shih}, A.~Y. and {Sick}, J. and {Silbiger}, M.~T. and {Singanamalla}, S. and {Singer}, L.~P. and {Sladen}, P.~H. and {Sooley}, K.~A. and {Sornarajah}, S. and {Streicher}, O. and {Teuben}, P. and {Thomas}, S.~W. and {Tremblay}, G.~R. and {Turner}, J.~E.~H. and {Terr{\'o}n}, V. and {van Kerkwijk}, M.~H. and {de la Vega}, A. and {Watkins}, L.~L. and {Weaver}, B.~A. and {Whitmore}, J.~B. and {Woillez}, J. and {Zabalza}, V. and {Astropy Contributors}},
        title = "{The Astropy Project: Building an Open-science Project and Status of the v2.0 Core Package}",
      journal = {\aj},
         year = 2018,
        month = sep,
       volume = {156},
       number = {3},
          eid = {123},
        pages = {123},
          doi = {10.3847/1538-3881/aabc4f},
archivePrefix = {arXiv},
       eprint = {1801.02634},
 primaryClass = {astro-ph.IM},
       adsurl = {https://ui.adsabs.harvard.edu/abs/2018AJ....156..123A}
}

@BOOK{Frank2002,
       author = {{Frank}, Juhan and {King}, Andrew and {Raine}, Derek J.},
        title = "{Accretion Power in Astrophysics: Third Edition}",
         year = 2002,
       adsurl = {https://ui.adsabs.harvard.edu/abs/2002apa..book.....F}
}

\end{document}